\documentclass{article}

\usepackage{arxiv}

\usepackage[utf8]{inputenc} 
\usepackage[T1]{fontenc}    
\usepackage[hidelinks]{hyperref}      
\usepackage{url}            
\usepackage{booktabs}       
\usepackage{amsfonts}       
\usepackage{nicefrac}       
\usepackage{microtype}      
\usepackage{lipsum}		
\usepackage{graphicx}
\usepackage{natbib}
\usepackage{doi}
\usepackage{amsmath}

\usepackage{algorithmic}
\usepackage{longtable}
\usepackage{multirow}
\usepackage[x11names]{xcolor}
\usepackage{cleveref}
\usepackage[acronym,toc]{glossaries}
\usepackage[inline]{enumitem}
\setlist{nosep}
\usepackage{xcolor}
\usepackage{listings}
\usepackage[table]{xcolor}

\usepackage{subcaption}
\usepackage{arydshln}

\usepackage{tabularx}

\usepackage{cleveref}
\usepackage[acronym,toc]{glossaries}
\usepackage[inline]{enumitem}
\newenvironment{inlinelist}{\begin{enumerate*}[label=\emph{(\roman{*})}]}{\end{enumerate*}}

\newcommand{\allparticipants}{17} 
\newcommand{\complete}{10} 
\newcommand{\partially}{2} 
\newcommand{\onlydemographics}{4} 
\newcommand{\contactedparticipants}{137} 
\newcommand{\susmean}{67} 
\newcommand{\susstd}{11.35} 

\usepackage{fontawesome}
\newcounter{findingcounter}
\newcommand{\finding}[1]{%
    \refstepcounter{findingcounter}%
    \textcolor{cyan!50!black}{\textbf{\faSearch~Finding~\thefindingcounter:}}~\textbf{#1}\par%
    \label{finding:\thefindingcounter}%
}

\definecolor{promptgray}{rgb}{0.95, 0.95, 0.96}
\definecolor{rolecolor}{rgb}{0.1, 0.1, 0.5}
\definecolor{commentgray}{rgb}{0.4, 0.4, 0.4}

\definecolor{scholarblue}{rgb}{0.1, 0.4, 0.8}
\definecolor{operatorcolor}{rgb}{0.8, 0.1, 0.1}

\lstdefinestyle{llmprompt}{
    backgroundcolor=\color{promptgray},
    basicstyle=\ttfamily\small\color{black},
    breakatwhitespace=false,         
    breaklines=true,                 
    captionpos=b,                    
    commentstyle=\color{commentgray}\itshape,
    escapeinside={(*}{*)},          
    frame=single,
    framerule=0pt,
    framesep=10pt,                   
    keepspaces=true,
    keywordstyle=\color{rolecolor}\bfseries,
    keywords={System:, User:, Assistant:, Role:},
    morecomment=[l]{//},             
    showspaces=false,                
    showstringspaces=false,
    showtabs=false,                  
    tabsize=2
}

\lstdefinestyle{scholar}{
    backgroundcolor=\color{gray!5},
    basicstyle=\ttfamily\small\color{black},
    breaklines=true,
    frame=leftline,          
    framerule=2pt,
    rulecolor=\color{scholarblue},
    xleftmargin=10pt,
    keywords={AND, OR, NOT, author:, source:, label:, intitle:, filetype:},
    keywordstyle=\color{operatorcolor}\bfseries,
    morestring=[b]",
    stringstyle=\color{scholarblue},
    showstringspaces=false
}

\title{Beyond the Syntax: Do Security Experts Trust LLMs for NIDS Rule Engineering?}

\author{{Lorenzo Di Filippo} \\
	Sorbonne University\\
	4 Pl. Jussieu\\
	Paris, France, 75005 \\
	\texttt{Lorenzo.Di-Filippo@lip6.fr} \\
	\And
	{Enkeleda Bardhi} \\
	Delft University of Technology\\
	Van Mourik Broekmanweg 5\\
	Delft, The Netherlands, 2628 XE \\
	\texttt{E.Bardhi-1@tudelft.nl} \\
	\And
	{Andrea Agiollo} \\
	Delft University of Technology\\
	Van Mourik Broekmanweg 5\\
	Delft, The Netherlands, 2628 XE \\
	\texttt{A.Agiollo-1@tudelft.nl} \\
    \And
	{Alessandro Palma} \\
	Sapienza University of Rome\\
	Via Ariosto 25\\
	Rome, Italy, 00185 \\
	\texttt{palma@diag.uniroma1.it} \\
    \And
	{Silvia Bonomi} \\
	Sapienza University of Rome\\
	Via Ariosto 25\\
	Rome, Italy, 00185 \\
	\texttt{bonomi@diag.uniroma1.it} \\
    \And
   {Fernando Kuipers} \\
	Delft University of Technology\\
	Van Mourik Broekmanweg 5\\
	Delft, The Netherlands, 2628 XE \\
	\texttt{F.A.Kuipers@tudelft.nl} \\
}

\date{}

\renewcommand{\shorttitle}{\textit{Beyond the Syntax}}

\hypersetup{
pdftitle={A template for the arxiv style},
pdfsubject={q-bio.NC, q-bio.QM},
pdfauthor={David S.~Hippocampus, Elias D.~Striatum},
pdfkeywords={First keyword, Second keyword, More},
}

\begin{document}
\maketitle

\begin{abstract}
As network threats evolve, manual NIDS rule engineering has become a critical operational bottleneck.
While Large Language Models (LLMs) show promise for automating this process, their ability to produce production-ready rules remains unvalidated. 
This paper presents a human-centered investigation into LLM-based NIDS rule engineering, formalizing a grounded generation framework and evaluating it through a user study with 10 domain experts.

Our evaluation reveals a syntax-semantics paradox: although LLMs generate syntactically correct rules, experts find them only partially deployable due to low specificity and logic hallucinations in 12\% of cases. 
While the system received a favorable SUS score of 67, practitioners remain skeptical of its autonomous capabilities, viewing LLMs as support tools for drafting and verification rather than independent generators.
Finally, our statistical analysis indicates that while large-scale models (($\geq 70B$) consistently produce syntactically valid rules, small models ($\leq 4B$) are largely ineffective for IDS rule generation.
\end{abstract}


\newacronym{nids}{NIDS}{Network Intrusion Detection System}
\newacronym{soc}{SOCs}{Security Operations Centers}

\newacronym{ml}{ML}{Machine Learning}
\newacronym{llm}{LLM}{Large Language Models}

\newacronym{poc}{PoC}{Proof-of-Concept}
\newacronym{cve}{CVE}{Common Vulnerabilities and Exposures}
\newacronym{cot}{CoT}{Chain-of-Thought}

\newacronym{sus}{SUS}{System Usability Scale}

\section{Introduction}
\label{sec:intro}

In the era of distributed and agentic systems where cyber threats evolve at unprecedented speed, \glspl{nids} serve as the first line of defense, continuously monitoring network traffic to detect malicious activities, thus being the cornerstone of modern network defense~\cite{KhraisatCybersecurity2019,vermeer2023alert,ZhangIBSZSO24,BardhiJISLCK25}. 
Open-source solutions like Snort\footnote{\url{https://www.snort.org}} and Suricata\footnote{\url{https://suricata.io}} have established as industry standards, distinguished by their robust packet processing capabilities, extensive rule communities, and multi-threaded architecture. 
%
These systems rely on signature-based detection engines, where network traffic is analyzed against databases of predefined rules. 
Each rule comprises a header specifying protocol, IP addresses and ports, alongside an option section encoding detection logic. 
While this engine delivers high-fidelity detection of known threats, its efficacy remains contingent upon the quality, timeliness, and comprehensive coverage of deployed rulesets.
These qualities, in turn, rely on significant human effort to create and maintain such rulesets~\cite{10.1145/3719027.3744800,goyal2023sometimes}.
This is further exacerbated by the escalating complexity of contemporary threat landscapes, including polymorphic malware, zero-day exploits, and rapidly evolving attack infrastructure.
This complexity has transformed ruleset maintenance into a critical operational bottleneck. 
As network traffic volumes continue their exponential growth, manual rule engineering emerges not merely as labor-intensive, but fundamentally unscalable~\cite{10.1145/3576915.3616631,vermeer2023alert}. 
This challenge is well-documented as Security Operation Center (SOC) fatigue, wherein security analysts confront alert fatigue from high false positive rates, syntactic complexity of rule languages, and chronic personnel shortages~\cite{vermeer2023alert,alahmadi202299}.

Traditionally, \gls{nids} rule generation constituted a highly specialized task: it demanded deep protocol knowledge and attack semantics, as well as experience in rule syntax, since even trivial violations render rules non-functional. 
Limited expert bandwidth has created persistent vulnerability windows, as organizations struggle to maintain parity with threat evolution.
To address this limitation, the research community has pursued two evolutionary trajectories. 
First, early efforts leveraged traditional \gls{ml} techniques to extract signatures from labeled traffic datasets, framing rule generation as supervised pattern recognition~\cite{wang2024incorporating,10646671,10646673,10.1145/3576915.3616631,10646725}. 
More recently, the advent of \glspl{llm} has catalyzed a paradigm shift, with studies exploring diverse applications including data fine-tuning for rule engineering~\cite{papoutsis2025rulexploit}, synthetic data generation for anomaly detection~\cite{ali2024next}, enhanced rule explainability~\cite{alnahdi2024towards}, and direct rule generation from vulnerability contexts~\cite{mitra2025falcon}.
Despite these technical advances, significant gaps remain in understanding how \gls{llm}s can be effectively integrated into operational \gls{nids} workflows. 
Although automated systems exhibit syntactic correctness and promising quantitative performance on benchmark datasets, their practical utility as perceived and validated by domain experts has yet to be empirically assessed and, to the best of the authors’ knowledge, no prior work has examined security professionals’ trust in \gls{llm}-generated rules or how they interact with such outputs.

To fill this gap in the literature, this paper proposes a human-centered investigation of \gls{llm}-based \gls{nids} rule generation. 
We systematically pursue the following research questions: 
\begin{inlinelist}
    \item \textbf{RQ1}: To what extent do \gls{nids} experts deem \gls{llm}-generated rules semantically correct and deployable?
    \item \textbf{RQ2}: What interaction patterns emerge when experts utilize \gls{llm}-assisted rule engineering systems, and how do these patterns correlate with perceived rule quality?
    \item \textbf{RQ3}: What is the overall usability of existing human-in-the-loop \gls{llm} functions for \gls{nids} rule engineering?
    \item \textbf{RQ4}: What are the best engines and tools for a NIDS rule engineering agent to construct valid rules?
\end{inlinelist}

To answer these questions, we adopt a four-phase methodology. 
First, we conduct a comprehensive literature review of \gls{llm}-\gls{nids} intersections,
formalizing prevalent system architectures and identifying common inputs. 
Second, we validate the system components and inputs through a pre-study with three domain experts. 
Thirdly, we generate a set of IDS rules using the formalized LLM-based agent system which components are grounded on the outcome of the second phase.
Finally, we execute a rigorous two-fold evaluation of the generated rule, including a statistical analysis of the generated rules and a user study with security practitioners to investigate
\begin{inlinelist}
    \item their semantic assessment of \gls{llm}-generated rules,
    \item their level of trust towards the system,
    \item the practical deployability of \gls{llm}-generated rules, and
    \item the usability of the \gls{llm}-based agent.
\end{inlinelist}
This way, this paper contributes:
\begin{itemize}
    \item a formalization of existing \gls{llm}-based rule engineering functionalities grounded in literature and expert pre-study; 
    \item a first empirical characterization of expert trust in \gls{llm}-generated \gls{nids} rules, revealing critical gaps between syntactic validity and semantic deployability;
    \item novel insights into expert-\gls{llm} interaction patterns, cognitive load under network complexity, and usability barriers.
\end{itemize}

This paper is structured as follows: \Cref{sec:methodology} details the methodology, including system formalization (\Cref{sec:system_formalization}), user study (\Cref{sec:user_study}), and evaluation (\Cref{sec:evaluation}), along with participant recruitment; related work is in \Cref{sec:related_work}; conclusions are in \Cref{sec:conclusion}.

\noindent {\bf Notions:}
In this paper, we focus on \acrfull{nids}, hereafter referred to as \textit{IDS} for brevity.
Additionally, we distinguish between \textit{rule generation} -- the specific task of producing signature syntax -- and \textit{rule engineering}---the holistic process of expert review, modification, and validation. 
In the remainder of this work, these terms are applied according to the specific technical task being discussed.
Lastly, we refer to the involved individuals as \textit{participants} in the context of our experimental results, though they represent professional \textit{security experts} and \emph{rule engineers}.

\section{Methodology}
\label{sec:methodology}

\subsection{Process Overview}
\label{ssec:process_overview}

\subsubsection{System Formalization (\Cref{sec:system_formalization})}
As a first step of the methodology, we formalize a \gls{llm}-based \gls{nids} rule generation system.
Formalizing the system is necessary to ensure a consistent and reproducible basis for evaluation. 
It provides a well-defined interface, workflow, and behavior, enabling controlled assessment of expert judgments on rule correctness and deployability (RQ1), systematic analysis of interaction patterns (RQ2), and rigorous usability evaluation (RQ3).
This formalization reduces confounding factors and strengthens the validity and interpretability of the study findings.

To achieve a rigorous system formalization, we conducted a comprehensive literature review and a preliminary survey to refine our understanding of the system and study design. 
Initially, we provide a deep understanding of the core components of an \gls{llm}-based rule generation system (RQ4).
Consequently, we validate the essential system inputs through the expertise of three professional participants via a pre-study survey.
Such a validation encompasses the intricacies of rule anatomy, automation, human judgment, and the inherent challenges of the rule generation process.
With both the literature review and the insights from the pre-study survey, we formalize the system to be tested via a human-centric study.

\subsubsection{User Study (\Cref{sec:user_study})}
We design a user study aimed at assessing both the functional impact of the proposed system and usability, the quality of the \gls{llm}-generated rules, while also exploring the extent to which LLMs can support IDS rule engineering in practice. 
The study was structured to answer the research questions of this paper, mainly focusing on RQ1, RQ2 and RQ3.
Participants are asked to complete a set of representative tasks involving IDS rule validation, creation and modification, including the evaluation of LLM-generated explanations to IDS rules.
These tasks were designed to reflect realistic scenarios derived from the pre-study, and employed a combination of quantitative and qualitative measures, allowing to assess both performance improvements and the practical usefulness of LLM support.
Additionally, participants provided open feedback through statement-based and open-ended questionnaires, enabling a comprehensive evaluation of both performance and user experience, as well as deeper insights into when and how LLM assistance is perceived as beneficial or limiting.

\subsubsection{Evaluation (\Cref{sec:evaluation})}
Using the design choices elaborated previously, we contacted \contactedparticipants{} participants to complete the user study via an online survey platform. 
We received \allparticipants{} responses and used a subset of \complete{} participants given their elevated expertise for the results analysis. 
The evaluation is two-fold, including:
\begin{inlinelist}
    \item the complete statistical analysis  (\Cref{ssec:statistical_analysis}) on the rule syntax (\Cref{sssec:syntax_performance}) and rule quality (\Cref{sssec:rule_quality_analysis});
    \item the human-centered evaluation deriving as a result of \complete{} participants responses from the user study (\Cref{ssec:human_evaluation}), evaluating rule quality from humans perspective (\Cref{sssec:human_rule_quality}), system utility via a SUS-based questionnaire (\Cref{sssec:sus_evaluation}), and both statement-based and open-ended questionnaires (\Cref{sssec:statement_evaluation} and \Cref{sssec:open_evaluation}, respectively).
\end{inlinelist}

\subsection{Participant Recruitment}
\label{ssec:participant_recruitment}

To ensure the user study validity, we employed a multi-pronged recruitment strategy designed to engage high-level experts in IDS, SOC and rule engineering. 
First, we utilized purposive sampling by directly contacting established professionals within the authors' professional networks and academic circles.
This was supplemented by a targeted literature-based approach, where we reached out to the authors of works in the field of automated IDS generation (mentioned in \Cref{ssec:literature_review} and elaborated in \Cref{sec:related_work}), ensuring that the study included individuals with a deep theoretical and practical understanding of the domain.
To broaden our reach beyond immediate circles, we performed systematic discovery using two distinct methods:
\begin{inlinelist}
    \item a curated search via LLMs to identify active researchers and practitioners at the intersection of Cybersecurity, IDS, AI and LLMs. 
    \item keyword-based queries within Google Scholar to identify prolific contributors to IDS signature research.
\end{inlinelist}
Appendix~\ref{app:recruitment_details} reports specific details of the recruitment phase.
This hybrid approach allowed us to assemble a diverse cohort of experts (\contactedparticipants{}), spanning both academia and industry SOC environments, thereby capturing a comprehensive range of perspectives.

\section{System Formalization}
\label{sec:system_formalization}

\subsection{Literature Review}
\label{ssec:literature_review}
Via the literature review we collect insights on how the current research integrate LLMs into the IDS rule generation process, with a particular focus on system definition, inputs, validation and inclusion of human-in-the-loop.
We reviewed 11 primary works (detailed in~\Cref{sec:related_work} and compared in Appendix~\cref{app:compairison_sys}) tailored to automatic \gls{llm}-based IDS rule generation systems.
Our analysis revealed that the standard pipeline typically includes an input layer, a core \gls{llm} agent featuring specialized modules for pipeline-specific tasks, and diverse prompting techniques.
A subset of these works~\cite{moreno2025leveraging,mitra2025falcon} integrates post-generation components, such as syntax checkers or human-based feedback mechanisms.

The data inputs utilized across these systems vary significantly, reflecting different perspectives on what constitutes relevant information for rule generation. 
The most prevalent inputs are vulnerability specifications~\cite{hu2024llm,papoutsis2025rulexploit}, \gls{poc} exploits~\cite{lian2025rulemaster+,papoutsis2025rulexploit}, and network traffic captures (PCAPs)~\cite{du2025harnessing, li2025gridai}.
Specialized frameworks also leverage honeypot captures~\cite{balasubramanian2024hex2sign}, malicious payloads~\cite{hu2024llm}, or Cyber Threat Intelligence (CTI) and SIEM logs~\cite{wang2025rulepilot, mitra2025falcon}.

Regarding the \glspl{llm} landscape, the field is dominated by GPT-based models (GPT-3.5, GPT-4o, GPT-4o-mini) \cite{achiam2023gpt} and the Llama family \cite{touvron2023llama,grattafiori2024llama}.
Other models utilized include Claude \cite{claude}, Gemini \cite{gemini}, Falcon \cite{falcon}, Mistral \cite{mistral}, Qwen \cite{qwen}, Granite \cite{granite}, Phi4 \cite{phifour}, and DeepSeek-V3 \cite{deepseek,deepseekv3}, while one approach focuses on fine-tuning BERT-based models rather than using general-purpose LLMs~\cite{balasubramanian2024hex2sign}.

Role Prompting is the most prevalent technique, used to define the LLM's persona during rule generation. 
\gls{cot} is frequently applied to handle complex reasoning tasks~\cite{hu2024llm, wang2025rulepilot}. 
Additional methods include Zero-shot/Few-shot prompting~\cite{moreno2025leveraging} and the use of instructional templates~\cite{papoutsis2025rulexploit}. 

The majority of the analyzed systems do not include a dedicated syntax checker or corrector. 
Among those that do, Suricata parsing is the standard for validation~\cite{papoutsis2025rulexploit, moreno2025leveraging}. 
Correction is typically handled through customized correctors, LLM-based re-prompting for error fixes, or grammar checks specific to SIEM requirements.

Lastly, there is a significant lack of human-in-the-loop approaches, as ten out of the eleven papers report no human feedback mechanism. 
Only one system~\cite{mitra2025falcon} explicitly incorporates human feedback into its autonomous mining and rule generation process.

\subsection{Pre-study Survey}
\label{ssec:prestudy_survey}
Inspired by the findings of the literature review, the pre-study survey is designed to gather human perspective on the rule generation, system inputs and rule engineering process.
We recruited three participants for this study, selected specifically for their extensive professional experience in rule engineering.
\begin{table}[ht]
\centering
\scalebox{0.85}{
\begin{tabular}{c l c l}
    \toprule
    \textbf{ID} & \textbf{Role} & \textbf{Exp (Yrs)} & \textbf{Primary IDS} \\
    \midrule

    \rowcolor{gray!10}
    P1 & Security Researcher & 10+ & Suricata, Zeek \\
    \rowcolor{gray!10}
    & (former SOC Security Analyst) & & and Proprietary IDS \\
    
    P7 & SOC Security Analyst & 1--2 & Suricata \\
    & (former Security Analyst) & & \\
    
    \rowcolor{gray!10}
    P11 & SOC Threat Hunter & 3--5 & Snort, Suricata \\
    \rowcolor{gray!10}
    & & & and Proprietary IDS \\
    \bottomrule
\end{tabular}
}
\caption{Participant demographics for pre-study survey ($N=3$). P1 and P7 are a subset of the user study participants, while P11 did not participate in the final user study.}
\label{tab:preliminary_demographics}
\end{table}
\Cref{tab:preliminary_demographics} details the demographics of the survey respondents.
The participants represent diverse professional backgrounds, including the private SOCs, academia (P1 in~\Cref{tab:participants}), individuals formerly affiliated with government-based security authorities (P7 in~\Cref{tab:participants}) and industry-based experienced participant (P11). 
While the respondents' primary expertise lies with the Suricata IDS engine, they also reported proficiency with Snort, Zeek, and various proprietary IDS solutions.

\subsubsection{Rule Generation Considerations} When asked about the typical initialization of the IDS rule-writing process, all participants emphasized the need of characterizing the attack surface and define the specific detection objective. 
Furthermore, respondents noted that the strategic placement of the IDS within the network architecture is a critical factor.
One participant specifically highlighted that the directionality of network traffic must be accurately reflected within the rule logic to ensure operational efficacy, which we later consider during the rule quality evaluation in~\Cref{sssec:human_rule_quality}.

\subsubsection{System Inputs} When evaluating the information required to draft effective rules, there was a unanimous consensus on the importance of \gls{cve} descriptions. 
The majority of participants (two out of three) identified network topology, device-specific \gls{cve} inventories, and \glspl{poc} of known vulnerabilities as essential inputs for the generation process.

\subsubsection{Rule Engineering Process}\label{sssec:rule_engineering}
A standard Snort or Suricata rule structure consists of a rule header and subsequent rule options.
Through a combination of open-ended and multiple-choice questions, we gathered insights into header specificity (e.g., the explicit definition of IP addresses and ports), the importance of various option fields, and the potential for automation.
Practitioners prioritize maximum specificity in rule headers to increase detection accuracy, typically opting for specific IPs, ports, or network variables (e.g., \textsc{home\_net}) whenever the attack details allow. 
They transition to generic values (e.g., \textsc{any}) primarily when targeting broad behavioral patterns or when services run on non-standard, user-configured ports. 
Instead, when asked about the relevance of options fields, practitioners consistently identify \textsc{content} as the most critical rule option, valuing its flexibility in matching specific malicious patterns and customizing detection. 
Options that provide context and operational stability -- such as \textsc{msg} for alert interpretation and \textsc{flow} for managing computational overhead -- are also highly prioritized.
Furthermore, technical precision options like payload positioning (e.g., \textsc{offset}, \textsc{depth}) and protocol-specific fields (e.g., \textsc{http.uri}, \textsc{dns.query}) are essential for refining detection and minimizing false positives.
Defining when and how to use such fields represents a complex task, as we show in \Cref{sssec:rule_quality_analysis}. 
Practitioners see automation as a tool to generate rule skeletons, compile metadata, and suggest headers from attack keywords.

\subsection{Formalization}
\label{ssec:formalization}
\begin{figure}[h!]
    \centering
    \includegraphics[width=1\linewidth]{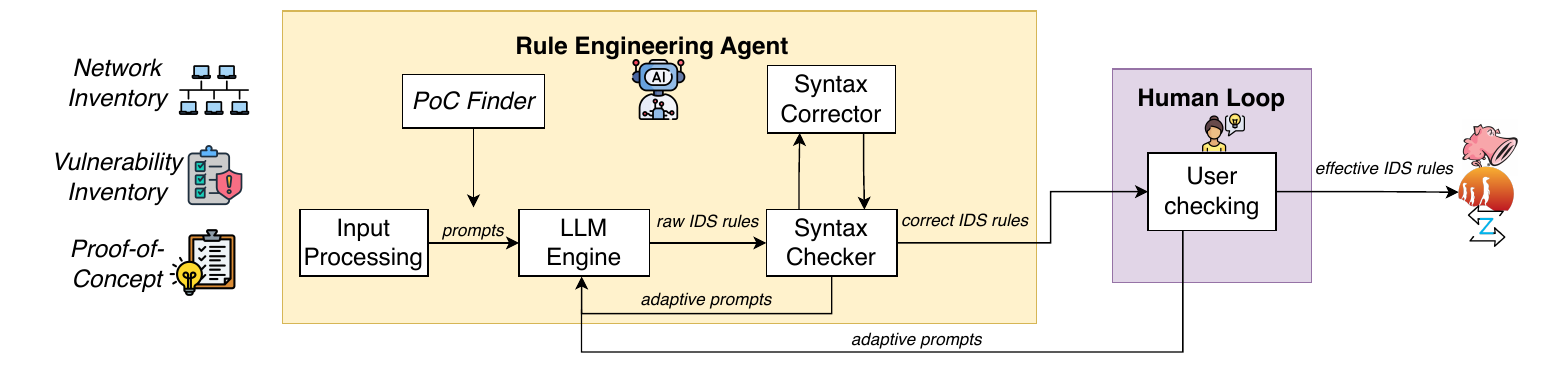}
    \caption{System formalization.}
    \label{fig:system_formalization}
\end{figure}
Based on the findings from our literature review and the pre-study survey, we formalized the system architecture illustrated in~\Cref{fig:system_formalization}.
The system comprises two primary components: an \textit{rule engineering agent}, which leverages multi-source input data and specialized engines to generate syntactically valid IDS rules, and the \textit{human loop}, which facilitates adaptive prompting to the LLM engine for iterative rule refinement.

The system ingest data from a network inventory, which provides critical context regarding device types, active services, and topological relationships.
Furthermore, the rule engineering agent is integrated with a vulnerability inventory---comprising detailed vulnerability specifications mapped to specific network devices retrieved from existing benchmark~\cite{palma_behind_2025} and the NIST catalogs (CVE\footnote{\url{https://www.cve.org/}}, CPE\footnote{\url{https://nvd.nist.gov/products/cpe}}).
Finally, where available, the agent incorporates \gls{poc} exploits associated with these vulnerabilities to ground the IDS rule generation process in realistic attack vectors.

The internal workflow of the rule engineering agent begins by preprocessing the aforementioned inputs into structured JSON templates to ensure consistent prompt engineering. 
A dedicated \gls{poc} finder tool identifies and associates relevant exploit code with the identified vulnerabilities.
This contextual data is then provided to the LLM engine using a combination of role-based and few-shot prompting techniques.
The LLM engine itself is implemented via the Ollama framework~\cite{ollama2023}, hosting a variety of open-source models for comparative benchmarking.
To ensure technical viability, the raw IDS rules generated by the LLM undergo a multi-stage validation process: 
\begin{inlinelist}
    \item \textit{syntax verification}: a syntax checker validates the rules against a native compiler of the target IDS tool;
    \item \textit{automated correction}: if a rule fails validation, it is forwarded to a Syntax Corrector module. This module utilizes adaptive prompting to provide the LLM engine with error feedback, enabling autonomous structural repairs. 
\end{inlinelist}

To enrich existing functionalities of the system, the final stage incorporates a human-in-the-loop verification mechanism.  
This module allows domain experts to perform proactive modifications via adaptive prompting.
By interacting directly with the LLM engine, the user can mitigate hallucinations and technical inaccuracies, ensuring that the outputted rules are not only syntactically sound but also semantically effective. 
The system ultimately produces deployment-ready IDS rules.

\noindent {\bf System Scope and Design Decisions.}
The formalized system is designed to encapsulate the common features of state-of-the-art frameworks and is not intended to supersede existing specialized tools.
While a human-in-the-loop component is absent in the majority of contemporary literature, it constitutes the primary focus of this work. 
Given that the central objective of this paper is a human-centric user study, we explicitly include the human loop component enabling participants to interact with the formalized system and provide feedback on the generated rules.

As established in~\Cref{ssec:literature_review}, system inputs vary significantly across existing frameworks. 
Our system formalization strategically prioritizes experts' observations from the pre-study survey (see \Cref{ssec:prestudy_survey}), including vulnerability specifications and \gls{poc} exploits.
This selection is designed to maximize the system's utility in real-world environments where network traffic packet captures (PCAPs) may be unavailable or suboptimal, specifically favoring our approach for the following reasons:
\begin{inlinelist}
    \item by focusing on vulnerability data rather than live traffic, the system remains functional although direct access to sensitive raw network streams might lack;
    \item utilizing a \gls{cve} inventory allows the system to generate rules for potential threats before they manifest in a network, offering a more proactive posture compared to the reactive nature of traffic-based analysis;
    \item prioritizing \gls{poc} logic over PCAPs from common benchmarking datasets mitigates the risk of \gls{llm} memorization \cite{tirumala2022memorization,leybzon2024learning} bias, ensuring the model generates rules through reasoning rather than retrieving training data; and,
    \item our approach leverages high-density informational inputs that fit efficiently within \gls{llm} context windows, reducing computational overhead and token latency compared to long-context PCAP processing.
\end{inlinelist}
To preserve essential network context without the weight of raw traffic, we utilize network topology information to map relationships between devices and services. 
This representation provides a more refined and compact alternative for \gls{llm} processing while maintaining high informational density. 
Furthermore, the synthesis of \gls{poc} data with topological inventories yields a more comprehensive technical understanding of an exploit’s mechanics, offering greater depth for rule engineering than the initial connection data typically extracted from traffic captures.

Finally, the selection of the Ollama framework for the LLM engine was driven by several strategic considerations.
First, we prioritized a local implementation over API-based services to facilitate the benchmarking of open-source models, thereby ensuring research reproducibility and eliminating the financial overhead of proprietary models.
Second, similar to other privacy-sensitive domains, IDS rule engineering is ideally performed using local or fine-tuned models~\cite{balasubramanian2024hex2sign, lin2026rulellm} to prevent the leakage of sensitive infrastructure data. 
Lastly, utilizing local models prevents the effect where provided prompts are used for subsequent model training, thus eliminating potential biases in our evaluation.
\section{User Study}
\label{sec:user_study}

\subsection{Study Design}
\label{ssec:study_design}
The user study is structured into three primary phases: 
\begin{inlinelist}
    \item \textit{Ethics, consent and demographics}
    \item \textit{Rule evaluation and cognitive load analysis} and
    \item \textit{Interactive system evaluation}.
\end{inlinelist}

\subsubsection{Ethics, Consent, and Demographics}
The study was conducted in strict accordance with the ethical guidelines of the lead authors' institution and received formal approval from the Human Research Ethics Committee. 
All participants are provided with informed consent detailing the study objectives, data anonymization procedures, and their right to withdraw.
To safeguard professional standing, all qualitative feedback and metadata were pseudonymized and stored on secure institutional servers.
To contextualize the results, we collected demographic data and professional background information.
This allowed for a stratified analysis based on expertise, work environment, and familiarity with IDS rule engineering.
Key metrics included primary role, organization size, years of experience, and self-assessed proficiency with complex rule structures.
Further information on participant demographics is included in~\Cref{sapp:demographic_questions}.

\subsubsection{Rule Evaluation and Cognitive Load Analysis}
The second phase assesses the quality of LLM-generated rules. 
While a rule may be syntactically valid, the quality evaluation ensures it accurately encodes detection policies and effectively discriminates between malicious and benign traffic. 
This phase pursues the following objectives:
\begin{inlinelist}
    \item Quantifying the quality and deployment-readiness of generated rules and LLM-generated explanations (\Cref{sssec:rule_quality_analysis}); 
    \item Evaluating rule engineering difficulty as a function of cognitive load (\Cref{sssec:human_rule_quality}). 
\end{inlinelist}
Candidate rules were sampled from the highest-performing models (see \Cref{sssec:syntax_performance}), filtered from hallucinations and bad explanations. 
To mitigate selection bias, participants performed evaluations blindly through a structured interface.

\noindent {\bf Evaluation Metrics.}
The evaluation interface presented the Snort rule, the corresponding network context (including topology, assigned CVEs, PoCs and network details), and a structured response form.
The central evaluation metric is a binary adoption decision (Yes/No), reflecting whether the rule is considered correct by a user or not.
Here, the concept of correctness refers to the possibility of adopting the rule in the presented scenario.
Following each decision, participants reported their confidence on a 5-point Likert scale (1: uncertain, 5: certain) to measure the calibration between perceived and actual correctness.
Additionally, the utility of LLM explanations was evaluated based on their contribution to rule comprehension by answering the question \textit{Does the explanation provided enhance the comprehension of the IDS rules?} with answers from ``not at all'' to ``extremely''.
Finally, for rejected rules, participants estimated the corrective effort required on a 5-point scale.

\noindent {\bf Cognitive Load Analysis.}
As confirmed in the pre-study survey, analysts must not interpret rules in isolation, but rather in relation to network topology, asset roles, and the distribution of vulnerabilities. 
Such contextual richness introduces varying degrees of cognitive demand, which we model in our experimental design.
To simulate real-world conditions, we modeled three network scenarios with increasing structural complexity. 
These scenarios—comprising 5, 10, and 18 nodes with corresponding \gls{cve} densities—were derived from established benchmarks~\cite{palma_behind_2025}.
The smallest scenario models a simple, minimally segmented environment, while the intermediate and largest scenarios represent progressively more complex enterprise and multi-segment infrastructures with broader services and vulnerabilities. 
To maintain tractability, we limited each node to at most two CVEs, avoiding excessive combinatorial complexity.
Appendix~\ref{ssec:cognitive_load} depicts more details on the cognitive load analysis.
%


\subsubsection{Interactive System Evaluation}

\begin{figure}[h!]
    \centering
    \includegraphics[width=0.75\linewidth]{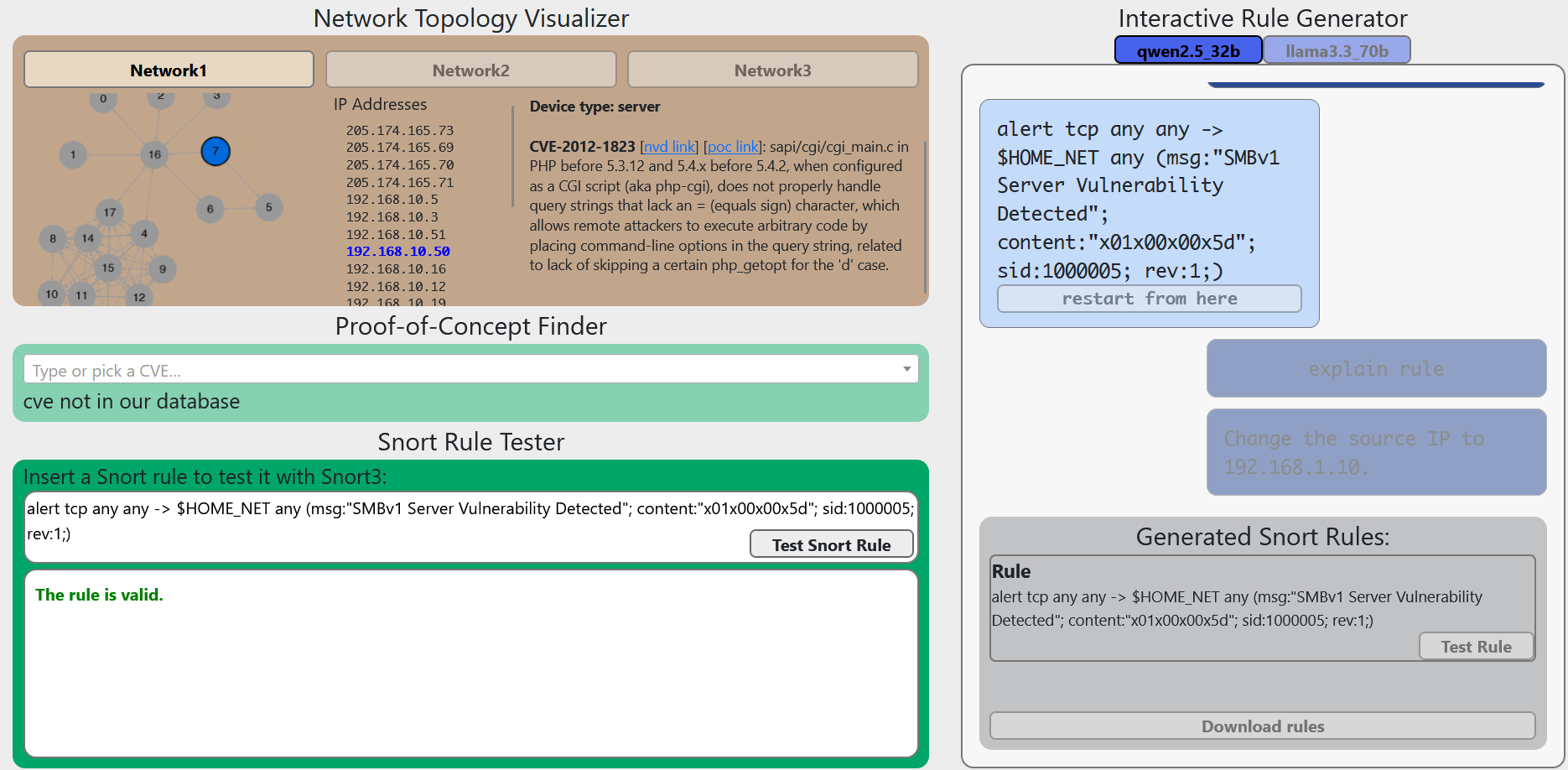}
    \caption{Overview of the system interface.}
    \label{fig:iface}
\end{figure}

The final phase evaluates the formalized system through direct user interaction. 
After a standardized tutorial, participants used a workflow-oriented interface (see \Cref{fig:iface}) to perform iterative rule engineering. 
The system integrates selected \glspl{llm} and support network visualization, \gls{cve} inspection, rule validation and chat-based interface.
Such a setup enables assessment of system usability, user interaction with LLM agents, and the readiness of rules for real-world adoption.

\noindent {\bf Operational Tasks.}
The tasks enabled on the system mirror realistic SOC workflows.
While the system supports free exploration for some tasks (e.g., syntax checking, rule writing, network exploration, PoC and CVE finder, and Snort runs), certain resource-intensive tasks (e.g., generating and modifying complex rules) were provided as curated examples. 
Specifically, we provided examples to \emph{generate} new rules, \emph{explain} generated rules, and \emph{modify} the rules to update IP addresses and ports or add specific fields of the IDS rule.
These tasks are indeed the most performed changes during rule engineering according to existing literature~\cite{vermeer2023alert} and pre-study results.
This approach ensures participants can evaluate high-parameter models (e.g., DeepSeek-R1 70B) without the latency associated with real-time generation.
The free usage allowed participants to explore the interface using their own strategies, helping identify usability issues, interaction patterns, and perceived utility. 

\noindent {\bf Usability Analysis.}
Following the interaction, participants completed the System Usability Scale (SUS)~\cite{Bangor29072008}. 
This quantitative measure was supplemented by structured 5-point scales assessing efficiency, effectiveness, and learnability~\cite{butt2018efficiency}, alongside open-ended questions to capture subjective user experiences and technical bottlenecks.
The combination of quantitative and qualitative data provides a comprehensive evaluation of the system's usability, highlighting strengths, weaknesses, and areas for improvement from the perspective of professional IDS rule engineers (see \Cref{sssec:sus_evaluation}). 

\subsection{Online Platform}
\label{ssec:online_platform}
To accommodate the demands of this study while ensuring comfort for the participants, we developed a centralized web-based platform to host the study. 
This eliminated the need for local software installation or complex environment configurations (e.g., Docker). 
The platform was built using Python and Ngrok for backend functions, and a modern HTML/CSS/JavaScript frontend, ensuring a seamless transition between the survey components and the interactive system interface.
We hosted the platform frontend using PythonAnywhere\footnote{\url{https://www.pythonanywhere.com/}}, while the backend and system run on private server machines.


\section{Evaluation}\label{sec:evaluation}

%
In this section we evaluate the LLM-based rule engineering agent over six different network scenarios and 27 different \glspl{llm} engines. 
More in detail, we consider the following network scenarios:
\begin{inlinelist}
    \item one small synthetic network (SS),
    \item one medium-sized synthetic network (SM), and
    \item the real world network used to collect the CIC-IDS 2017 dataset (CI)~\cite{sharafaldin2018toward}.
\end{inlinelist}
For the CIC-IDS network, we construct a total of four different combinations, including one or two CVEs for each of the devices in the network, and either if the CVEs match a corresponding PoC or not.
The obtained four combinations are:
\begin{inlinelist}
    \item CIC-IDS with 1 CVE per device and no PoCs (CI$_{1}$),
    \item CIC-IDS with 1 CVE per device and PoCs (CI$_{1}^{P}$),
    \item CIC-IDS with 2 CVEs per device and no PoCs (CI$_{2}$), and
    \item CIC-IDS with 2 CVEs per device and PoCs (CI$_{2}^{P}$).
\end{inlinelist}
For the detailed description of the generation process used for the two synthetic networks, we refer the reader to \Cref{app:net_gen}.
We consider these six network scenarios -- two synthetics and four CIC-IDS -- to account for the different levels of complexity that may arise when experts need to write IDS rules, thus mimicking the increase in cognitive load that they face in practice.
Indeed, the combination of network size and vulnerabilities introduces interaction effects (e.g., exploit chains) that exponentially increase the analytical complexity beyond a simple additive effect.
The assessment scenarios thus provide a controlled and representative environment for the proposed study. 
For each combination of network scenario and LLM engine, we run our agent 30 times for statistical significance\footnote{The number of iterations is empirically a good balance between statistical significance and resource consumption.}, aiming at generating as many IDS rules as possible. 
%
To keep the computational cost under control, the IDS rule generation process is terminated if the agent fails to produce any syntactically correct rule within an 30-hour time window. 
%
%
All rules were used for our statistical analysis in \Cref{ssec:statistical_analysis}, while only the rules generated with the statistically best performing LLM engine -- i.e., DeepSeek-R1 70B (see \Cref{app:best_model_selection}) -- were used for our human-centered evaluation in \Cref{ssec:human_evaluation}.

\subsection{Statistical Analysis}\label{ssec:statistical_analysis}
We first analyze the rules generated by the LLM-based agent from a statistical perspective, analyzing their syntax validity, semantic content and their proneness to hallucinations.

\subsubsection{Syntax Evaluation}\label{sssec:syntax_performance}

We first filter rules that violate basic Snort structure, yielding 2871 Snort-looking candidates. 
Out of the LLM engines considered, models with fewer than 4B parameters failed to generate any valid syntax (see \Cref{tab:generated_rules}), highlighting the complexity of automatic NIDS rule generation. 
We use a dashed line in \Cref{tab:generated_rules} to distinguish between models: those above successfully generated rules across topology, while those below failed.

%

\finding{Small LLMs ($\leq$4B parameters) are ineffective for IDS rule generation.}\label{find_1}

\begin{table}[htbp]
    \centering
    \scalebox{0.75}{
        \begin{tabular}{lccccccc}
            \toprule
            \textbf{Model} & CI$_{1}$ & CI$_{1}^{P}$ & CI$_{2}$ & CI$_{2}^{P}$ & SS & SM & Total \\
            \textbf{Rules} & \textbf{461/756} & \textbf{396/560} & \textbf{404/604} & \textbf{236/362} & \textbf{181/305} & \textbf{185/284} & \textbf{1863/287}1 \\
            \midrule
            Qwen2.5 32B & 56/86 & \textbf{60/92}$^{\ast}$ & 43/67 & 17/26 & \textbf{42/56}$^{\ast}$ & 28/40 & \textbf{246/367}$^{\ast}$ \\
            Llama3.3 70B & \textbf{46/51}$^\dagger$ & 78/83 & 39/45 & 34/42 & 28/35 & 24/29 & \textbf{249/285}$^\dagger$ \\
            DeepSeek-R1 70B & 42/48 & 42/49 & \textbf{63/71}$^{\ast}$ & 12/13 & 9/15 & \textbf{51/62}$^{\ast}$ & 219/258 \\
            Llama3.1 70B & 37/52 & 57/68 & 33/40 & \textbf{39/47}$^\dagger$$^{\ast}$ & 15/27 & 16/22 & 197/256 \\
            Qwen2.5 14B & \textbf{48/90}$^{\ast}$ & 30/70 & 33/57 & 11/26 & 7/8 & 2/5 & 131/256 \\
            Qwen2.5 7B & 49/89 & 37/58 & 42/63 & 18/27 & 8/10 & \textbf{7/8}$^\dagger$ & 161/255 \\
            Qwen 32B & 20/54 & 10/39 & 13/40 & 9/22 & 11/20 & 10/17 & 73/192 \\
            Phi4 & 33/52 & 20/27 & 14/23 & 22/31 & 5/14 & 7/23 & 101/170 \\
            Llama3.1 8B & 18/32 & 0/0 & 38/50 & 21/43 & 8/12 & 14/24 & 99/161 \\
            Qwen2.5 72B & 25/30 & 33/39 & \textbf{10/10}$^\dagger$ & 22/37 & 10/18 & 11/26 & 111/160 \\
            DeepSeek-R1 8B & 19/40 & 12/15 & 22/33 & 16/25 & \textbf{10/11}$^\dagger$ & 9/19 & 88/143 \\
            \hdashline
            DeepSeek-R1 14B & 29/40 & \textbf{14/14}$^\dagger$ & 30/40 & 0/0 & 20/30 & 6/9 & 99/133 \\
            Gemma2 27B & 14/27 & 1/2 & 10/26 & 4/5 & 6/22 & 0/0 & 35/82 \\
            Gemma2 9B & 8/27 & 0/0 & 9/24 & 0/0 & 2/14 & 0/0 & 19/65 \\
            DeepSeek-R1 32B & 16/18 & 2/2 & 5/5 & 0/0 & 0/7 & 0/0 & 23/32 \\
            Qwen2.5 1.5B & 0/17 & 0/0 & 0/6 & 0/0 & 0/4 & 0/0 & 0/27 \\
            DeepSeek-V2 16B & 1/3 & 0/2 & 0/2 & 5/7 & 0/2 & 0/0 & 6/16 \\
            Phi3 14B & 0/0 & 0/0 & 0/2 & 5/6 & 0/0 & 0/0 & 5/8 \\
            Qwen 72B & 0/0 & 0/0 & 0/0 & 1/5 & 0/0 & 0/0 & 1/5 \\
            Llama3.2 1B & 0/0 & 0/0 & 0/0 & 0/0 & 0/0 & 0/0 & 0/0 \\
            Llama3.2 3B & 0/0 & 0/0 & 0/0 & 0/0 & 0/0 & 0/0 & 0/0 \\
            Qwen 1.8B & 0/0 & 0/0 & 0/0 & 0/0 & 0/0 & 0/0 & 0/0 \\
            Gemma2 2B & 0/0 & 0/0 & 0/0 & 0/0 & 0/0 & 0/0 & 0/0 \\
            DeepSeek-R1 1.5B & 0/0 & 0/0 & 0/0 & 0/0 & 0/0 & 0/0 & 0/0 \\
            Qwen 4B & 0/0 & 0/0 & 0/0 & 0/0 & 0/0 & 0/0 & 0/0 \\
            Qwen 7B & 0/0 & 0/0 & 0/0 & 0/0 & 0/0 & 0/0 & 0/0 \\
            Phi3 3.8B & 0/0 & 0/0 & 0/0 & 0/0 & 0/0 & 0/0 & 0/0 \\
            \bottomrule
        \end{tabular}
    }
    \caption{Number of syntactically accurate rules out of the total generated by each LLM engine across all scenarios. Legend: SS - small synthetic network; SM - medium-sized synthetic network; and CI - real world network used to collect the CIC-IDS 2017 dataset. For each topology, $^\ast$ denotes the largest number of generated rules, while $^\dagger$ denotes the highest percentage of correct rules.} 
    \label{tab:generated_rules}
\end{table}

\begin{figure*}[t] 
    \centering
    \begin{subfigure}[t]{0.24\textwidth} 
        \centering
        \includegraphics[width=\textwidth]{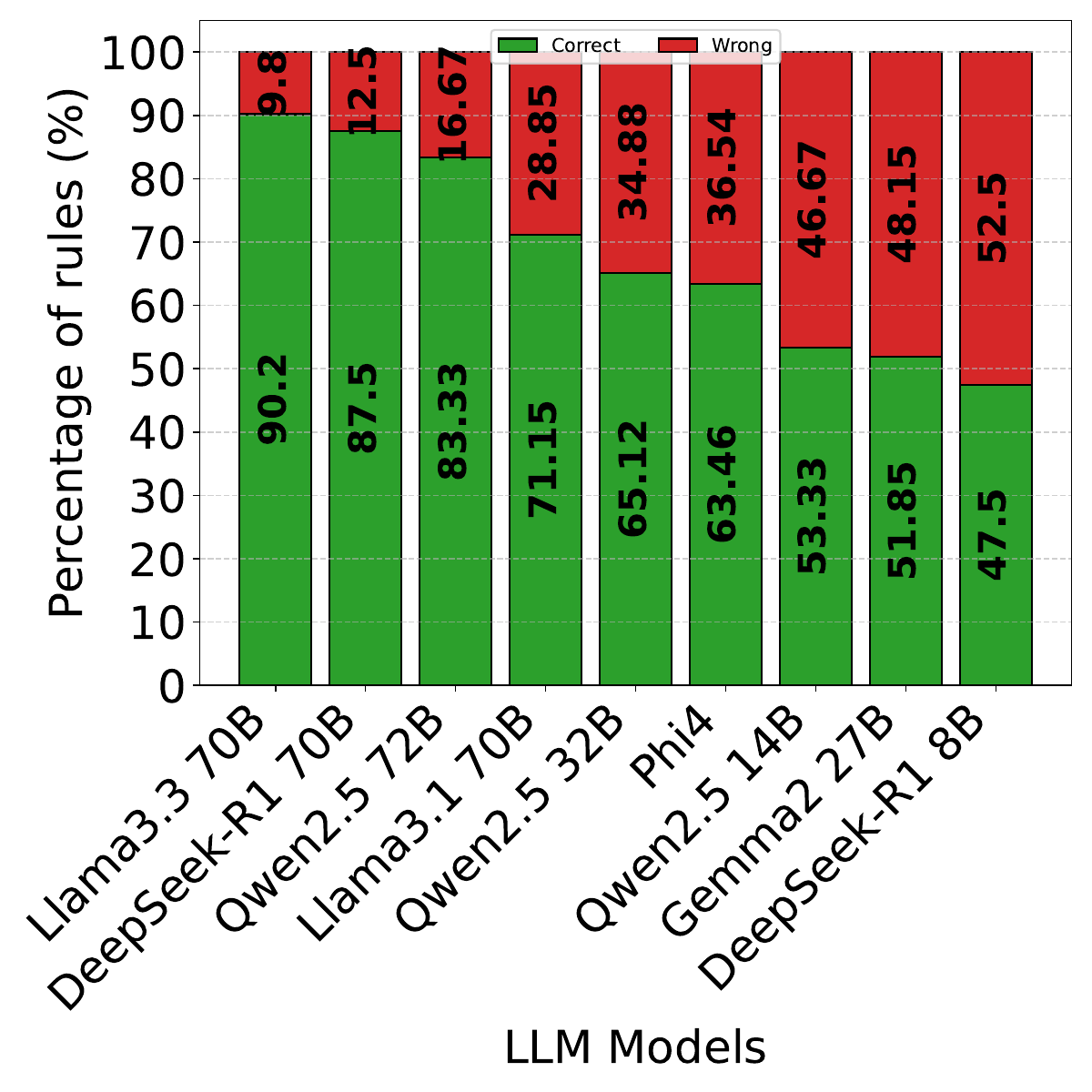} 
        \caption{Fraction of syntactically valid rules by LLM engines over CI$_{1}$. 
        }
        \label{subfig:syntax_correct_cicids1cve_percentage}
    \end{subfigure}
    \hfill
    \begin{subfigure}[t]{0.24\textwidth}
        \centering
        \includegraphics[width=\textwidth]{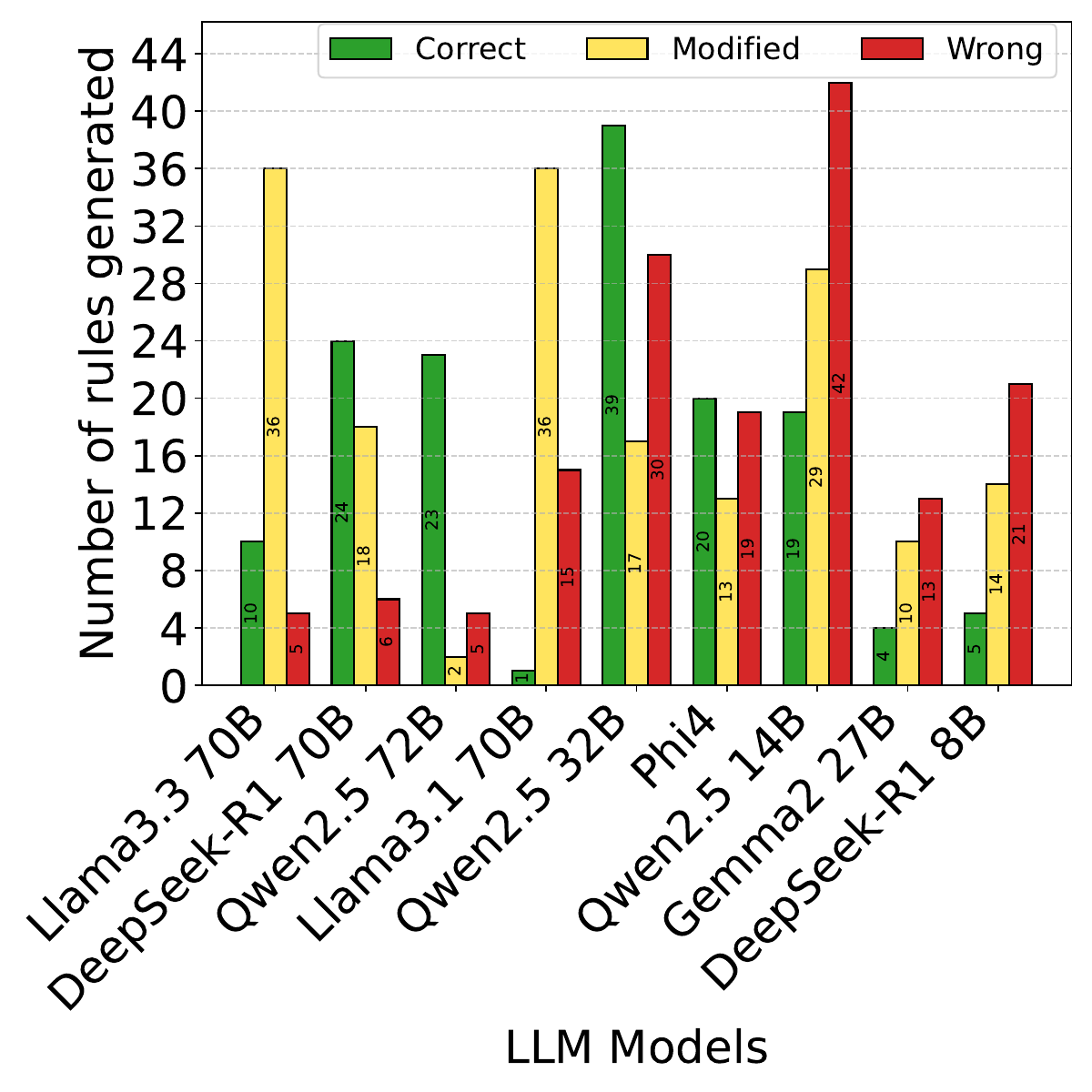}
        \caption{Fraction of generated rules that require corrections by LLM engines over CI$_{1}$. 
        }
        \label{subfig:cicids_1_cve_require_loop_percentage}
    \end{subfigure}
    \hfill
    \begin{subfigure}[t]{0.24\textwidth}
        \centering
        \includegraphics[width=\textwidth]{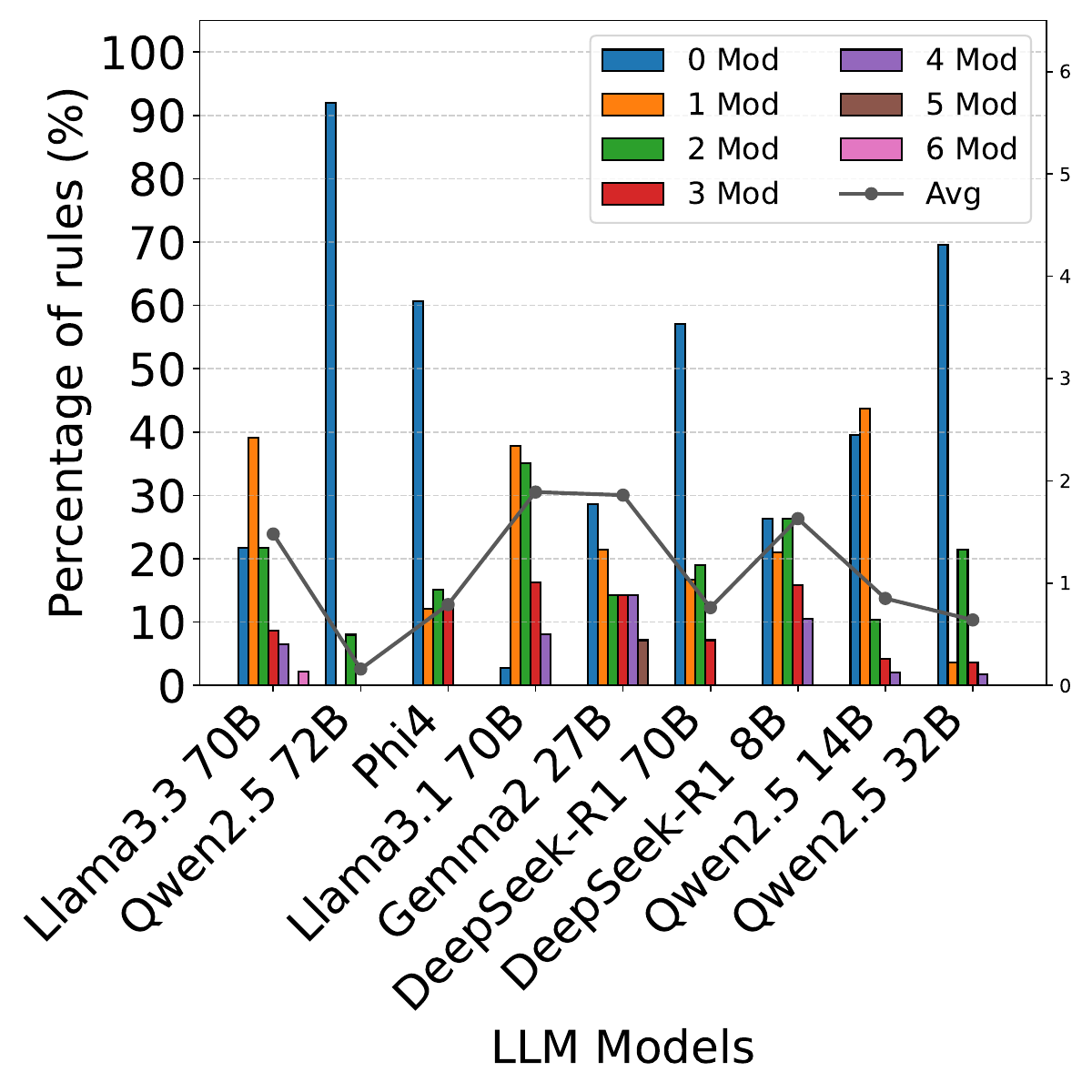}
        \caption{Distribution of correction trials required to generate a rule by LLM engines over CI$_{1}$. 
        }
        \label{subfig:avg_modifications}
    \end{subfigure}
    \hfill
    \begin{subfigure}[t]{0.24\textwidth}
        \centering
        \includegraphics[width=\textwidth]{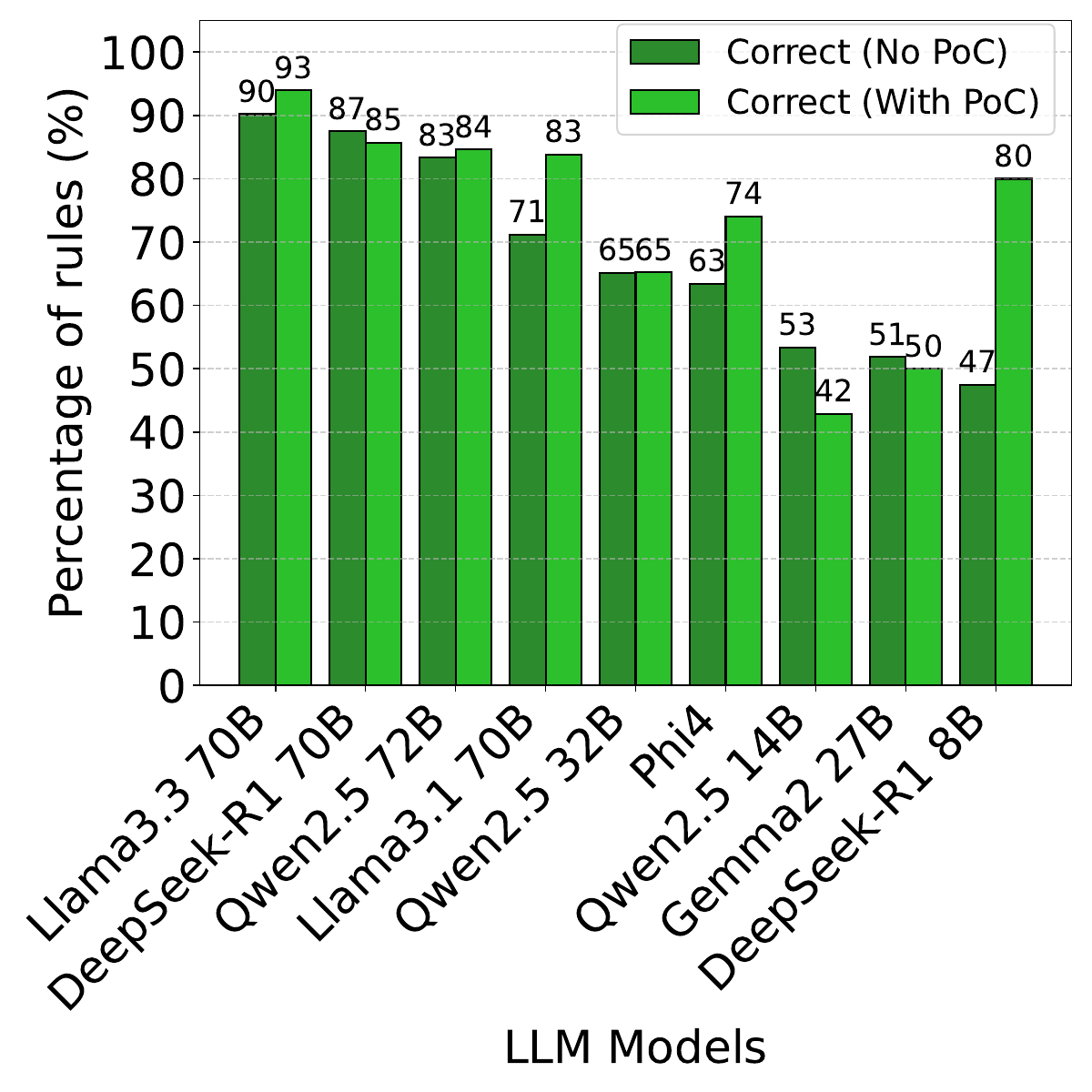}
        \caption{Fraction of syntactically valid rules when including PoCs by LLM engines over CI$_{1}$ and CI$_{1}^P$.
        }
        \label{subfig:poc_ablation}
    \end{subfigure}
    \hfill
    \caption{Syntactical correctness analysis of the generated Snort rules leveraging different LLM engines. LLM-based agents enable the effective construction of valid IDS rules.}
    \label{fig:main_figure_syntax}
\end{figure*}

Subsequently, we compile each rule in Snort, testing how many of them result in compilation issues.
\Cref{subfig:syntax_correct_cicids1cve_percentage} shows the fraction of syntactically valid rules generated by each LLM engine for CI$_{1}$---additional results across the other  scenarios are reported in Appendix~\ref{app:syntax_validity_all_networks} and highlight similar findings.
While smaller models like Gemma 2 9B and Llama 3.2 1B struggle, larger models achieve significant proficiency: Llama 3.1 70B reached 77\% accuracy, while DeepSeek-R1 70B and Llama 3.3 70B reached 85\% and 87\% accuracy, respectively. 
Furthermore, a Pearson correlation of 0.5991 confirms a positive relation between model size and syntactic validity.
Therefore, these results highlight how larger LLM engines have the necessary understanding of the NIDS grammar to generate usable rules.

\finding{Large-scale models ($\geq$70B) demonstrate high syntactic proficiency, with success rates up to 90\%.}

To determine if this proficiency is LLM-intrinsic or correction-dependent -- and thus connected to the agent's tool usage --, we track rules that were generated immediately correct versus those requiring a feedback loop to be corrected. 
\Cref{subfig:cicids_1_cve_require_loop_percentage} shows such fractions for each LLM engine for the CI$_{1}$ scenario, while additional results across the other scenarios are reported in Appendix~\ref{app:syntax_validity_all_networks} with similar findings.
54.85\% of valid rules require the corrector loop and this necessity persists regardless of model family.
For example, only 21.74\% of the syntactically best performing model (i.e., Llama-3.3 70B) rules were valid on the first attempt (see \Cref{subfig:cicids_1_cve_require_loop_percentage}).

\finding{Syntax-checking and correction modules are essential. Even the best LLMs struggle to produce compliant signatures without feedback.}

To further analyze the importance of the rule checking and correction feedback loop, we measure the number of correction loop iterations that each generated rule requires before being correctly output.
As shown in \Cref{subfig:avg_modifications}, even the best performing LLM  engines require a non-negligible amount of rule corrections to generate valid rules, with results being similar across engines.
On average, DeepSeek-R1 70B requires 0.76 corrections (increasing to 1.78 for rules requiring at least one fix).
These results highlight a bimodal distribution where rules are either immediately valid or require significant effort to be fixed, with some rules requiring up to four corrections before being correctly output.
Furthermore, we also note how this number represents a lower bound, as we stop attempting to correct a rule after five consecutive failed updates.

\finding{Generated rules tend to be either immediately valid or cumbersome to fix.}

Finally, we analyze the impact that the inclusion of Proof-of-Concept (PoC) data has on rule generation.
To this end, we compute the fraction of valid rules generated whenever the PoCs are (not) available across each LLM engine (see \Cref{subfig:poc_ablation}).
The inclusion of \gls{poc} information yields a moderate enhancement in output quality as the mean syntactic accuracy increases from 67.23\% to 74.83\% when PoC context is provided.
Moreover, we observe a moderate positive correlation (Pearson = 0.6619, Spearman = 0.7160) indicating that engines performing well without PoCs also tend to perform well with PoCs. 
Lastly, we note that larger models, which can better handle long and complex contexts, benefit from PoC information, whereas models with limited context capacity are negatively affected by the substantial increase in input length, which rises on average from approximately 7,500 (without PoCs) to more than 58,000 (without PoCs) tokens per generated rule.

\finding{PoCs improve rule generation quality only when the LLM engine can handle large context windows.}

To answer {\bf RQ4}, DeepSeek-R1 70B and Llama3.1 70B emerge as superior engines (85-87\% syntactic success), augmented by syntax checkers, PoC exploits, and vulnerability inventories; small models prove ineffective.
As a complementary analysis, the fraction of hallucinated rules is in Appendix \ref{subapp:halluciantions}.

\subsubsection{Rule Quality Analysis}\label{sssec:rule_quality_analysis}

\begin{figure*}[t] 
    \centering
    \begin{subfigure}[t]{0.24\textwidth} 
        \centering
        \includegraphics[width=\textwidth]{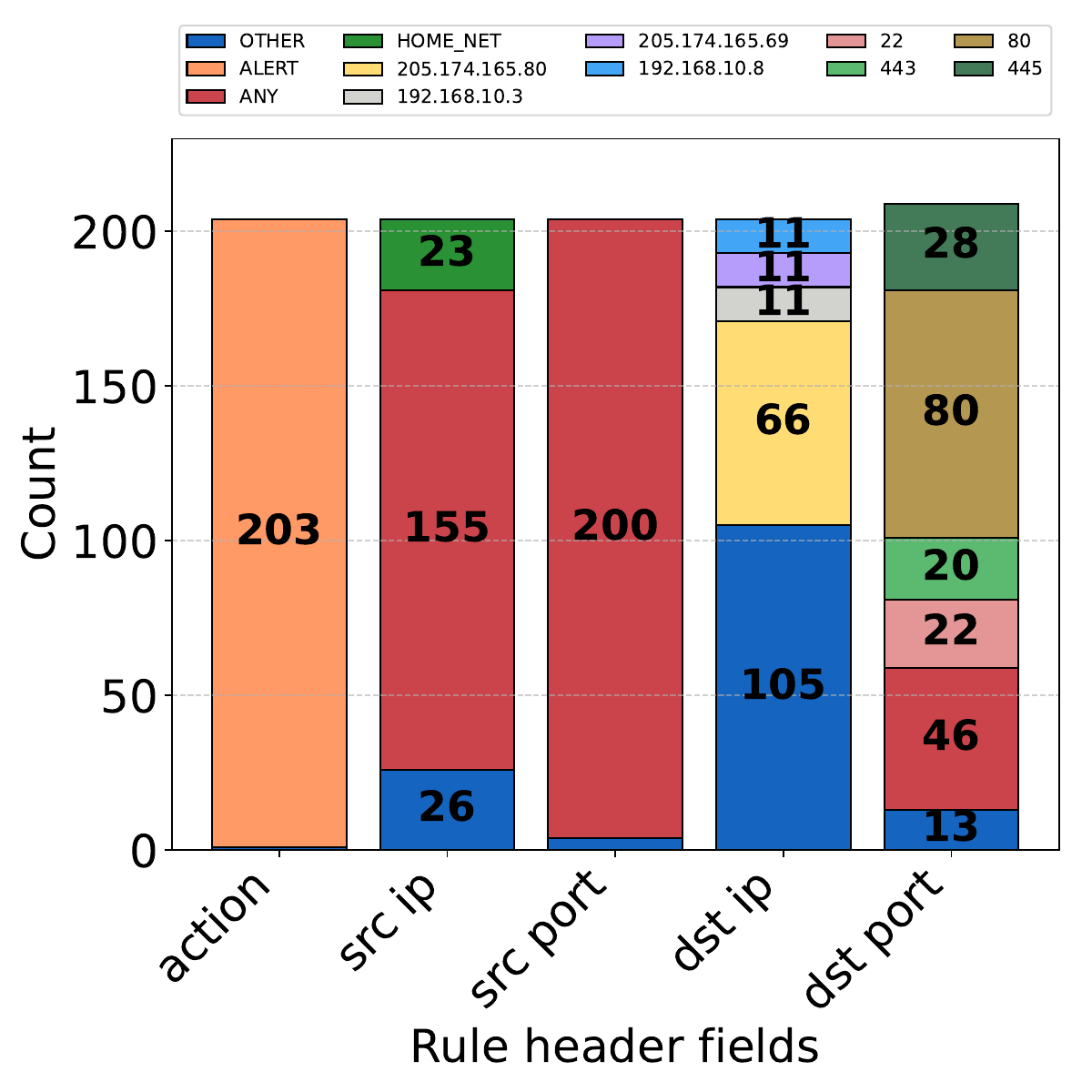} 
        \caption{Distribution of fields used in rule headers (aggregate metric across LLM engines over CI$_1$). 
        }
        \label{subfig:header_distribution}
    \end{subfigure}
    \hfill
    \begin{subfigure}[t]{0.24\textwidth}
        \centering
        \includegraphics[width=\textwidth]{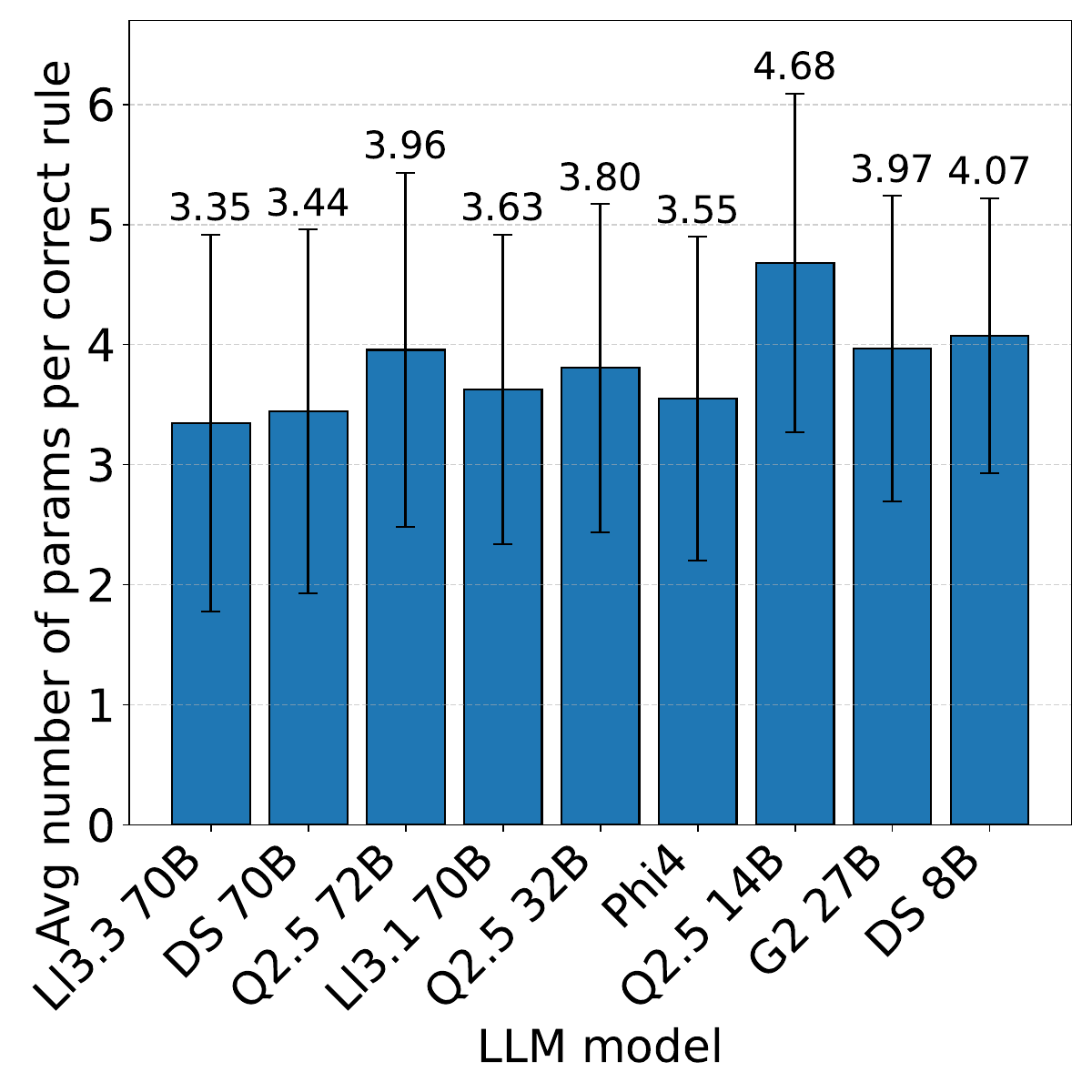}
        \caption{Average length -- i.e., used body fields -- of the generated rules across LLM engines.
        }
        \label{subfig:avg_rule_length}
    \end{subfigure}
    \hfill
    \begin{subfigure}[t]{0.24\textwidth}
        \centering
        \includegraphics[width=\textwidth]{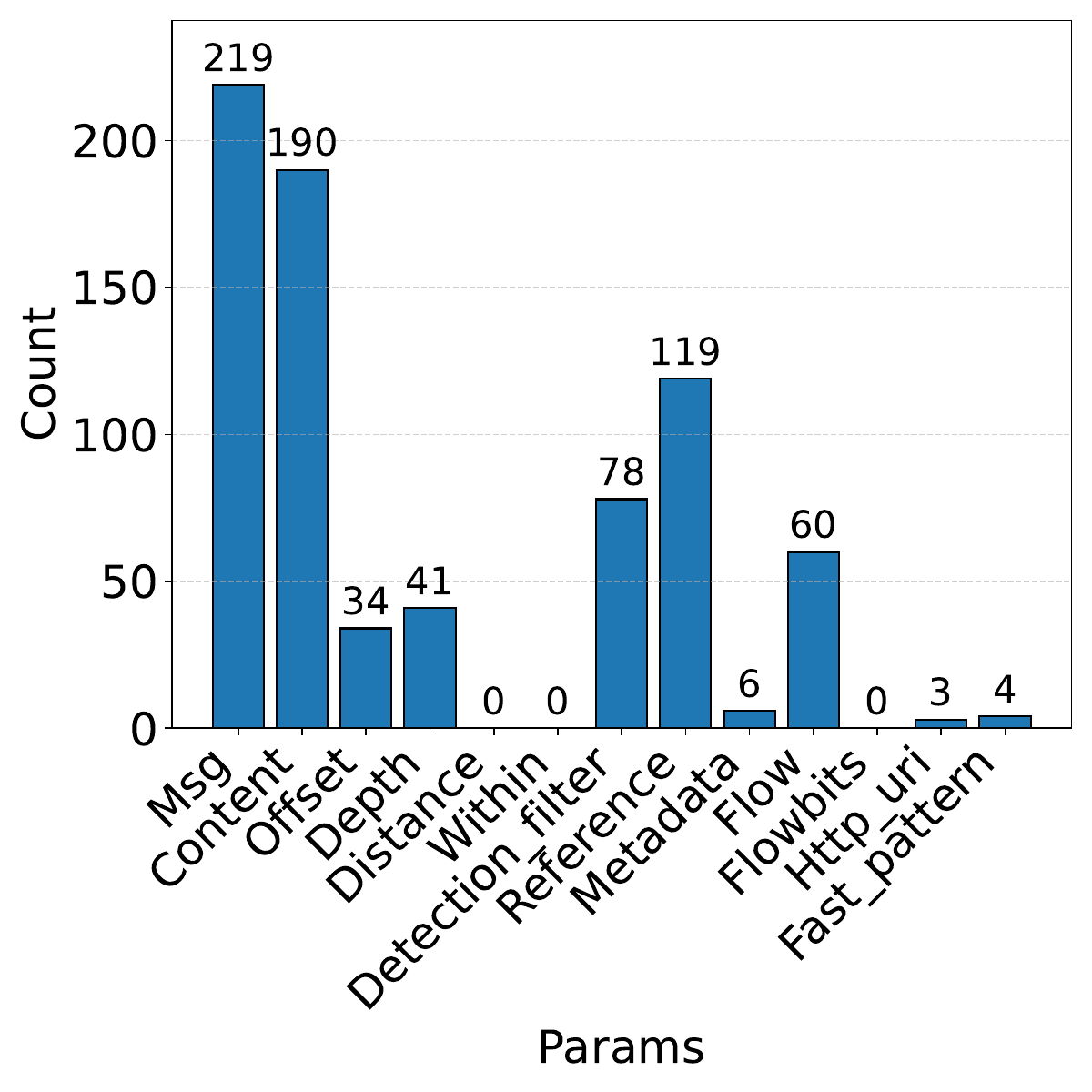}
        \caption{Distribution of the rule body fields used in the rules generated by the DeepSeek-R1 70B engine. 
        }
        \label{subfig:deepseek_field_dist}
    \end{subfigure}
    \hfill
    \begin{subfigure}[t]{0.24\textwidth}
        \centering
        \includegraphics[width=\textwidth]{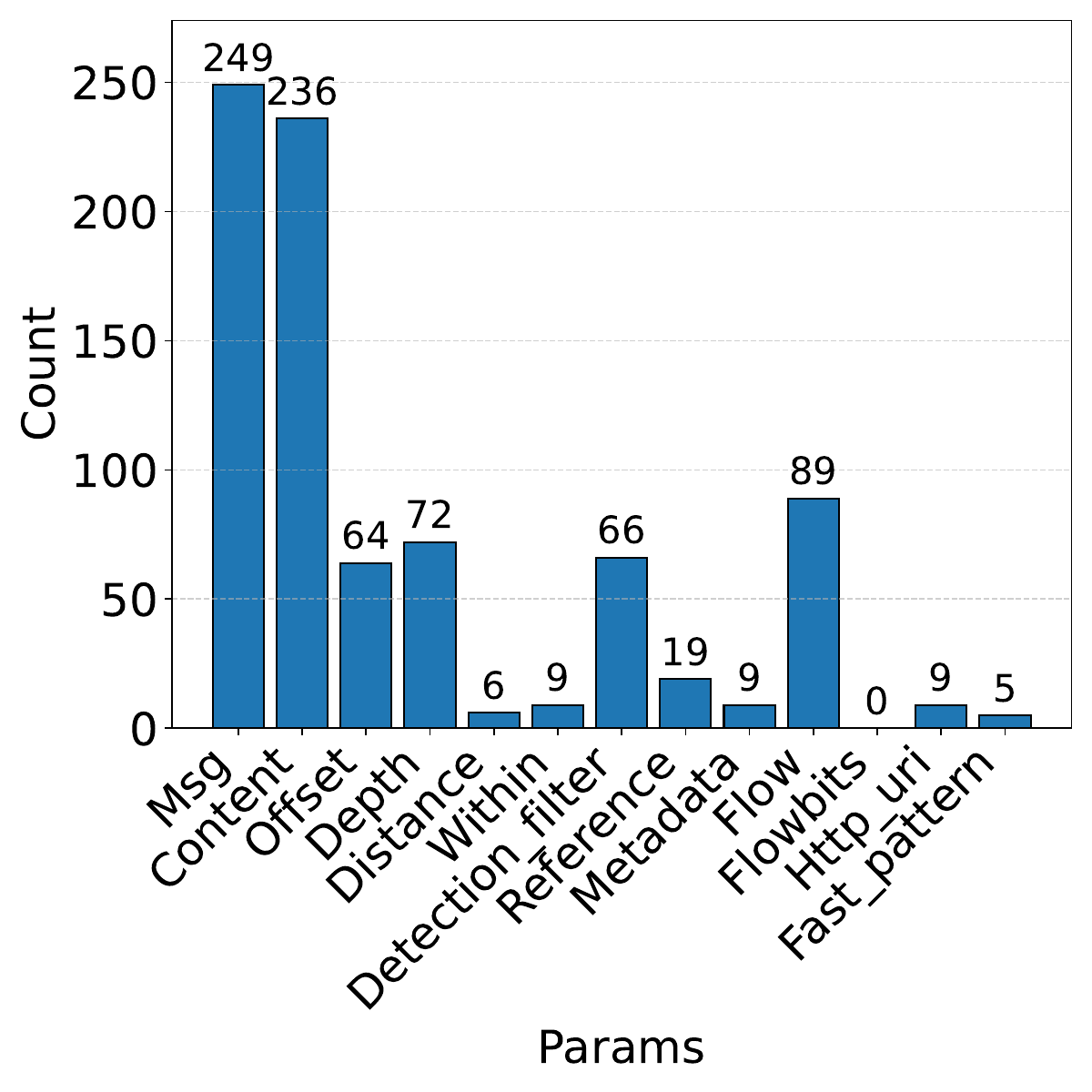}
        \caption{Distribution of the rule body fields used in the rules generated by the Llama-3.3 70B engine. 
        }
        \label{subfig:llama_field_dist}
    \end{subfigure}
    \hfill
    \caption{Semantical content analysis of the generated rules across LLM engines. LLM-based agents rely heavily on content and port/protocol matching to generate rules.}
    \label{fig:main_figure_semantic}
\end{figure*}

We statistically analyze the distribution of header and body fields to asses the quality of generated rules, referencing the criteria established in our pre-study (see~\Cref{ssec:prestudy_survey}).
\Cref{subfig:header_distribution} presents the field distribution over the generated rule headers for the CI$_{1}$ scenario aggregated over all engines.
Most headers (87.87\%) default to generic alerting actions and use the \textsc{any} keyword for source IPs and ports to map general threats for which the source is unknown. 
Conversely, the destination port is the most specific field, with 71.77\% of rules correctly targeting ports 22, 80, 443, or 445.
This specificity highlights the agent's ability to reason about the most common network protocols -- such as HTTP at port 80 or SSH at port 22 --, and the LLM engine's tendency to generate rules focusing on where the traffic is landing.
Therefore, the agent aims to match specific characteristics of an attack with the network protocol it expects to see---e.g., generating a rule against an ownCloud remote improper authentication attempt as defined in CVE-2023-49105, requires grounding the attack on HTTPS and thus on port 443.
Meanwhile, it seems that the agent avoids enforcing specific constraints on the target destination -- with 57\% of destination IPs being unspecified/scenario-dependent -- or blacklisting some potential source IP threats.
These headers patterns mirror the ones in the Emerging Threats ruleset\footnote{\href{https://rules.emergingthreats.net/open/}{https://rules.emergingthreats.net/open/}} available to Snort and Suricata practitioners -- where most headers are alert-driven and heavily rely on \textsc{any} and placeholders like \textsc{external\_net}, thus suggesting that open-source LLMs are trained on this ruleset and back our intuition on memorization made in Appendix~\ref{subapp:halluciantions}.
Further analysis of the rule quality is depicted in Appendix~\ref{app:header_distribution_appendix}.

\finding{While 54.66\% of the generated rules rely on general variables such as \textsc{any}, models demonstrate an inherent ability to reason about specific network protocol ports.}

While we will rely on the user study of \Cref{ssec:human_evaluation} as the definitive test of rule quality, we here analyze the generated rule bodies to provide preliminary findings on their specificity and reliability.
We analyze the average length of rule bodies across LLM engines in \Cref{subfig:avg_rule_length} where results show stability across models, ranging from three to five fields.
This distribution is quite far from the average length that characterize rule bodies of commonly available rule databases\footnotemark[5], where rules adopt several fields like payload positioning (e.g., \textsc{offset}, \textsc{depth}) and protocol-specific fields (e.g., \textsc{http.uri}, \textsc{dns.query}) to ensure rule specificity (as highlighted by the pre-study insights of \Cref{sssec:rule_engineering}). 
Therefore, the generated rules seem to lack the level of specificity that SOC experts require~\cite{teuwen2025ruling}, hinting at LLM not being able to process the fine-grained information required to write very specific rules. 
Similarly, LLMs do not seem to posses the reasoning capability to compose multiple fields of the rule body together to improve the rule detection scope.
These intuitions were backed during our user study, where human experts mentioned the lack of specificity of prompted rules and their worry towards possible false alerts (see \Cref{ssec:human_evaluation}).

Additionally, in \Cref{fig:main_figure_semantic}, we visualize the distribution of the fields used in the bodies of rules generated using two of the best-performing LLM engines, DeepSeek-R1 70B (\Cref{subfig:deepseek_field_dist}) and Llama-3.3 70B (\Cref{subfig:llama_field_dist}).
The most common parameters are the message and content fields, which are almost always used---the former is used to provide a description of the generated rule, while the latter provides the actual detection framework, aiming to match the specific traffic content.
The results show that, while large models sometimes enable the agent to generate rules containing specific fields for detection, such as \textsc{distance} and \textsc{within}, the vast majority of generated rules rely on pure content matching.
Once again, these intuitions were backed during our user study (see \Cref{ssec:human_evaluation}).
These results highlight the intrinsic complexity of the IDS rule engineering task, where rule precision requires complex reasoning abilities about several aspects of the attack behavior across several parameters that current LLMs lack.

\finding{Most generated rules rely on pure content matching, with specific fields such as \textsc{distance} and \textsc{within} -- valuable fields for rule specificity -- rarely used.}\label{finding:specificity_stats}

To answer {\bf RQ1}, these findings reveal that while large-scale LLMs achieve up to 90\% syntactic validity, NIDS experts deem only 37.5\% of generated rules semantically correct and deployable, primarily due to insufficient specificity, heavy reliance on content matching, and logic hallucinations.

\subsection{Human Evaluation}\label{ssec:human_evaluation}

Here we examine the outcomes of presenting a group of human expert SOC analysts with the rules produced by the top-performing agent -- namely, DeepSeek-R1 70B -- to evaluate their quality.
We refer the reader to Appendix~\ref{app:best_model_selection} for the detailed description of how DeepSeek-R1 70B was selected as the best performing engine across the used ones.

\subsubsection{Participants Recruitment}
The participant recruitment process was conducted in multiple phases to maximize the response rate.
Initially, we reached out to 137 potential participants, establishing a 14-day window for survey completion. 
Following the initial outreach, a reminder was sent to all candidates after 11 days to encourage participation before the closing date.
By the primary deadline, 14 responses had been recorded. Two days post-deadline, a final solicitation was directed toward the remaining non-respondents to capture any late submissions.
Ultimately, we secured a total of \complete{} completed surveys, while \partially{} participants completed the survey only partially and \onlydemographics{} participants filled only demographics information. 
The subsequent analysis is based on the data provided by the respondents who partially or fully complete the study. 
%
\Cref{tab:participants} shows the participants demographic information and background.
\begin{table}[ht]
\centering
\rowcolors{2}{gray!10}{white} 
\scalebox{0.8}{
\begin{tabular}{lllll}
\toprule
\textbf{ID} & \textbf{Sector} & \textbf{Role} & \textbf{Experience} & \textbf{Primary IDS Engine} \\ 
\midrule
P1  & Cybersecurity & Security Researcher & 10+ years & Suricata, Forescout eyeInspect\\
P2  & Cybersecurity & Professor & 1 year & Zeek/Bro\\
P3  & Cybersecurity & Professor & 1 year & Snort\\
P4  & Information Security & Security Researcher & 10+ years & Snort, Suricata, Zeek/Bro \\
P5  & Cybersecurity & PhD Student & 3-5 years & Suricata\\
P6  & Cybersecurity & Security Researcher & 1 year & Snort, Suricata\\
P7  & Cybersecurity & Security Researcher & 1-3 years & Snort, Suricata\\
P8  & IT & Detection Engineer & 5-10 years & Suricata, Zeek/Bro\\
P9  & Cybersecurity & Security Researcher & 1-3 years & Snort, Suricata, Zeek/Bro\\
P10  & Cybersecurity & Security Engineer & 1 year & Snort\\
\bottomrule
\end{tabular}
}
\caption{Participant demographics and background.}
\label{tab:participants}
\end{table}

\subsubsection{Rule Quality Evaluation}\label{sssec:human_rule_quality}
\begin{figure*}[t]
    \centering
    \begin{subfigure}[t]{0.24\linewidth} 
        \centering
        \includegraphics[width=\textwidth]{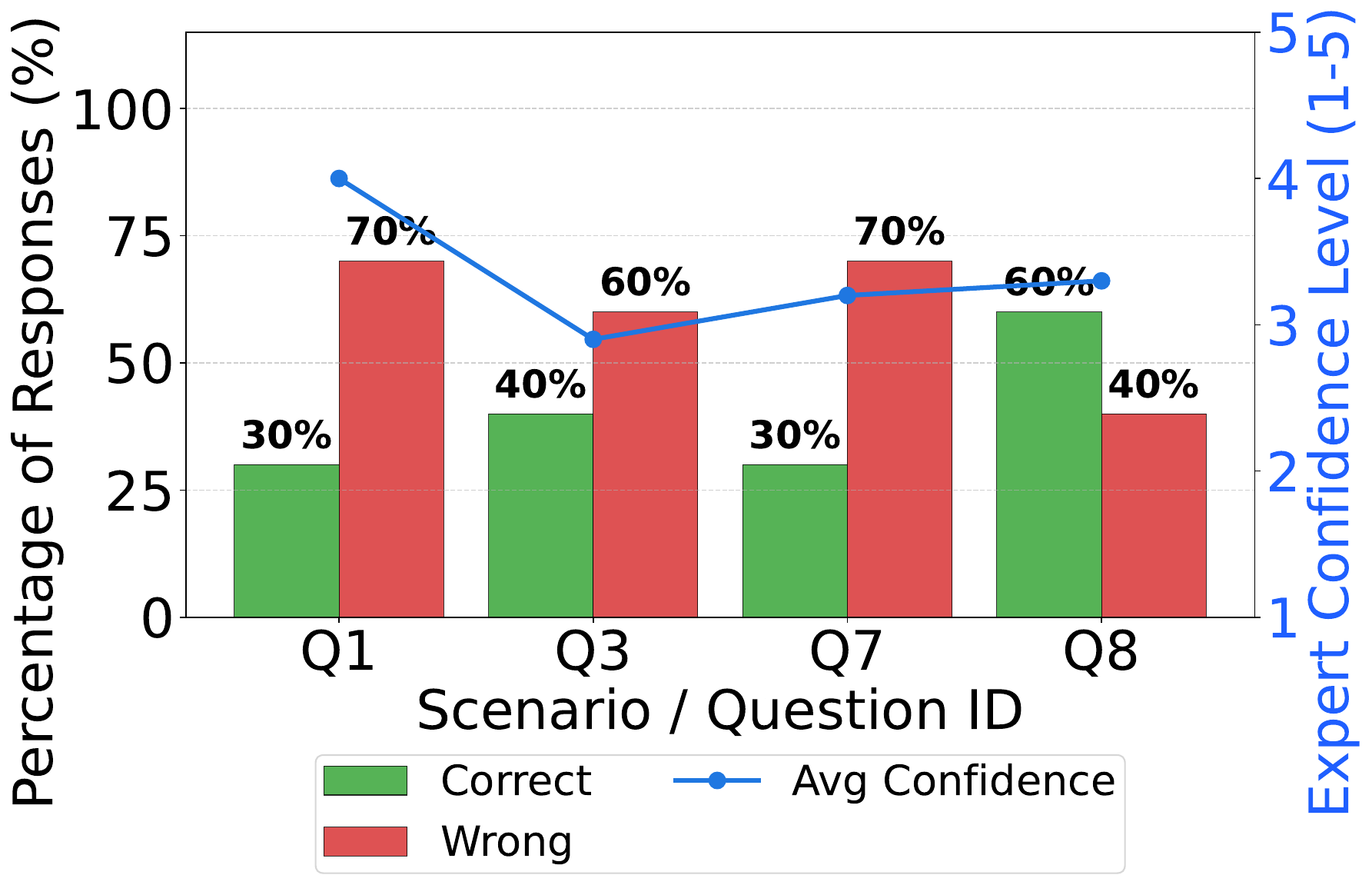} 
        \caption{Expert validation of rule correctness across scenarios.}
        \label{fig:correctnes_confidence}
    \end{subfigure}
    \hfill
    \begin{subfigure}[t]{0.24\linewidth}
        \centering
        \includegraphics[width=\textwidth]{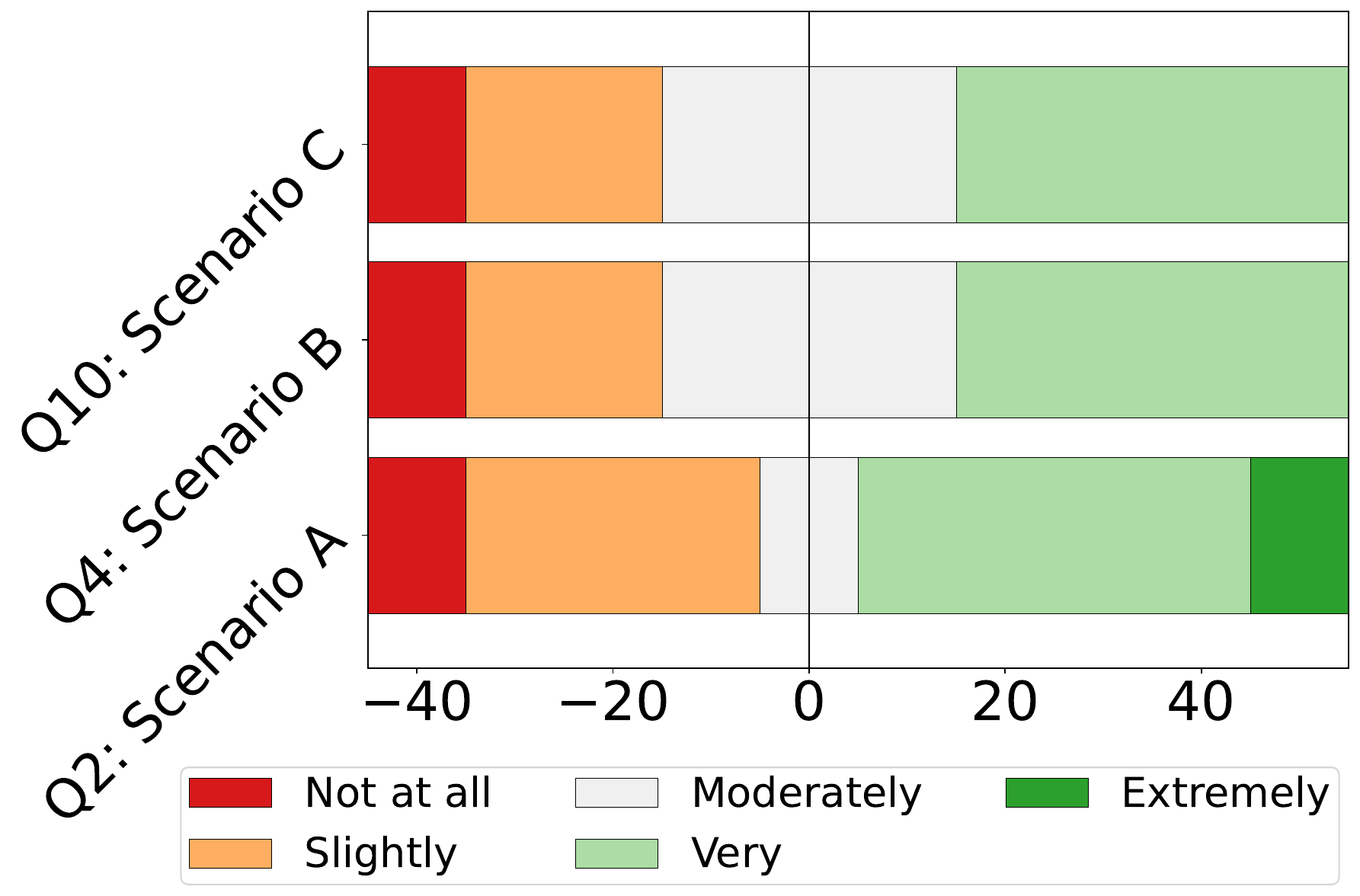}
        \caption{Utility of LLM explanations (Likert Scale).}
        \label{subfig:utility_explanations}
    \end{subfigure}
    \hfill
    \begin{subfigure}[t]{0.24\linewidth} 
        \centering
        \includegraphics[width=\textwidth]{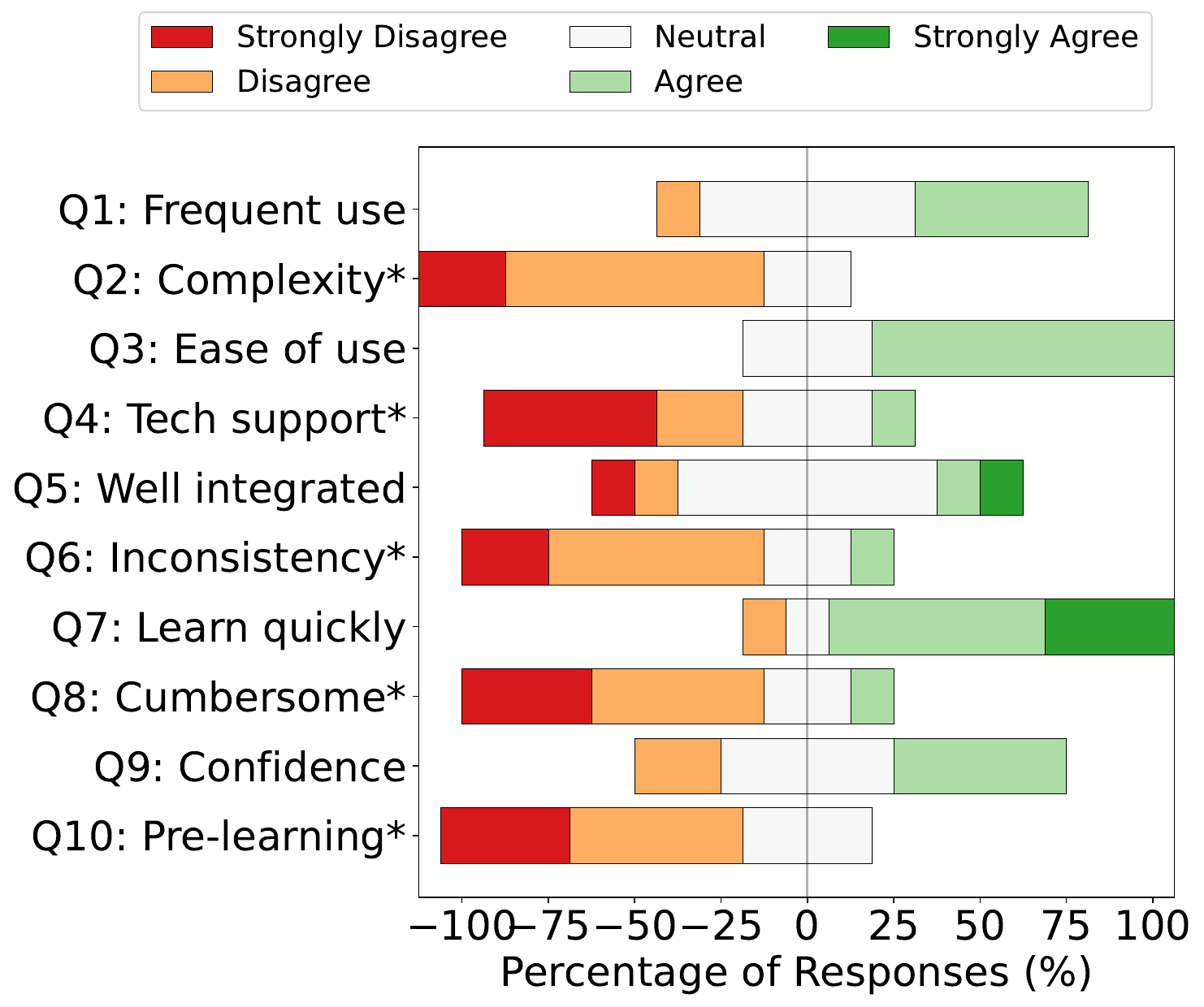} 
        \caption{SUS questionnaire results (Likert scale).}
        \label{fig:sus_diverging_likert}
    \end{subfigure}
    \hfill
    \begin{subfigure}[t]{0.24\linewidth}
        \centering
        \includegraphics[width=\textwidth]{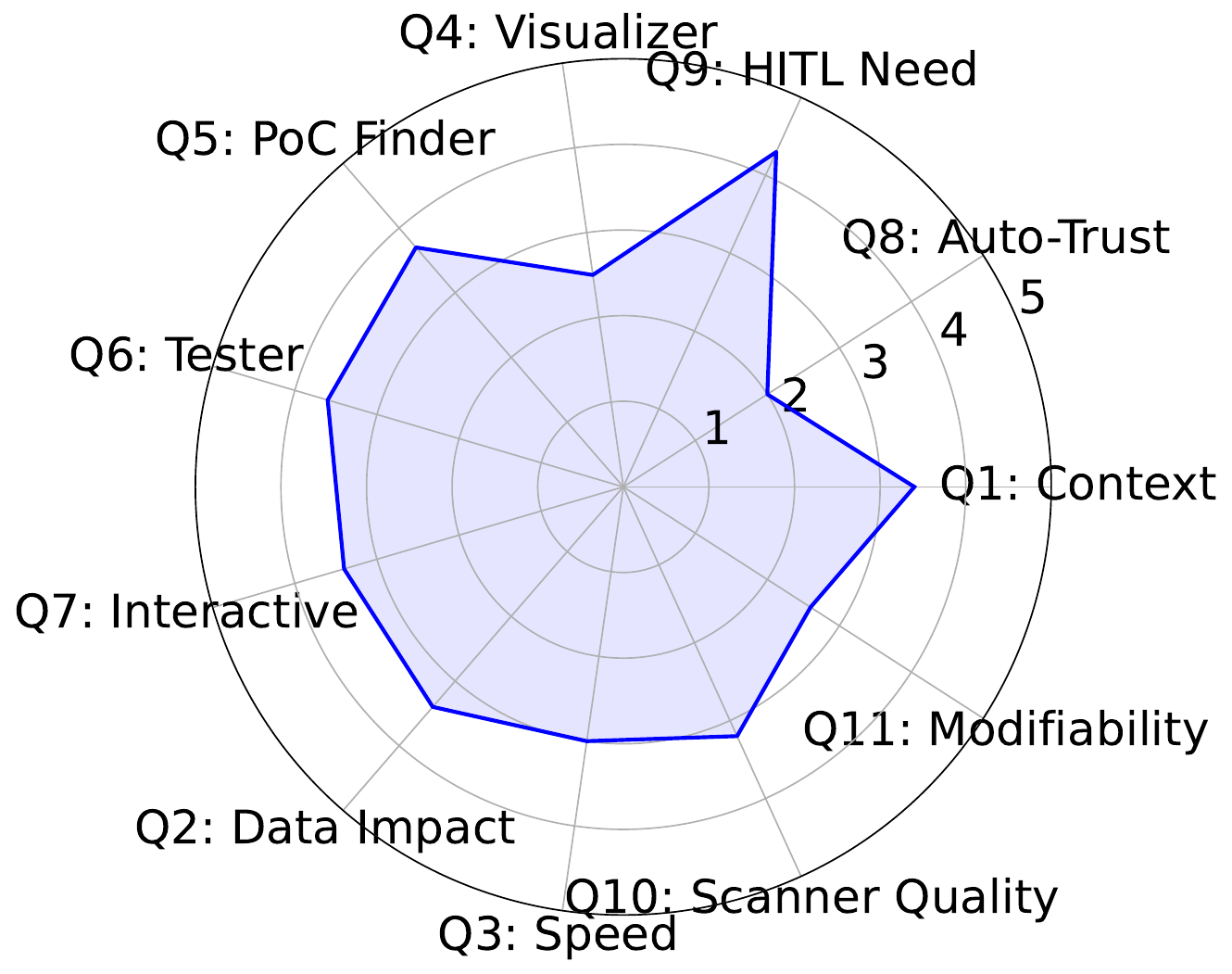}
        \caption{Radar-like chart for 11 statements human evaluation.}
        \label{fig:radar_statement_based}
    \end{subfigure}
    \hfill
     \caption{Human evaluation results summarizing (a) the expert validation of rule correctness, (b) the perceived utility of LLM-generated explanations across scenarios, (c) the SUS questionnaire responses, and (d) the statement-based ratings on rule correctness and the perceived utility of LLM-generated explanations.}
\end{figure*}

To evaluate the technical fidelity of the generated rules, experts audited specific signatures across four distinct scenarios of varying complexity---i.e., topology size, number of \glspl{cve} per device, and number of \glspl{poc}.
The results, summarized in~\Cref{fig:correctnes_confidence}, show that the experts maintained a high degree of critical rigor.
When evaluating SS, despite the 70\% of ``Wrong'' rate (indicating the expert identified a flaw or a need for modification), the average confidence was at its peak ($\mu = 4.0$). 
This suggests that the flaws in the \gls{llm}-generated rules in simpler contexts are transparent and easily caught by human operators.
For example, P6 highlights that ``JumpServer runs on port 2222 and not 22'', while P4 identifies that a generated rule ``detects SMB, while the message says it's about SSH brute force'', thus highlighting contextual inconsistencies of the generated rules.
This feedback also points to additional hallucinations \cite{ye2023cognitive,alansari2026large} that are harder to measure than what was done in Appendix \ref{subapp:halluciantions}, also highlighting the complexity of the rule validation task.
Additional insights into the perceived quality of LLM-generated explanations, including participants’ generally positive feedback, are discussed in \Cref{subapp:llm_explanation}.
%

\finding{Over small-scale scenarios, human users can easily pinpoint issues with generated rules independently of their level of experience.}

P1 also highlights the low specificity of the generated rule, mentioning how ``the payload matching has low specificity'' and that ``there is no check to test that the traffic to be matched is indeed SMB v1, no check for the SMB message type and no check for offset/position at which the bytes to be tested should be at. The condition is so generic that it could trigger in many cases of file transfers.''
Backing our intuition of Finding \ref{finding:specificity_stats}, rule specificity represents a fundamental source of complaint for the human experts across the considered scenarios, with P10 mentioning that ``the rule looks for a specific byte sequence in the SMB traffic without containing an additional context or conditions that could help refine the detection of malicious activity'', P7 stating ``the rule seems to be a bit too general'' and questioning if ``100 alerts within the span of 300 seconds are enough to cause a DoS attack?'', and P3 claiming that ``the rule looks to be too generic to be useful'', thus backing our intuitive findings of \Cref{sssec:rule_quality_analysis}.

\finding{Human users stress the issue of generating specific rules that avoid triggering over normal traffic.}

As scenarios progressed to more complex topologies (Q3 and Q7 in~\Cref{fig:correctnes_confidence}), we observed a slight dip in both accuracy and confidence, which is consistent with the increased cognitive load required to verify multi-node \gls{cve} mappings.
For example, P8 mentions that they are ``not confident if the rule will be useful'', while P5 states that ``the rule appears correct, but \textit{may} trigger false positives when web servers receive more requests from a single host for benign reasons'', thus showing reasonable doubts about their validation of the rule.
This uncertainty is likely to be attributed to the complexity of dealing with a large source of information over an increasingly complex topology, which once again highlights the complexity of the task at hand and the usefulness that an automatic assistant system would have in practice. 
This intuition is backed by the increased correctness (60\%) reached in the final scenario (Q8), where confidence stabilizes at $\mu = 3.3$.
Here, a few participants mention how the rules ``seems to be correctly written'' (P7), ``correctly uses detection filter and the direction is also correct'' (P6) and that the rule ``should work fine'' (P3).
In this context, human struggle to elaborate the completeness of the provided information and rather focus on sub-context of the scenario or generated rule---e.g., ``The content \textsc{union select} is valid'' (P10), ``I am not sure where the string used in the content match originates from'' (P5) and ``I don't think the tracking will work'' (P8).

\finding{On average, human users tend to struggle to analyze in depth the generated rules when the context information grows.}

Interestingly, considering the set of highly expert human users as the participants with self-reported experience of at least 3 years with an IDS engine -- i.e., P1, P4, P5, and P8 -- the perception of the generated rule quality is harsher.
Indeed, these highly expert participants evaluated negatively 11 out of 16 questions  of the generated rules with a high average confidence score $\mu = 4$. 
%
This is true for both small scenarios and large contexts.
For example, in Q7, P1 states that ``while the rule is trying to detect multiple SSH connections, the specific SSH banner is unlikely to be incorrect'' and that ``UDP port 445 is used for SMB over QUIC, which is encrypted, while SMBv1 traffic would happen over TCP'', thus pointing our an hallucination of the agent which is not capable of managing properly the protocol-level knowledge.
Similarly, P4 identifies an issue with the written rule as ``it should contain \textsc{track by} to make it work'', thus stressing once again the lack of specificity.
Therefore, it seems that skilled human experts can pinpoint issues affecting the generated rules even within large contexts, possibly due to an acquired cognitive bias.

\finding{Expert proficiency is a critical factor in rule validation: skilled participants successfully navigate large context volumes to identify issues, while less-skilled users struggle to maintain accuracy under the same conditions.}


In the second part of the study, the focus of questions Q2, Q4 and Q10 is quantifying how much the explanations provided by the \gls{llm} actually helped experts understand the generated IDS rules.
The data, depicted in~\Cref{subfig:utility_explanations} (and~\Cref{tab:explanation_utility}), indicates a high level of consistency in the perceived value of the explanations.
Across all three scenarios (A for the SS topology, B for the SM topology and C for CI), the average confidence/utility score remained steady at approximately 3.0. 
Specifically, in Q2, 50\% of participants rated the explanation utility at a 4 or higher score, suggesting that the textual context significantly lowers the cognitive barrier.
However, as the scenarios became more complex in Q4 and Q10, the distribution became even more concentrated in the middle-to-high range.
A strong majority of participants (70\%) clustered in the ``Moderately'' to ``Very'' helpful categories.
%

%
Having understood the general level of quality of rules that can be generated using \gls{llm} engines, we here aim at understanding to what extent human experts would be willing to deploy the automatically engineered rules in practice.
More in detail, for each of the generated rules tested in the user study, we ask the user if he would be confident deploying such a rule as is.
Consequently, we ask to motivate their choice, aiming to gather some insight into the adjustments required to use the generated rules.
Overall, the participants expressed a rather negative sentiment on the adoption of the automatically generated rules, with them answering ``No'' to the adoption question more than 70\% of the times.
The results are pretty equally distributed across network scenarios, highlighting how the gap between agentic AI definition and its practical adoption is still large. 
%
More in detail, most experts pinpoint some logical issues in the detection process of the rules designed by the LLM-agent which hinder their adoptions, such as ``the content match would have to be adjusted to \textsc{union select}'' and ``the depth would have to be increased'' (P5) or ``the rule misses the detection logic'' (P1), and ``the targeted port being wrong'' and ``the traffic direction being incorrect'' (P6).
As such, most experts agree on the necessity of high human effort to adopt such rules with P5 mentioning how ``essentially, a rule engineer would have to research the vulnerability from the ground up before he can start fixing the rule'', that ``the most work would be to identify the correct sequence to match the malicious behavior in the payload'' (P1) and that ``how to use this rule to detect an attack in practice would require very high effort to figure out'' (P4).
These results are reflected also on the numerical level of effort expressed by the participants, with most of them voting from "Moderate" to "Very High" effort ($\mu = 4.25$).
%
Furthermore, human experts tend to highlight once again the lack of specificity of the engineered rules, causing large amounts of false alerts.
In particular, P3 mentions that some rules would ``only give false positives'' and how rules ``would need to include many more checks to be useful'', while P7 highlights that a rule ``seems to be way to general'' and how it ``would cause a lot of noise''.

Lastly, there exist a few instances of generated rules that were positively assessed by human experts, such as P9 mentioning that ``it would take low to moderate effort to properly modify and make this rule production-ready, mostly tuning rather than rewriting from scratch'', thus suggesting that a few of the generated rules might be used effectively with some effort, thus providing some hope for future investigation.
Appendix \ref{app:avg_modifications_appendix} provides a more detailed analysis of the effort required to correct the LLM-generated rules.

\finding{Expert evaluation indicates that automatically engineered rules require substantial refinement before practical adoption; only 37.5\% of the rules received positive assessments, and even these were met with neutral confidence ($\mu = 3$), suggesting significant skepticism regarding their immediate utility.}
%



\subsubsection{SUS-based System Evaluation}
\label{sssec:sus_evaluation}
To quantitatively assess the user experience of \gls{llm}-based rule generation system, we employed the \gls{sus}-based evaluation~\cite{Bangor29072008}, a robust and industry-standard metric for perceived usability. 
Participants rated ten alternating positive and negative statements on a 5-point Likert scale, ranging from Strongly Disagree to Strongly Agree.

The system achieved a mean SUS score of \susmean{} (standard deviation of \susstd{}) with a 95\% confidence interval of [59.96, 74.04], calculated over the raw scores presented in \Cref{fig:sus_diverging_likert} and Appendix~\ref{subapp:sus_results}, thus satisfying the requirement to consider the system well usable (above 67)~\cite{Bangor29072008,bangor2009determining}.
%
%
%
To better understand the drivers of the system's perceived quality, we analyzed the results across three primary dimensions: \textbf{usability} (Q1, Q2, Q3, Q6, Q7, Q8 and Q9), \textbf{learnability} (Q4 and Q10), and \textbf{integration} (Q5).
The full set of questions is available in Appendix~\ref{ssec:usability_questions}.

In terms of \textbf{usability}, the responses indicate high degree of perceived usability.
Specifically, Q3 (``Ease of use'') received the strongest positive consensus, with 70\% of participants providing a rating of 4.
This is complemented by the results of  Q9 (``Confidence''), where no participant reported a lack of confidence while interacting with the rule generation interface.
The low agreement with  Q2 (``Complexity'') -- where 8 out of 10 participants disagreed that the system was unnecessarily complex -- suggests that the underlying LLM-driven system is presented through a sufficiently abstracted and user-friendly interface, which professionals would easily use.

In terms of \textbf{learnability}, the system demonstrated strong learnability markers, which is critical for tools intended for expert security workflows.
Q7 (``Learn quickly'') showed a significant positive skew, with 8 out of 10 participants ``Agreeing'' or ``Strongly Agreeing'' that most users would adapt to the system rapidly.
This is further supported by Q10 (``Pre-learning''), where the majority of users disagreed with the notion that a significant amount of prior learning was required before they could effectively use the platform.

Lastly, the \textbf{integration} results for Q5 (``Well integrated'') were generally positive, while showing more variance than the usability metrics.
However, the consistent disagreement with Q6 (``Inconsistency'') suggests that the participants found the system's behavior predictable, a vital requirement of the reliability of automated IDS rule engineering.

\finding{Most participants find the LLM-based rule generation agent intuitive and easy to adapt to: 70\% rate its usability positively, and 80\% agree it is not overly complex. Similarly, 80\% believe most users would quickly adapt to it.}

To answer {\bf RQ3}, the LLM-assisted system exhibits favorable usability, with experts rating it efficient and learnable for drafting/explaining rules, yet emphasizing the necessity of human oversight to mitigate limitations in semantic precision and false positives.


\subsubsection{Statement-based Questions Evaluation}
\label{sssec:statement_evaluation}
While the previous \gls{sus}-based evaluation studies the usability of the system, this questionnaire assesses the technical viability and expert trust in an \gls{llm}-based IDS rule generation system.
The raw score distributions are depicted in \Cref{subapp:statement_results}.
Participants perceptions are evaluated across three dimensions: \textbf{trust and autonomy}, \textbf{tool utility} and \textbf{operational impact}.
The results reveal a clear boundary regarding system autonomy. 
While participants expressed moderate confidence in the \gls{llm}'s contextual awareness (Q1, $\mu = 3.40$), there is a near-unanimous consensus on the necessity of a human-in-the-loop (Q9, $\mu = 4.30$), with over 60\% of participants providing the high ratings.
This cautious stance is further evidenced by the low trust in a fully autonomous ``suggest-and-deploy'' mode (Q8, $\mu = 2.00$). 
Notably, 70\% of participants rated their trust for full autonomy at 2 or lower.
Participants identified the internal verification and enrichment modules as the most valuable aspects of the system.
In particular, the \textit{PoC finder} (Q5, $\mu = 3.70$) and the \textit{Snort rule tester} (Q6, $\mu = 3.60$) result as primary drivers of utility.
The high valuation of the interactive generator (Q7, $\mu = 3.40$) indicate that experts value the verification loop, mainly for the ability to test and iteratively refine rules within the same interface.
The inclusion of vulnerability context -- i.e., \gls{poc} and \gls{cve} data -- is confirmed to significantly improve confidence in the final rules (Q2, $\mu = 3.40$).

Regarding the deployment, experts are optimistic about the system's ability to reduce time-to-deploy (Q3, $\mu = 3.00$), although responses remained distributed. 
Interestingly, when comparing rule quality to traditional scanners (Q10, $\mu = 3.20$), the majority of experts (60\%) provided a neutral score, indicating that they view \gls{llm}-based generation as a complementary technology that offers flexibility rather than a complete replacement for legacy logic. 
Finally, the relative ease of modifying \gls{llm}-generated rules (Q11, $\mu = 2.60$) is rated highly, confirming that the system provides a manageable starting point for human operators.

\finding{Experts perceive the system as a high-utility force multiplier for rule drafting and verification, but reject full autonomy in favor of a human-in-the-loop workflow grounded in verifiable vulnerability data.}

\subsubsection{Open-ended Questions Evaluation}
\label{sssec:open_evaluation}
Lastly, we aimed at identifying the feelings of the human experts towards the usage of LLM-based agent for engineering IDS rules.
To this end, we defined a set of open questions that focus on the users' experience with rule writing, LLMs, and their desires for a rule engineering automated assistant. See Appendix~\ref{ssec:open_questions} for the full list of questions.
Here we analyze the feedback gathered from such questions.
The participants experienced with rule engineering backed the results obtained during our preliminary study, with several experts mentioning that either
\begin{inlinelist}
    \item rely on existing rules, adapting them to the context at hand if necessary -- e.g., ``mostly adapting an existing rule that does a similar job'' (P1) and ``using a rule template'' (P10) --, or
    \item construct a rule signature from scratch given a pre-defined context ---e.g., ``using Wireshark and manual writing of the signature'' (P4) and ``I already have a scenario in mind ... and I will derive multiple rules for it'' (P5).
\end{inlinelist}
In this context, few of the prompted experts mentioned having used LLMs in the past in their rule-writing process mainly to check the syntax validity of a rule -- e.g., P1: ``deal with the task of checking the syntax'' and P5: ``understand syntax features'' -- or explaining complex rule templates---e.g., ``I have used LLMs to explain existing rules'' (P5) and ``Explaining complex rules'' (P8).
However, no human expert mentioned being successful in leveraging LLMs to generate a valid and reliable rule from scratch.
Thus, we highlight how human experts seem to be aware of the LLM limits when it comes to engineer IDS rules from scratch, as testified by P1 explicitly mentioning that he sometimes used LLMs in the past, but expressly stating how ``the detection characteristics were provided by me''.

\finding{Human experts seem to be aware of the limits of generative AI tools for IDS rule engineering, with most of them using LLMs only sporadically to check syntax validity or provide further insights on already existing rules.}

The above findings are backed by the overall feeling when answering the question ``if this tool was integrated into your daily SOC workflow, how would it change your team's priorities?''.
Indeed, while some participants were positive about the system adoption, mentioning how such tool could ``make creating rules easier'' (P4) or how ``the team could spend less time writing and tuning basic IDS rules'' (P10), several participants were cautious about the adoption of the full system in an en-to-end rule generation fashion.
For example, few experts focused on the fundamental principles that the system provides to understand the context and help the SOC team while not being interested in the complete automatic rule generation with P5 mentioning that ``it could be valuable in understanding the network topology and identifying vulnerabilities'' but also that they do not ``think the LLM module would be used a lot'' or how ``it could be a nice tool to give priority to the team on what to focus on'' (P4) or to ``relieve from syntax rule checking'' (P1).
To this end, P3 response is emblematic ``it would make it less fun, because the cognitive task will be mostly automated away'', as it highlights the adversity of SOC experts against completely automated pipelines.
Finally, one participant expressed a fairly negative judgment, mentioning that ``at the moment, [the system] wouldn't [be helpful]'' (P8).

\finding{Experts see the possible benefits of introducing assistance tool, but are skeptical of automating end-to-end the rule engineering pipeline.}

To back the above, there is the feedback we received from the participants while answering to the questions
\begin{inlinelist}
    \item\label{q_teach_llm} ``if you could 'teach' the LLM one specific nuance about your unique network environment that is not in the IP/Service list, what would it be'', and 
    \item\label{q_missing_info} ``what specific information was missing from the system's output that would have helped you make a faster decision?''.
\end{inlinelist}
In this context, rather than focusing on tools or components that would help improve the end-to-end automatic engineering of rules, human experts focused on asking for components of the systems that would help the human analyst understanding the context at hand, the actual performance of the automated rules or the impact they might have.
For example, to answer \ref{q_missing_info}, P5 mentions how ``it would be nice to have some sort of CVE prioritization'', while P1 highlights the need to add a tool to measure the ``detection performance with sample attack PCAPs'', and P10 asks for the addition of ``suggestions of mitigation strategies and impact assessment''.
Similarly, in answering \ref{q_teach_llm} P1 mentions ``what traffic it can expect to observe and what traffic is not being captured at all'', while P10 suggests ``the operational context of devices and services within the network, for example: device criticality and role and the past incidents patterns'', once again highlighting the human perspective on sound and complete understanding of the given context rather than perfectionist rule engineering perspective.
This behaviour is likely rooted in human experts having clearly in mind the risks that a completely automatic rule engineering system would bring as they deeply value the human-in-the-loop perspective.
For example, when answering the question ``who is 'responsible' for a false negative if an LLM-generated rule is accepted by an expert but fails in production: the model, the system designer, or the human reviewer?'', most participants agree on suggesting the human reviewer as the bearer of most responsibilities, with P3 stating ``all of the above, but in practice the human reviewer'' while P10 and P1 mention ``the human reviewer bears the most direct responsibility'' and ``I think the human reviewer would ultimately be responsible''.
These answers suggests how human analysts envision a human-in-the-loop partially-automated rule engineering system, where the human supervises the rule engineering process and suggests possible modifications or improvements rather than completely relying on AI tools---thus also backing our system vision of \Cref{sec:system_formalization}.
In this context, it is interesting to note P1's answer, as they also highlight how ``over time, the system might influence the way human reviewers approach detection'', implicitly raising concerns about the practical usage of end-to-end automatic rule engineering.
P1's answer seem to suggest how over time the overly reliance on fully automated rule engineering pipelines may render the human analysts more prone to accept generated rules without thorough testing, and thus raising reliability issues.

\finding{Participants tend to stress the need for carefully designing human-driven systems rather than end-to-end automatic rule engineering perspectives, as they have clearly in mind the importance of the rule engineering task and the human analyst's responsibilities.}

To answer {\bf RQ2}, experts predominantly engage in iterative human-in-the-loop refinement via adaptive prompting and network visualization, correlating higher perceived rule quality with PoC integration and topology context, though cognitive load rises with network complexity, favoring verification over autonomous generation.

\section{Related Work}
\label{sec:related_work}

\noindent {\bf User study on NIDS Rule Engineering}
\label{ssec:user_study_rule_engineering}
The automatic generation of IDS rules has garnered significant attention in recent years and recent research suggests that it is a challenging process.
Although not all of the works focus specifically on rule engineering, they provide insights on how \glspl{soc} operate in practice.  
Among them, Teuwen at al.~\cite{teuwen2025ruling} investigate the design of IDS rules within \gls{soc}, identifying key design principles consolidated through interviews with experienced \gls{soc} rule designers, while alert alchemy~\cite{vermeer2023alert} examines the processes involved in the creation, management, and acquisition of IDS rules, successively deeply looking into the evolution of rulesets in SOC production environments by analyzing over 130 thousands of rules and 62 million alerts~\cite{vermeer2022ruling}.
Differently, Jansen et al.~\cite{jansen2024comparative} tackles the practical issues with the creation of graph-based IDS rules as opposed to log-based detection rules via a qualitative user study, while Alahmadi et al.~\cite{alahmadi202299} quantified the burden of alert fatigue -- 99\% of the generated alerts are false positives -- via interviews and observations of practitioners.
Finally, Singh et al.~\cite{singh2025llms} studied real-world queries of LLMs from SOC analysts to discover that \glspl{llm} are integral part of the \gls{soc} as cognitive aids.

While these studies investigate the operational realities of rule engineering and nascent \gls{llm} integration, they leave critical questions unanswered about expert-\gls{llm} collaboration in the domain of \gls{nids} rule generation. 
Our study breaks new ground by revealing to what extent security professionals are willing to interact with, validate, and refine \gls{llm}-generate rules in realistic scenarios.\\

\noindent {\bf LLMs for NIDS Rule Engineering}
\label{ssec:llms_rule_engineering}
With the advent of \glspl{llm}, their use for \gls{nids} tasks has been increasing.
Our work, builds upon these works and provides a better understanding of the practicality of \glspl{llm} for \gls{nids} rule engineering via the user study.
Some of these works introduce an \gls{llm}-driven framework that automatically generates rules from code, like RuleMaster+~\cite{lian2025rulemaster+} -- using \gls{poc} --, RulePilot~\cite{wang2025rulepilot}, and RuleXploit~\cite{papoutsis2025rulexploit} -- using exploit samples.
In these cases, LLMs perform a deep code analysis and identify features from exploit payloads.
Other works leverage LLMs as agents, like Hu et al.~\cite{hu2024llm}, who extract threat features from multiple input sources and transform them into actionable signatures, and GRIDAI~\cite{li2025gridai}, who focuses on generalizing existing signatures, and LLM4Rule~\cite{du2025harnessing}, which is a framework for generalization of the rule.
Differently, Moreno et al.~\cite{moreno2025leveraging} tackles the LLM-based rule generation in the Industrial Control Systems (ICS), via the integration of a domain-specific knowledge into the LLM prompting process, while Hex2Sign~\cite{balasubramanian2024hex2sign} focuses on the aspect of converting raw network telemetry into human-readable IDS signatures, with LI-NIDS~\cite{abdennebi2025li} generating \gls{nids} rules from numerical network flow features transforming network flow data into an abstract language that the LLM can process.
FALCON~\cite{mitra2025falcon} proposes the generation of rules from unstructured Cyber Threat Intelligence (CTI) while also providing a validation pipeline, finally RuleLLM~\cite{lin2026rulellm} prompts the LLMs for characterization of benign and malicious behavior to extracts distinctive features for both classes. 

\section{Conclusion}
\label{sec:conclusion}

This paper presents a human-centered investigation into the practical viability of \glspl{llm} for \gls{nids} rule engineering. 
We formalize a state-of-the-art grounded system and rely on a user study to gather experts feedback on the semantic correctness of LLM-generated rules and their adoptability.
Our findings reveal a critical syntax-semantics paradox: while LLMs construct syntactically valid rules, experts deem them only partially deployable due to their lack of specificity, over-reliance on content matching and susceptibility to hallucinations. 
Furthermore, while the system’s usability was rated favorably, a fundamental skepticism persists among practitioners. 
Experts currently view LLMs as auxiliary tools, useful for syntax validation and natural language explanation, rather than autonomous generators. 
This underscores a pivotal requirement for human-in-the-loop architectures that prioritize the analyst's agency and decision-making workflow over automation.
Ultimately, this work serves as a foundational assessment of LLMs in the rule engineering domain. 
By delineating LLMs limitations in this domain, we provide a roadmap to develop more robust, reliable, and human-aligned rule engineering workflows.

\bibliographystyle{plain}
\bibliography{references}  





\appendix
\section{Open Science}
\label{secapp:open_science}


The built framework leverages Ollama~\cite{ollama2023} for local LLM orchestration, following the three-phase workflow illustrated in Figure~\ref{fig:system_formalization}. 
Initially, the system provides the LLM with structured environmental context, including network topology, CVE identifiers, and node connectivity. 
Subsequently, the model synthesizes Snort-style IDS rules and corresponding technical justifications. 
Finally, each rule undergoes an automated verification loop where a syntax validator and the LLM iteratively identify and rectify errors, ensuring rule integrity through a traceable audit log of modifications.

Beyond enabling the replication of our results, this codebase serves as a modular foundation for future research in automated security policy generation. 
By releasing this framework, we provide a reusable and traceable environment for exploring the integration of LLMs into IDS rule generation automation.

\section{Ethical Considerations}
\label{secapp:ethical_considerations}
This study involves a human-centered evaluation of LLM-generated IDS rules, and we have taken several measures to ensure it adheres to the highest ethical standards. 
The recruitment process and the survey instrument were conducted in strict accordance with the ethical guidelines and internal protocols of the lead authors' institution. 
Prior to participation, all individuals were provided with comprehensive instructions regarding the nature and content of the study. 
We obtained explicit informed consent from each participant before they commenced the survey, ensuring they were aware of their right to withdraw at any point. 
Furthermore, the evaluation was designed to protect participant privacy; the survey did not solicit, collect, or store any sensitive personal information or personally identifiable information (PII). 
All data used in our analysis is anonymized and treated in aggregate to ensure that no individual respondent can be identified.

\section{Recruitment Details}
\label{app:recruitment_details}
To identify suitable participants for our study, we adopted a multi-stage recruitment strategy combining prompt-based discovery and structured academic search queries. 
This approach allowed us to target both academic and industry researchers working at the intersection of cybersecurity, intrusion detection systems, and LLMs.
\begin{lstlisting}[breaklines=true, style=llmprompt, caption={Recruitment Prompts}, label={lst:recruitment_prompt}]
Prompt 1: "Can you find list of researchers in academia and industry that on their Scholar or Scopus profiles have one of the following keywords or a combination of those: "Cybersecurity", "AI", "LLMs" and "IDS"?"
Prompt 2: "We will soon start a user study on the use of LLMs for Intrusion Detection Systems rule generation. For context, Snort and Suricata tools work based on rules that are usually written and generated by human analysts. In this study, we have tested different LLMs on the task of rule generation given several input informations such as the network topology, CVE list per each node in the topology, services for each node in the topology. Can you search for researchers (both academic and industry) that work in this domain?"
\end{lstlisting}

We first leveraged natural language prompts (Listing~\ref{lst:recruitment_prompt}) to identify researchers whose profiles explicitly mention relevant keywords such as \textit{Cybersecurity}, \textit{AI}, \textit{LLMs}, and \textit{Intrusion Detection Systems}. 
These prompts were designed to capture a broad yet relevant pool of candidates by querying publicly available academic and professional profiles. 
The second prompt further refined the search by focusing on researchers with demonstrated experience in applying LLMs to IDS rule generation, explicitly referencing tools such as Snort and Suricata and contextual inputs like CVE data and network topology.

\begin{lstlisting}[style=scholar, caption={Recruitment Google Scholar Queries}, label={lst:recruitment_scholar_queries}]
Query 1: "intitle:("Large Language Model" OR LLM) AND (IDS OR "Intrusion Detection") AND ("Rule Generation" OR "Signature Generation")"
Query 2: "label:network_security label:LLM"
\end{lstlisting}

To complement the prompt-based approach, we conducted structured searches on Google Scholar using carefully designed queries (Listing~\ref{lst:recruitment_scholar_queries}). 
Query 1 targeted publications explicitly mentioning LLMs in the context of intrusion detection and rule or signature generation. 
Query 2 leveraged Google Scholar labels to identify researchers associated with both network security and LLM-related topics.


\section{Rule Engineering Systems Comparison}
\label{app:compairison_sys}

\begin{table*}[ht] 
\centering
\small 
\rowcolors{2}{gray!10}{white} 
\renewcommand{\arraystretch}{1.4} 
\scalebox{0.75}{
\begin{tabularx}{\textwidth}{l c l X X c c c c}
\toprule
\textbf{Paper} & \textbf{Hum.} & \textbf{Prompting} & \textbf{Inputs} & \textbf{LLMs} & \textbf{Chk.} & \textbf{Cor.} & \textbf{LP} & \textbf{HP} \\
\midrule

\cite{hu2024llm} & $\times$ & CoT & Vuln. report, malicious payload, ET rules & GLM4 & Custom & Custom & $\times$ & $\times$ \\

\cite{lian2025rulemaster+} & $\times$ & Role & External KB, PoC & LLaMA, ChatGLM, Vicuna & $\times$ & $\times$ & $\times$ & $\times$ \\

\cite{du2025harnessing} & $\times$ & Role & PCAP & GPT-3.5, GLM-4, GPT-4o & $\times$ & $\times$ & $\times$ & $\times$ \\

\cite{moreno2025leveraging} & $\times$ & ZS, FS, CoT & Processed PCAPs & GPT-4o-mini, GPT-3o-mini, Claude, Gemini & Suricata & LLM & $\checkmark$ & $\times$ \\

\cite{balasubramanian2024hex2sign} & $\times$ & Fine-tuning & Honeypot captures & BERT-based models & $\times$ & $\times$ & $\times$ & $\times$ \\

\cite{abdennebi2025li} & $\times$ & Role & Custom prompt template & LLaMA, Falcon & $\times$ & $\times$ & $\times$ & $\times$ \\

\cite{mitra2025falcon} & $\checkmark$ & Role & CTI data, YARA rules & GPT-4o, Llama 3.3, Qwen 3, Mistral, Granite, Phi-4 & $\checkmark$ & $\checkmark$ & $\checkmark$ & $\checkmark$ \\

\cite{lin2026rulellm} & $\times$ & Fine-tuning & -- & -- & $\times$ & $\times$ & $\times$ & $\times$ \\

\cite{li2025gridai} & $\times$ & Role & PCAP & GPT-4.1, GLM-4 Flash & $\times$ & $\times$ & $\times$ & $\times$ \\

\cite{wang2025rulepilot} & $\times$ & CoT & SIEM information & GPT-4o, LLaMA-3, DeepSeek-V3 & SIEM & $\checkmark$ & $\times$ & $\times$ \\

\cite{papoutsis2025rulexploit} & $\times$ & Role+Temp & Exploit, CVE, rule & GPT-4o & Suricata & LLM & $\times$ & $\times$ \\

\bottomrule
\end{tabularx}
}
\caption{Comparative analysis of LLM-based IDS rule generation frameworks. Abbreviations: Hum. (Human Evaluation), Chk. (Syntax Checker), Cor. (Automated Corrector), LP (LLM Prompts), HP (Human Prompts), ZS/FS (Zero/Few-Shot).}
\caption{Comparison of LLM-based IDS rule generation approaches}
\label{tab:appendix_comparison}
\end{table*}

\Cref{tab:appendix_comparison} provides a comparative overview of recent approaches that leverage \gls{llm} for \gls{nids} rule generation. 
Most approaches rely on prompt engineering techniques such as role prompting or chain-of-thought (CoT), while only a few explore zero-shot or few-shot configurations. The input data varies across studies, ranging from structured network traffic (e.g., PCAP files) and vulnerability reports to unstructured cyber threat intelligence (CTI) and exploit descriptions. 
Regarding model usage, both proprietary and open-source LLMs are widely adopted, including GPT-4 variants, LLaMA-based models, and domain-specific or fine-tuned architectures. However, explicit syntax validation and correction mechanisms remain limited, with only a subset of methods incorporating dedicated parsers or grammar checks tailored to IDS frameworks such as Suricata or SIEM systems.
Human involvement is generally minimal, with most systems operating in an automated or semi-automated fashion. Only a few approaches incorporate human feedback loops or adaptive prompt refinement. Overall, the table highlights the growing diversity of LLM-based IDS rule generation methods while also revealing a lack of standardized evaluation pipelines and systematic human-in-the-loop integration.

\section{Synthetic Network Generation}\label{app:net_gen}
To ground our user study in realistic and controlled environments, we generated the synthetic vulnerable networks using the existing benchmark on vulnerable network generation~\cite{palma_behind_2025}.
It generates reachability graphs of any network topology and pairs them with vulnerability inventories derived from real-world CPE databases (e.g., Windows, MySQL, Apache).
By benchmarking network size, running services, operating systems, and hardware, the generated scenarios emulate enterprise segments (small office, mid-sized firm, multi-segment infrastructure), and the realism of which is validated against real networks via statistical analysis.

\section{Selecting Best LLM for User Study}\label{app:best_model_selection}
To enable the human-based evaluation of the generated rules, we here address the issue of selecting which rules to prompt experts with.
Indeed, in our experimental evaluation, we considered six different network scenarios and 18 LLM engines, generating a total of 1679 syntactically valid rules.
However, it would be impossible to evaluate all such rules given the difficulty of reaching highly skilled experts working in the domain.
Therefore, we here consider to identify one LLM engine as the best performing agent whose generated rules will be used for the human experts user study.
Meanwhile, we integrate all six network scenarios in the user study, with the aim of capturing the human confidence towards the LLM-based generated rules whenever the task complexity varies.

We identify two parallel approaches to identify the best performing LLM agent either relying on 
\begin{inlinelist}
    \item an aggregation of statistical metrics concerning the syntax and semantical quality of the generated rules, or
    \item an LLM-as-a-judge setting \cite{gu2025surveyllmasajudge,li2025generation} to compare a subset of the rules generated by each model over a round-robin tournament.
\end{inlinelist}
For the former, we consider as relevant information to identify the best performing LLM engines the following statistical parameters:
\begin{inlinelist}
    \item the percentage of syntactically correct rules written (higher percentage hints to the model having a good intuition about the Snort grammar),
    \item the average number of corrections required to generate a valid rule (as fewer required modifications hint higher level of comprehension by the LLM),
    \item the average number of parameters in a generated rule (more parameters should translate to a highly detailed and possibly precise rule),
    \item the average rule entropy (higher entropy identifies reacher and more diverse rules),
    \item the percentage of hallucinated rules (fewer hallucinations means the model can handle correctly all the network information), and 
    \item the size of the LLM engine (in practice, we would like to favour smaller models to ease real-world deployment).
\end{inlinelist}
As such we compute a statistically-driven metric as the weighted sum over the normalized values of these statistical parameters---e.g., we aggregate the rules generated by each engine across all network scenarios, compute the statistics, normalize them and sum.

For the latter, we run a round-robin LLM-as-a-judge tournament in which each candidate engine is compared against every other one. 
For each pair of engines, we randomly sample 11 rules from each model and compare them pairwise. 
Each comparison is evaluated independently by a judge model -- we provide the experimental results when using ChatGPT 4.5 nano as the judge, but similar results are obtained for ChatGPT 5.4 mini and ChatGPT 5.4 -- which select the preferable rule. 
The winning rule -- and thus the corresponding LLM engine -- receives one point, while the losing rule is assigned zero points. 
After all matches are completed, engines are ranked by their cumulative score across the tournament, and the top-ranked engine is selected.

\begin{figure}[htbp] 
    \centering
    \begin{subfigure}[t]{0.48\linewidth} 
        \centering
        \includegraphics[width=0.75\textwidth]{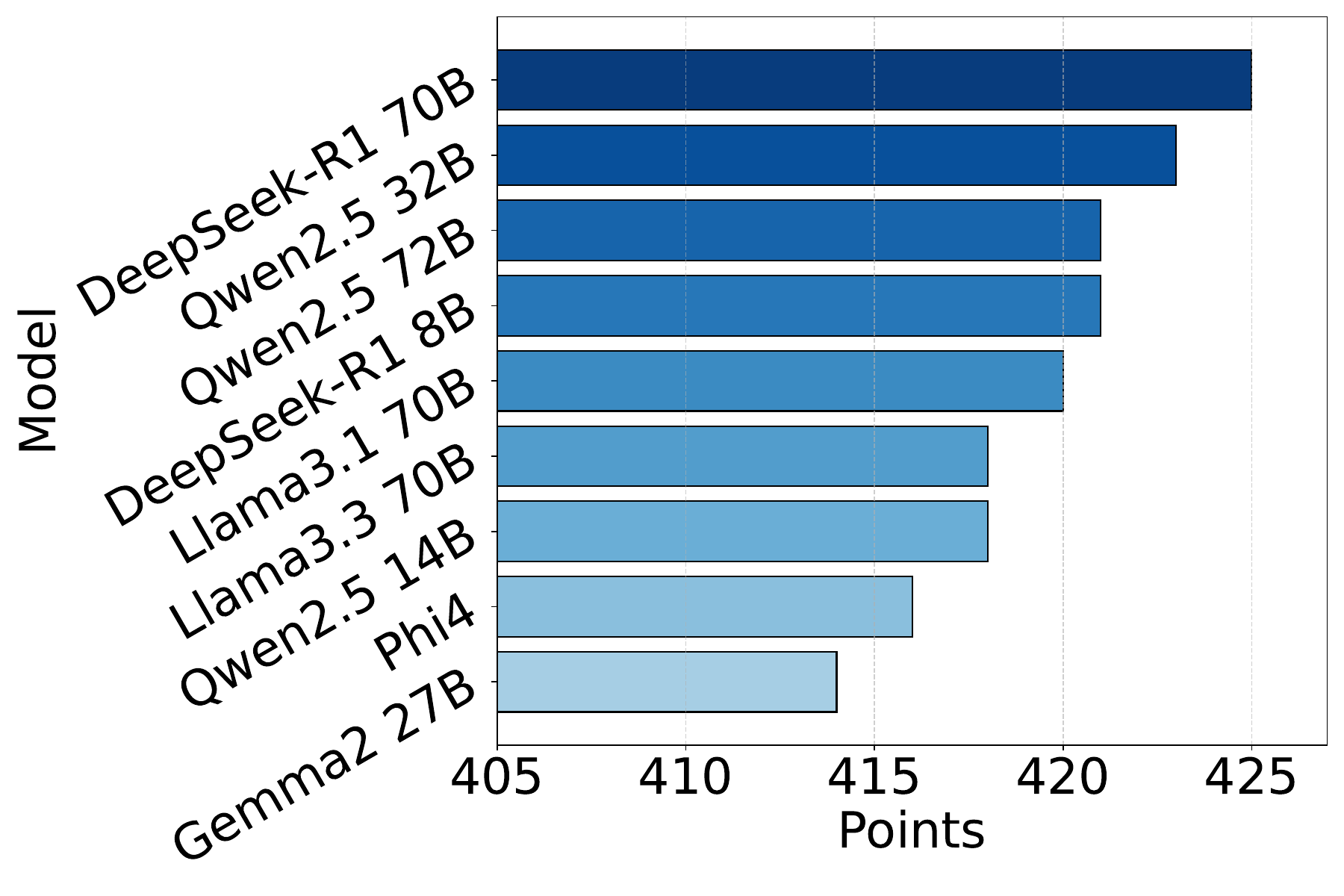} 
        \caption{LLM engine ranking results of the LLM-as-a-judge tournament to evaluate the quality of generated rules. DeepSeek-R1 70B is the model generating the best rules according to the judge.}
        \label{subfig:llm_judge_rank}
    \end{subfigure}
    \hfill
    \begin{subfigure}[t]{0.48\linewidth}
        \centering
        \includegraphics[width=0.75\textwidth]{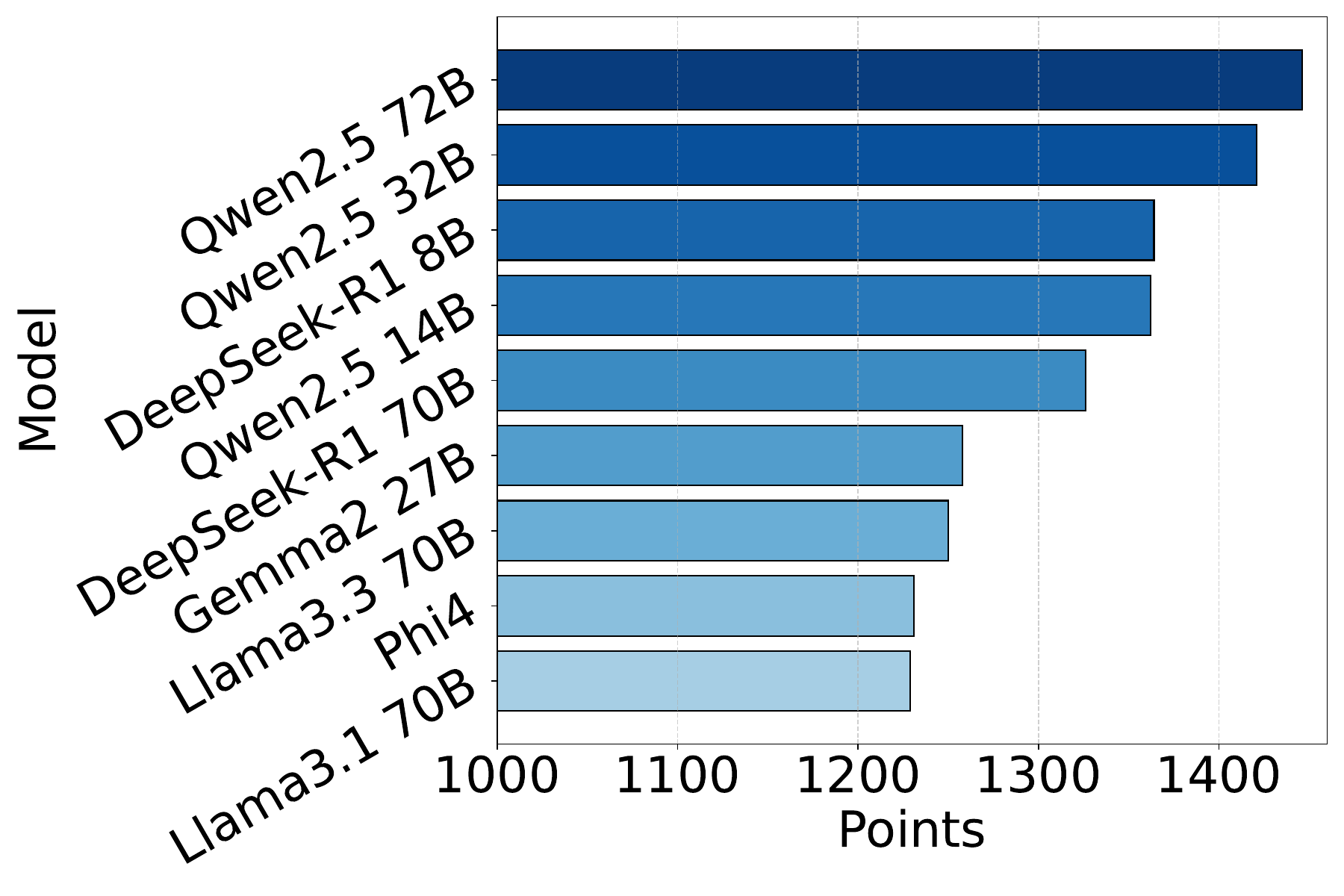}
        \caption{LLM engine ranking results of the statistical-based evaluation of the generated rules quality. DeepSeek-R1 70B is the model generating the best rules according to our statistical metric.}
        \label{subfig:statistical_rank}
    \end{subfigure}
    \hfill
    \caption{Analysis of the LLM engine to be used in our experts user-study. LLM engines are ranked according to two performance metrics to identify the best LLM-based agent.}
    \label{fig:ranking}
\end{figure}

\Cref{fig:ranking} provides the comparison between the engine ranking computed using the LLM-as-a-judge paradigm and the statistical metric we defined.
Both rankings prove how larger models are more effective engines for generating valid rules, with many of them performing comparably well.
Surprisingly, in the LLM-as-a-judge setting, most LLM engines perform similarly which highlight the difficulty of relying on LLMs to test the quality of given Snort rules.
Grounded on these results, we select DeepSeek-R1 70B as the LLM engine used to generate the rules to be tested in our human study.
We select DeepSeek-R1 70B as it is the best performing model for both ranking, its relatively lighter nature -- w.r.t. other large LLMs -- and its capability of leveraging Chain-of-Thought \cite{wei2022chain}.
Unsurprisingly, smaller engines provide to be less effective, as the generated rules are more vague when correct, being characterized by very few fields and general headers that are not usable in practice, characteristics that both the LLM judge and our metric grasp correctly. 
%


\section{User Study}
\label{app:user_study}

\subsection{Demographic Questions}
\label{sapp:demographic_questions}

The demographic information page of the online platform includes the following information:
\begin{itemize}[leftmargin=*]
    \item \textbf{Current Primary Role:} (e.g., SOC Analyst, Security Researcher, Threat Hunter, Network Engineer)
    \item \textbf{Sector:} (e.g., Academia, Finance, Critical Infrastructure/ICS, Government, Managed Security Service Provider).
    \item \textbf{Organization size:} (e.g., 1-10 employees, 10-50 employees, 50+ employees).
    \item \textbf{Age:} 18-24, 25-34, 35-44, 45-54, 55-64, 65+
    \item \textbf{Country/Region:} Relevant to check participation and interest in automation.
    \item \textbf{Education Level:} (e.g., MSc, PhD). This is standard for academic papers in this domain.
    \item \textbf{Years of Experience in Rule Engineering:} (e.g., <1, 1-3, 3-5, 5-10, 10+ years).
    \item \textbf{Primary IDS Engine Used and familiarity:} (Checkboxes for Snort, Suricata, Zeek/Bro, Other)
    \item \textbf{Frequency of Rule Writing:} (e.g., Daily, Weekly, Monthly, Rarely). This helps distinguish between people who manage the systems and those who actually craft the signatures.
    \item \textbf{Self-Reported Proficiency:} Ask them to rate their comfort level with writing complex signatures (e.g., using pcre, flowbits, or byte\_jump keywords) on a scale of 1-5.
\end{itemize}

\subsection{Usability Questions}
\label{ssec:usability_questions}
To assess the perceived usability and practical value of the proposed system, we designed a set of Likert-scale questions targeting key dimensions of user experience, trust, and operational effectiveness. These questions aim to capture expert judgments on multiple aspects of the system, including confidence in LLM capabilities, the impact of contextual security information (e.g., CVEs and proof-of-concepts), and the perceived benefits in terms of efficiency, such as reduction in time-to-deploy.
In addition, the questionnaire evaluates the usefulness of individual system components in supporting the rule writing workflow.
Finally, the questions explore comparative and practical dimensions, such as how the quality of LLM-generated rules is perceived relative to traditional automated tools, and the difficulty of modifying such rules in practice. All items are measured on a five-point Likert scale, enabling a structured and quantifiable assessment of usability and perceived system effectiveness across participants.
The questions are the following:
\begin{itemize}[leftmargin=*]
    \item How confident are you in an LLM's ability to understand complex network topologies with several security contexts (e.g., CVEs and PoCs) without human oversight?
    \item How much does the inclusion of vulnerability data (e.g., PoCs) change your confidence in the generated rules?
    \item How much do you think the system would reduce the time-to-deploy?
    \item How valuable is the network topology visualizer for the rule automation process?
    \item How valuable is the proof-of-concept finder for the rule automation process?
    \item How valuable is the Snort rule tester for the rule automation process?
    \item How valuable is the interactive rule generator for the rule automation process?
    \item How much would you trust this system to run in an autonomous ``suggest-and-deploy'' mode?
    \item How necessary do you feel a human-in-the-loop is in this system?
    \item How does the rule quality of this system compare to rules generated by traditional automated vulnerability scanners?
    \item How would you rate the difficulty of modifying LLM-generated rules? 
\end{itemize}

\subsection{Open-ended Questions}
\label{ssec:open_questions}
The following questions are designed to gather insights into your experience with NIDS rule creation and the use of LLMs in rule engineering. Your responses will help evaluate usability, effectiveness, and potential areas for improvement in such systems.
\begin{itemize}
    \item How do you currently approach NIDS rule creation? (e.g., manual writing, adapting templates, or using automated scripts?)
    \item Have you previously used any LLMs for rule engineering tasks? If so, how did you use them?
    \item Did you notice any instances where the LLM generated syntactically correct rules that were logically flawed for the specific topology?
    \item Did you leverage external tools/information when taking this user study that helped you writing the rules? If yes, please specify the source of information.
    \item What specific information was missing from the system's output that would have helped you make a faster decision?
    \item What `logic style' did you perceive in the rules generated? Did they feel overly aggressive (high false positives) or too conservative?
    \item If you could `teach' the LLM one specific nuance about your unique network environment that isn't in the IP/Service list, what would it be?
    \item Who is `responsible' for a false negative if an LLM-generated rule is accepted by an expert but fails in production: the model, the system designer, or the human reviewer?
    \item Describe the `friction' you felt while modifying the LLM's output. Was it easier to fix a bad rule or would it have been faster to write it from scratch?
    \item If this tool were integrated into your daily SOC (Security Operations Center) workflow, how would it change your team's priorities?
\end{itemize}

\subsection{Cognitive load}
\label{ssec:cognitive_load}
To assess the impact of task complexity on participants’ performance, we designed three network scenarios with progressively increasing structural and informational complexity (\Cref{tab:network_scenarios}). 
These scenarios were specifically constructed to induce different levels of cognitive load while remaining representative of realistic operational environments in intrusion detection settings.
The smallest scenario models a minimally segmented network, consisting of five nodes and a limited number of vulnerabilities. This configuration reflects environments such as small office or home office deployments, where the reduced diversity of assets and attack surface enables relatively straightforward reasoning about rule generation.
The intermediate scenario increases both the number of nodes and the distribution of vulnerabilities, approximating a typical enterprise network segment. In this setting, participants must account for a broader range of services and potential attack paths, thereby requiring more complex reasoning and integration of multiple pieces of information.
The largest scenario represents a multi-segment infrastructure with 18 nodes and a substantially expanded set of vulnerabilities. This configuration introduces layered complexity, where participants must simultaneously consider multiple hosts, services, and vulnerability interactions, significantly increasing the cognitive demands of the task.
To ensure that the scenarios remained analytically tractable while still challenging, we constrained the number of vulnerabilities to at most two CVEs per node. This design choice prevents an exponential increase in possible rule combinations, while still preserving sufficient complexity to meaningfully differentiate cognitive load across scenarios.
Overall, the three scenarios provide a controlled yet realistic progression of difficulty, enabling us to systematically evaluate how increasing cognitive load affects the rule generation process.
\begin{table}[htbp]
\centering
\scalebox{0.85}{
\begin{tabular}{|c|c|c|c|}
\hline
\textbf{Scenario} & \textbf{Network Size} & \textbf{CVEs per node} & \textbf{Total CVEs} \\ \hline
Scenario 1  & 5 nodes & 2 & $\approx 10$ \\ \hline
Scenario 2  & 10 nodes & 2 & $\approx 20$ \\ \hline
Scenario 3  & 18 nodes & 2 & $\approx 36$ \\ \hline
\end{tabular}
}
\caption{Network scenarios with increasing complexity used to evaluate cognitive load.}
\label{tab:network_scenarios}
\end{table}

\section{Complementary Analysis}

\subsection{Hallucinations}
\label{subapp:halluciantions}
\begin{figure}[t] 
    \centering
    \begin{subfigure}[t]{0.48\linewidth} 
        \centering
        \includegraphics[width=0.75\textwidth]{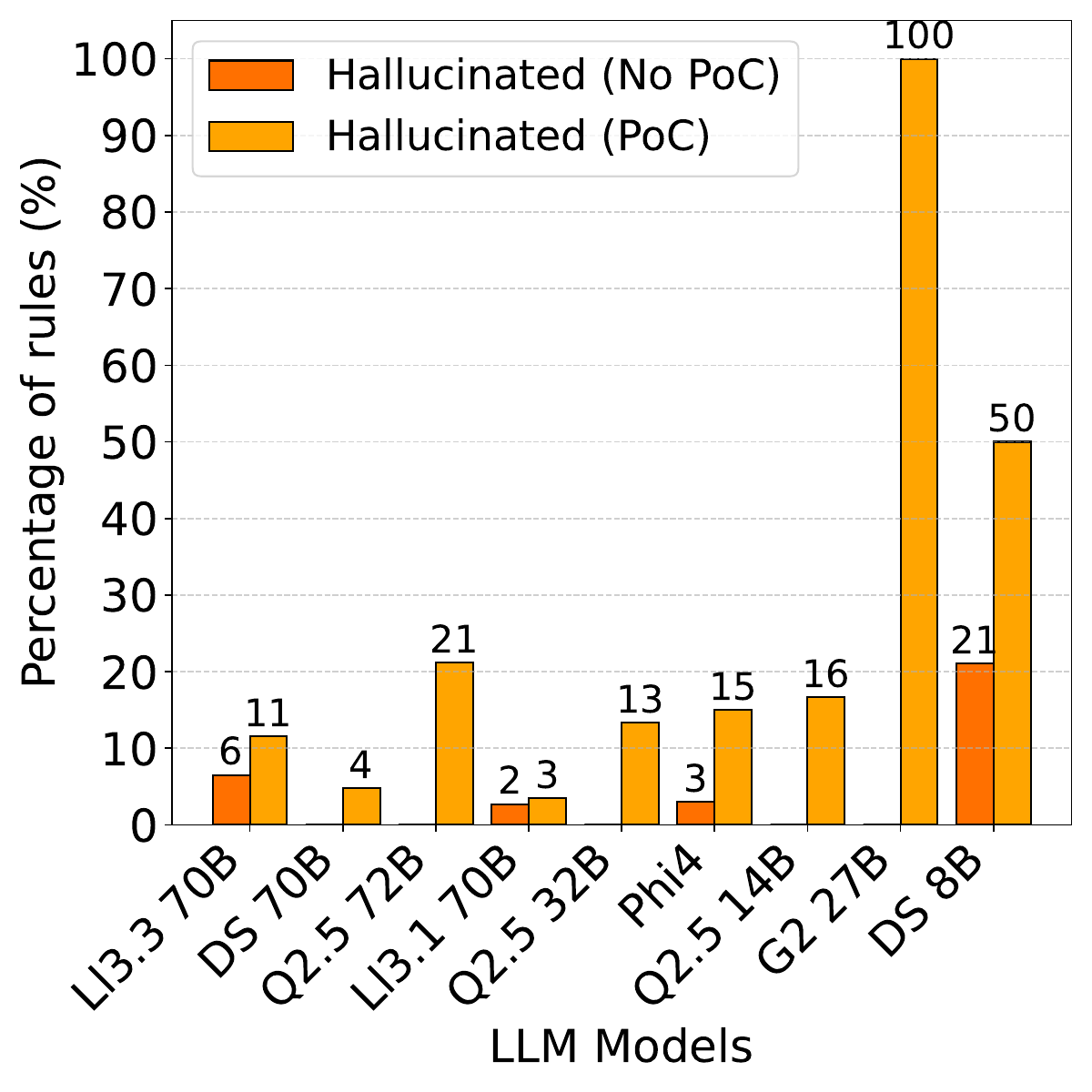} 
        \caption{Fraction of generated rules affected by easy-to-spot hallucinations for the best-performing LLM engines over CI$_{1}$ and CI$_{1}^P$.}
        \label{subfig:hallucinations_frequency}
    \end{subfigure}
    \hfill
    \begin{subfigure}[t]{0.48\linewidth}
        \centering
        \includegraphics[width=0.75\textwidth]{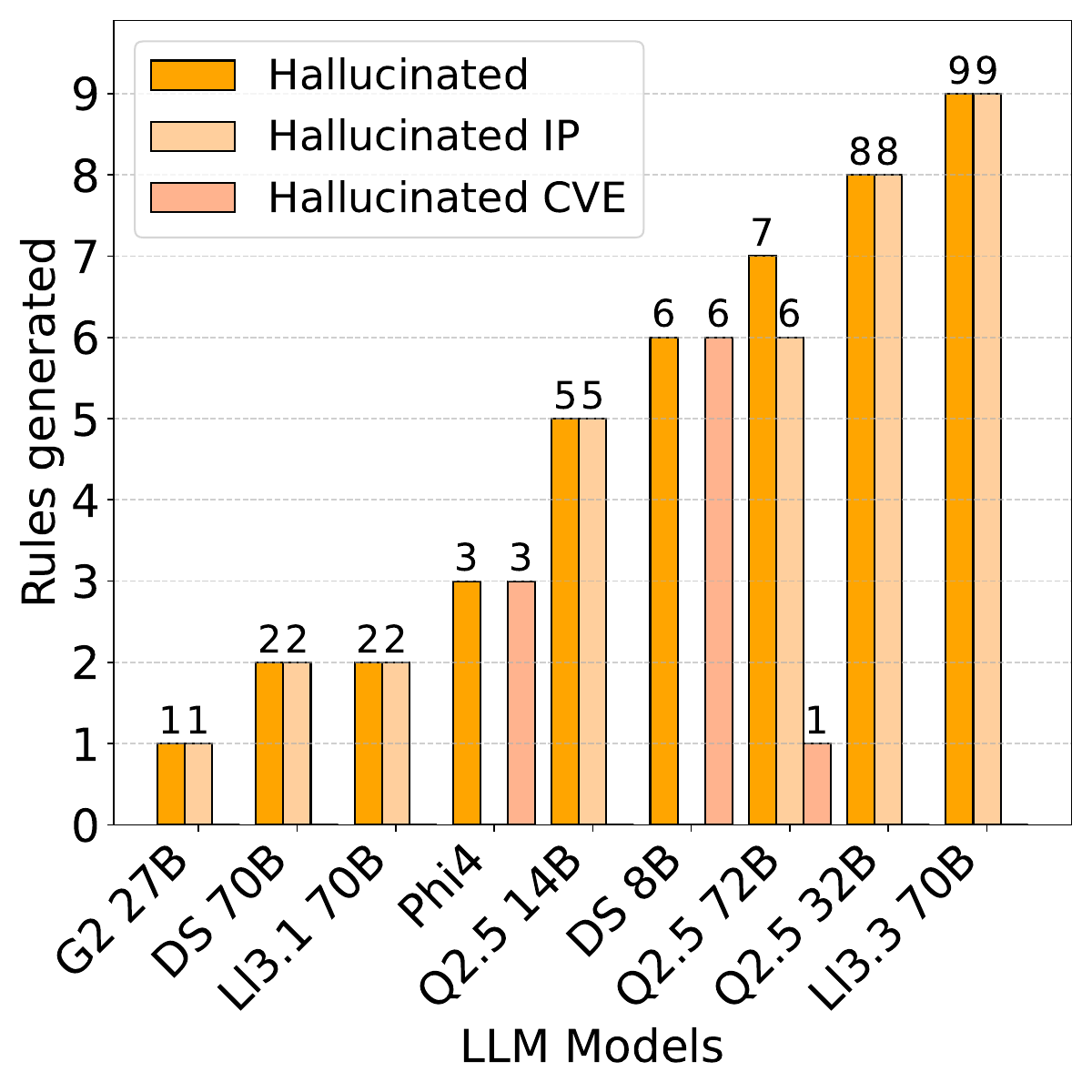}
        \caption{Distribution of hallucinations modality for the best-performing LLM engines over CI$_{1}^P$.}
        \label{subfig:hallucinations_type_1}
    \end{subfigure}
    \hfill
    \caption{Analysis of easy-to-spot hallucinations affecting generated rules. Most engines are robust against hallucinations whenever PoCs are not included. Most hallucinations are rooted on IP address management.}
    \label{fig:hallucinations1}
\end{figure}

LLMs are known to be prone to hallucinations \cite{ye2023cognitive,alansari2026large}, where the model confidently presents fabricated, factually incorrect, or nonsensical information.
Therefore, we here evaluate practically this issue and identify two easy-to-detect hallucinations that might affect the generated rules.
Specifically, we define as hallucinated rules, those rules:
\begin{inlinelist}
    \item that contain IP addresses that do not match any of the IP addresses given in the context (excluding common placeholders like HOMENET, EXTERNAL\_NET, etc.), and 
    \item whose description points to CVEs or PoCs that are not given in the context, nor that are logically linked to CVEs given in the context.
\end{inlinelist}
The ratio of hallucinated rules is quite small whenever we consider network scenarios where the PoC information is not available (see \Cref{fig:hallucinations1}), with almost no engine reaching more than 10\% of hallucinated rules and several achieving no hallucinations at all.
However, when PoCs are included, the hallucination fraction swiftly increases (as shown in~\Cref{subfig:hallucinations_frequency}).

The root cause of such an effect can be found in three concurrent phenomena:
\begin{inlinelist}
    \item providing the PoCs increases the contextual information that the LLM engine needs to process, thus increasing the risk of confusion whenever multiple CVEs and PoCs need to be take into account,
    \item some PoCs may contain examples of IP addresses or ports that are relevant for the CVE exploitation and which the model might reuse when writing the rule without adapting them to the given context, and
    \item while relying only on CVEs the agent can generate valid but generic rules -- as shown in \Cref{sssec:rule_quality_analysis} --, aggregating PoCs implicitly incentivizes the LLM engine to refine rules to target specificity, but LLMs struggle to translate the PoC information onto specific rule fields thus introducing hallucinations.
\end{inlinelist}
Therefore, the inclusion of PoC information correlates with a 10.10\% increase in hallucinations (rising from 2.81\% to 12.91\%), particularly in engines struggling with large context handling. Conversely, CVE-based prompts remain relatively stable with negligible hallucination rates.

%

We here also note that hallucinations may extend beyond the two modalities we considered in this subsection, including logical inconsistencies that are much more difficult to identify.
Our hallucination analysis is designed to point out the easy-to-spot mistakes, while we refer to \Cref{ssec:human_evaluation} for the identification of more ``sophisticated'' hallucinations that require human evaluation.

We further analyze the distribution of the two hallucination types -- i.e., non-existing IPs and unrelated CVEs -- in 
\Cref{subfig:hallucinations_type_1}.
Hallucinations are frequently rooted in the generation of rules containing unrelated IP addresses rather than CVEs.
We hypothesize that such a phenomenon is rooted on the memorization effect of LLMs \cite{tirumala2022memorization,leybzon2024learning,agiollo2024approximating}, where some models may have seen Snort rules during training and memorized part of their content. 
LLMs subject to memorization may reuse the information from previously seen rules, including IPs, without adapting them to the context, thus introducing hallucinations.
IPs are more common than CVEs in IDS rules available on the web, and thus, IP-based hallucinations are more common than CVE ones. 
%

\subsection{LLM Explanation}
\label{subapp:llm_explanation}

\Cref{tab:explanation_utility} reports participants’ perceptions of the utility of the explanations provided across the three scenarios. Overall, the results indicate a moderately positive assessment, with mean scores ranging from 3.00 to 3.10 on a five-point Likert scale.
For Scenario A (Q2), responses show a relatively broad distribution, with 50\% of participants selecting ratings in the positive range (4–5), while 40\% expressed lower to neutral perceptions (1-2). This scenario also exhibits the highest mean score (3.10), suggesting that explanations were perceived as slightly more useful in simpler contexts.
For Scenario B (Q4) and Scenario C (Q10), the distributions are nearly identical, with 70\% of participants selecting mid-to-high ratings (3–4) and no responses at the highest level (5). Both scenarios yield a mean of 3.00, indicating a stable but not strongly positive perception of explanation utility as scenario complexity increases.
Across all scenarios, responses tend to cluster around the central and moderately positive values (3–4), with relatively few extreme ratings. Notably, the absence or limited presence of the highest rating (5) suggests that while explanations were generally considered helpful, they may not have fully met expert expectations in terms of clarity, completeness, or actionable insight.
These results point to a consistent perception of moderate usefulness, with a slight decline in perceived utility as task complexity increases, potentially reflecting the growing cognitive demands associated with interpreting explanations in more complex scenarios.

\begin{table}[h!]
\centering
\scalebox{0.85}{
\begin{tabular}{l | ccccc | r}
\toprule
\textbf{Question} & \textbf{1} & \textbf{2} & \textbf{3} & \textbf{4} & \textbf{5} & \textbf{Mean} \\ 
\midrule
Q2 (Scenario A)  & 10\% & 30\% & 10\% & 40\% & 10\% & 3.10 \\
Q4 (Scenario B)  & 10\% & 20\% & 30\% & 40\% & 0\%  & 3.00 \\
Q10 (Scenario C) & 10\% & 20\% & 30\% & 40\% & 0\%  & 3.00 \\
\bottomrule
\end{tabular}}
\caption{Expert perception of explanation utility ($n=10$).}
\label{tab:explanation_utility}
\end{table}

\subsection{SUS-based Results}
\label{subapp:sus_results}
\Cref{tab:sus_raw_votes} presents the granular distribution of participant responses for each of the ten statements in the \gls{sus}, while \Cref{fig:sus_diverging_likert} depicts the diverging Likert scale plot. A total of \complete{} experts participated in this evaluation following the system interaction phase.

The responses are recorded on a 5-point Likert scale, where 1 indicates "Strongly Disagree" and 5 indicates "Strongly Agree." Note that the \gls{sus} consists of alternating positive and negative statements; for instance, a high score on Item 3 ("Ease of use") indicates a positive user experience, whereas a high score on Item 2 ("Complexity") would indicate a perceived usability barrier. These raw counts were used to calculate the final aggregate \gls{sus} score reported in the main body of the paper.


\begin{table}[ht]
\centering

\setlength{\tabcolsep}{10pt} 
\scalebox{0.85}{
\begin{tabular}{l | ccccc}
\toprule
\textbf{Question} & \textbf{1} & \textbf{2} & \textbf{3} & \textbf{4} & \textbf{5} \\ 
\midrule
Q1 (Frequent use)    & 0 & 1 & 5 & 4 & 0 \\
Q2 (Complexity*)     & 2 & 6 & 2 & 0 & 0 \\
Q3 (Ease of use)     & 0 & 0 & 3 & 7 & 0 \\
Q4 (Tech support*)   & 4 & 2 & 3 & 1 & 0 \\
Q5 (Well integrated) & 1 & 1 & 6 & 1 & 1 \\
Q6 (Inconsistency*)  & 2 & 5 & 2 & 1 & 0 \\
Q7 (Learn quickly)   & 0 & 1 & 1 & 5 & 3 \\
Q8 (Cumbersome*)     & 3 & 4 & 2 & 1 & 0 \\
Q9 (Confidence)      & 0 & 2 & 4 & 4 & 0 \\
Q10 (Pre-learning*)  & 3 & 4 & 3 & 0 & 0 \\
\bottomrule
\multicolumn{6}{l}{\small *Negative items; 1 = Strongly Disagree, 5 = Strongly Agree.}
\end{tabular}}
\caption{Distribution of participant responses for each SUS statement ($n=10$).}
\label{tab:sus_raw_votes}
\end{table}




\subsection{Statement-based Results}
\label{subapp:statement_results}
%
%

\Cref{tab:statement_raw_votes} reports the distribution of expert responses for the 11 statement-based questions used to assess the perceived professional viability of the proposed \gls{llm}-based IDS rule generation system. 

The statements cover both functional aspects of the system, such as the usefulness of the PoC Finder, Snort Rule Tester, and Interactive Generator, as well as trust-related aspects, including confidence in context-aware generation, willingness to rely on automated deployment, and the perceived need for human supervision.

Overall, the responses suggest a moderately positive perception of the system. The highest agreement is observed for the PoC Finder and Snort Rule Tester, suggesting that experts particularly valued components that provide supporting evidence and validation before deployment. In contrast, the lowest score is associated with the \textit{Suggest-and-Deploy} mode, showing limited trust in unsupervised automatic deployment. 
This result is consistent with the strong perceived importance of maintaining a human-in-the-loop process, especially in operational security environments where incorrect IDS rules may lead to false positives, missed detections, or deployment risks. 
The relatively moderate scores for time-to-deploy reduction and comparison with traditional scanners further suggest that experts view the system as a useful aid rather than a complete replacement for established workflows.

\begin{table}[ht]
\centering
\scalebox{0.85}{
    \begin{tabular}{l | ccccc | r}
    \toprule
    \textbf{Evaluation Statement} & \textbf{1} & \textbf{2} & \textbf{3} & \textbf{4} & \textbf{5} & \textbf{Mean} \\ 
    \midrule
    Q1. Confidence in LLM context-awareness  & 0 & 2 & 2 & 6 & 0 & 3.40 \\
    Q2. Impact of PoC data on confidence    & 1 & 1 & 3 & 3 & 2 & 3.40 \\
    Q3. Expected reduction in time-to-deploy & 0 & 3 & 5 & 1 & 1 & 3.00 \\
    Q4. Value of Topology Visualizer        & 3 & 2 & 2 & 3 & 0 & 2.50 \\
    Q5. Value of PoC Finder                 & 0 & 1 & 3 & 4 & 2 & 3.70 \\
    Q6. Value of Snort Rule Tester          & 0 & 1 & 3 & 5 & 1 & 3.60 \\
    Q7. Value of Interactive Generator      & 1 & 1 & 2 & 5 & 1 & 3.40 \\
    Q8. Trust in ``Suggest-and-Deploy'' mode  & 4 & 3 & 2 & 1 & 0 & 2.00 \\
    Q9. Human-in-the-loop necessity         & 0 & 1 & 6 & 3 & 0 & 4.30 \\
    Q10. Quality vs. Traditional Scanners    & 0 & 1 & 6 & 3 & 0 & 3.20 \\
    Q11. Difficulty of modifying rules* & 2 & 3 & 2 & 3 & 0 & 2.60 \\
    \bottomrule
    \multicolumn{7}{l}{\small *Lower score indicates easier modification.}
    \end{tabular}
}
\caption{Distribution of expert responses regarding system utility and trust ($n=10$).}
\label{tab:statement_raw_votes}
\end{table}



\subsection{Syntactic Validity of Generated Snort Rules Across Network Scenarios}
\label{app:syntax_validity_all_networks}

\begin{figure*}[ht] 
    \centering
    \begin{subfigure}[t]{0.3\linewidth} 
        \centering
        \includegraphics[width=0.85\textwidth]{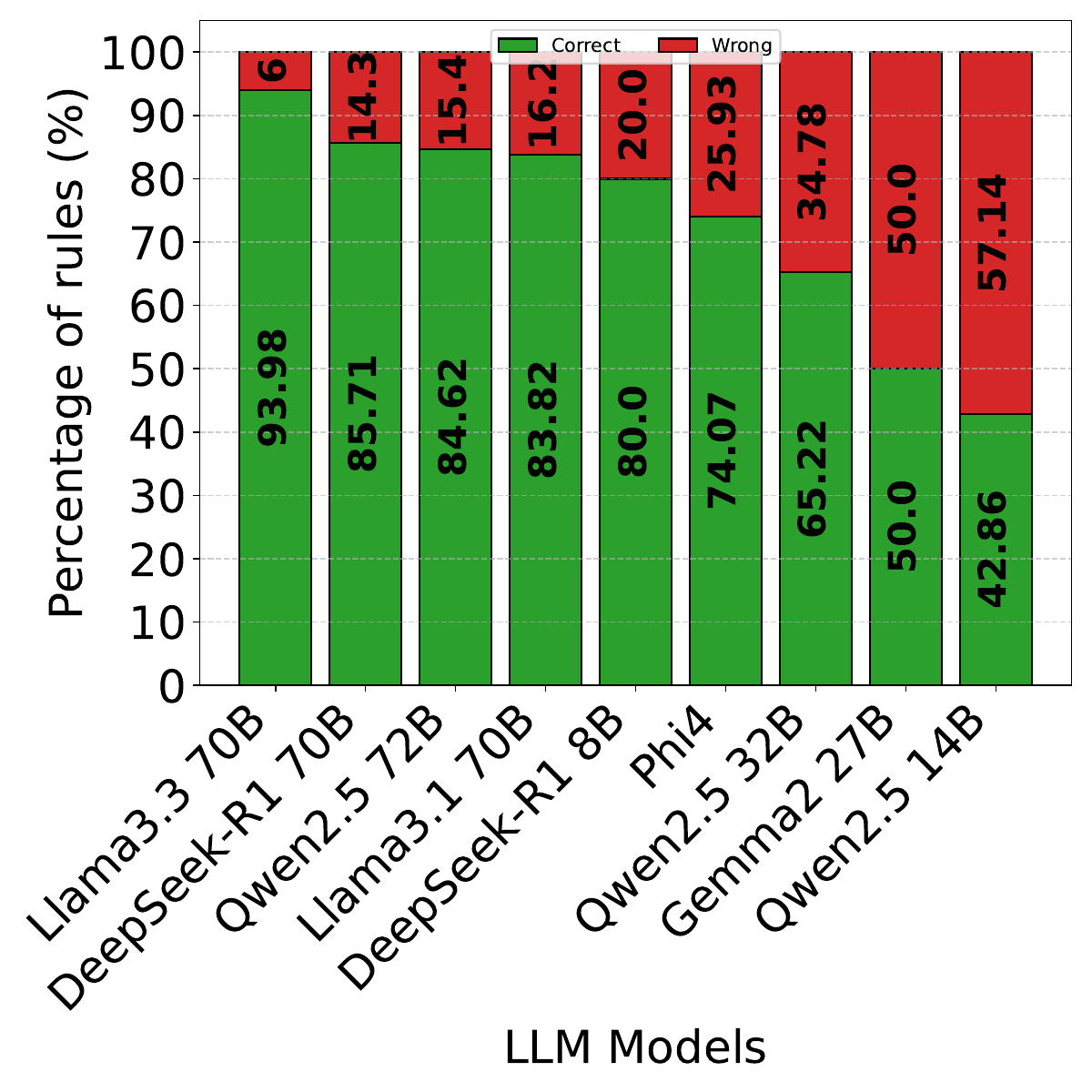} 
        \caption{Fraction of syntactically valid generated rules for each LLM engine generating more than one rule over the CIC-IDS with 1 CVE per device and PoCs scenario.}
        \label{subfig:syntax_correct_cicids1cvepoc_percentage}
    \end{subfigure}
    \hfill
    \begin{subfigure}[t]{0.3\linewidth}
        \centering
        \includegraphics[width=0.85\textwidth]{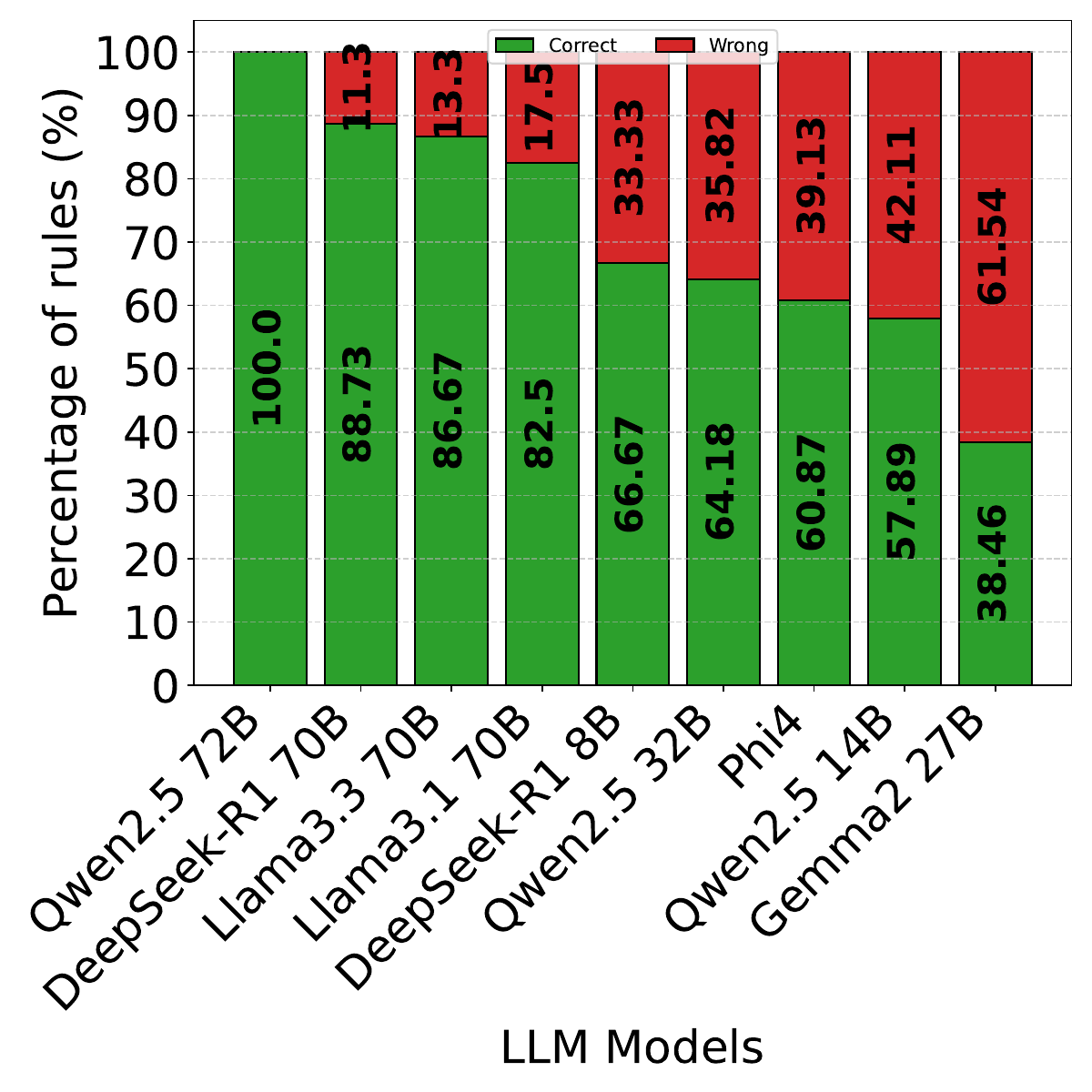}
        \caption{Fraction of syntactically valid generated rules for each LLM engine generating more than one rule over the CIC-IDS with 2 CVE per device and no PoCs scenario.}
        \label{subfig:syntax_correct_cicids2cve_percentage}
    \end{subfigure}
    \hfill
    \begin{subfigure}[t]{0.3\linewidth}
        \centering
        \includegraphics[width=0.85\textwidth]{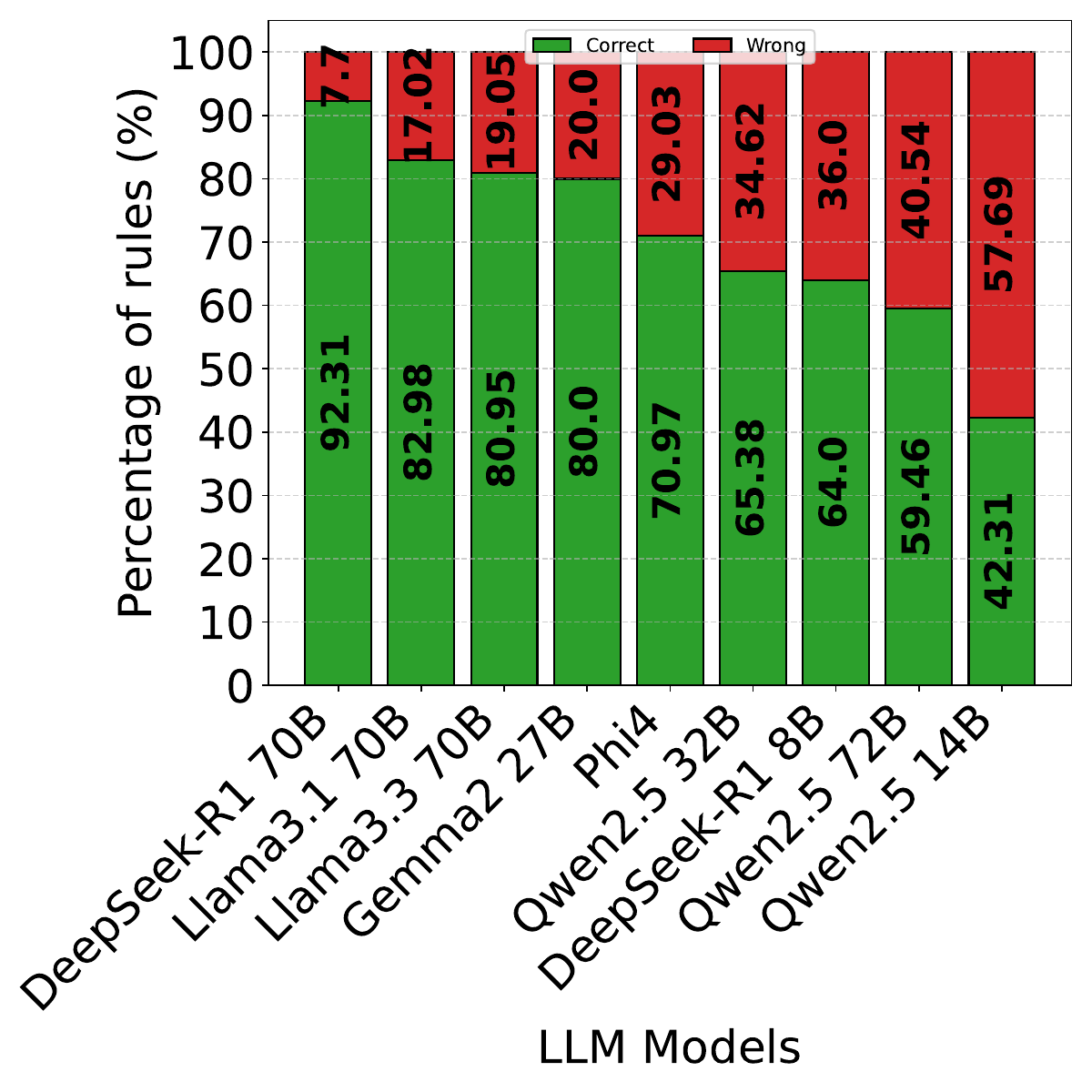}
        \caption{Fraction of syntactically valid generated rules for each LLM engine generating more than one rule over the CIC-IDS with 2 CVE per device and PoCs scenario.}
        \label{subfig:syntax_correct_cicids2cvepoc_percentage}
    \end{subfigure}
    \hfill
    \caption{Fraction of syntactically valid generated rules for each LLM engine generating more than one rule across the remaining CIC-IDS scenarios, complementing Figure~\ref{subfig:syntax_correct_cicids1cve_percentage}.}
    \label{fig:syntax_correct_all_topologies_percentage}
\end{figure*}


For each scenario, we compile the Snort-looking rules produced by each \gls{llm} engine and measure the fraction of rules that do not trigger compilation errors.  These plots therefore capture whether each model is able to generate rules that are structurally compatible with Snort syntax, independently of whether the rule is semantically effective for detecting the intended attack.

\Cref{fig:syntax_correct_all_topologies_percentage} complements the analysis in Section~\ref{ssec:statistical_analysis} by showing the percentage of syntactically valid and invalid rules for three additional CIC-IDS configurations. The first scenario considers one CVE per device with PoC information available, while the second and third scenarios consider two CVEs per device without and with PoC information, respectively. Across these settings, larger models generally achieve higher syntactic validity, although the ranking is not perfectly stable across scenarios.

Across the three scenarios, DeepSeek-R1 70B, Llama3.3 70B, Llama3.1 70B, and Qwen2.5 72B consistently emerge among the best-performing models, each maintaining syntactic validity above 80\%. This suggests that higher-capacity models are generally more reliable at producing Snort rules that conform to the expected syntax. At the same time, the results show that scenario complexity affects models differently. Although the strongest models remain relatively robust, syntactic correctness is not guaranteed across all engines or network configurations, especially for smaller or less consistent models.

\subsection{Rule Correction Effort Across Additional Configurations}
\label{app:avg_modifications_appendix}

\begin{figure*}[ht] 
    \centering
    \begin{subfigure}[t]{0.3\linewidth} 
        \centering
        \includegraphics[width=0.85\textwidth]{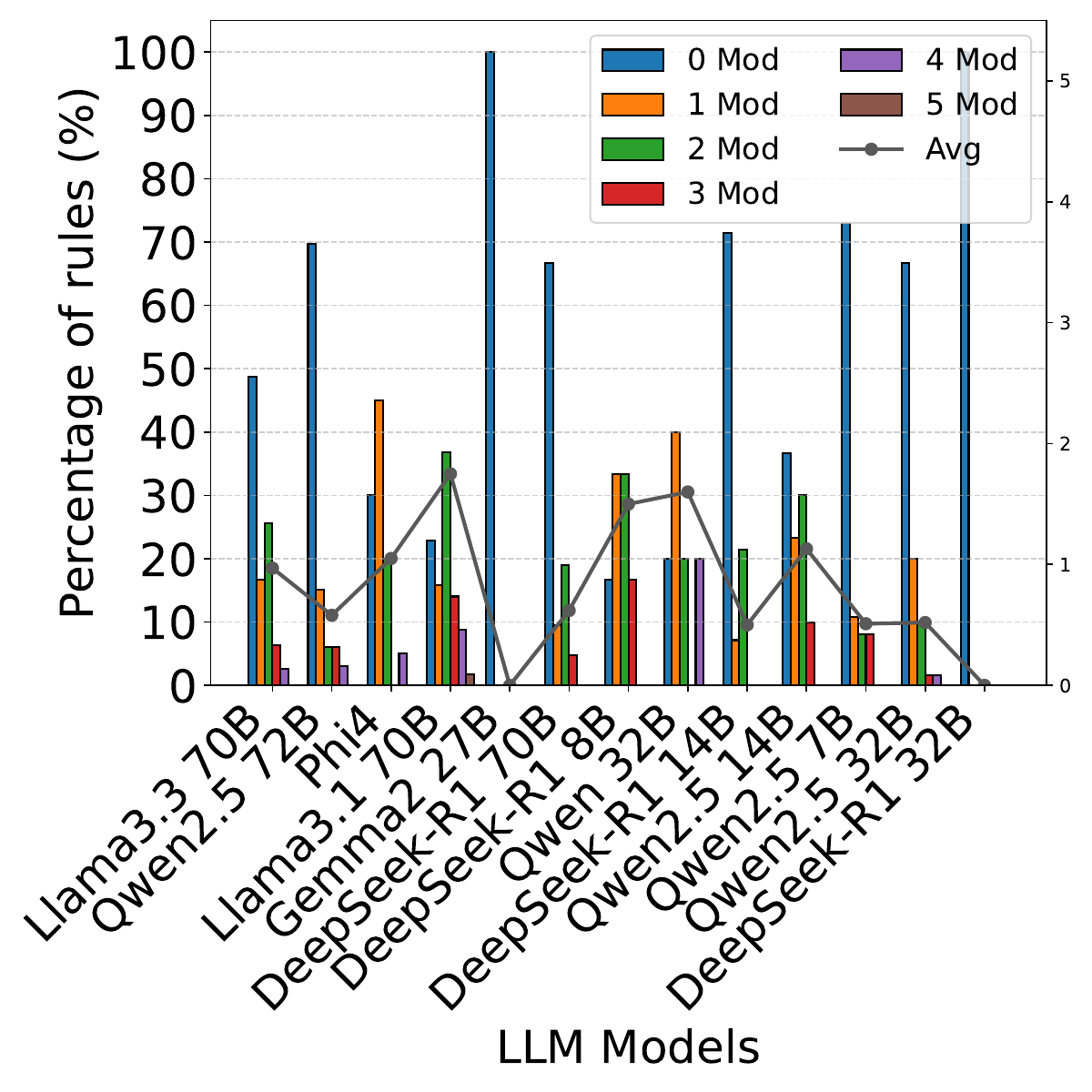} 
        \caption{Average number of corrections required to generate a rule for each LLM engine generating more than one rule over the CIC-IDS with 1 CVE per device and PoCs scenario. }
        \label{subfig:avg_modifications_1_poc}
    \end{subfigure}
    \hfill
    \begin{subfigure}[t]{0.3\linewidth}
        \centering
        \includegraphics[width=0.85\textwidth]{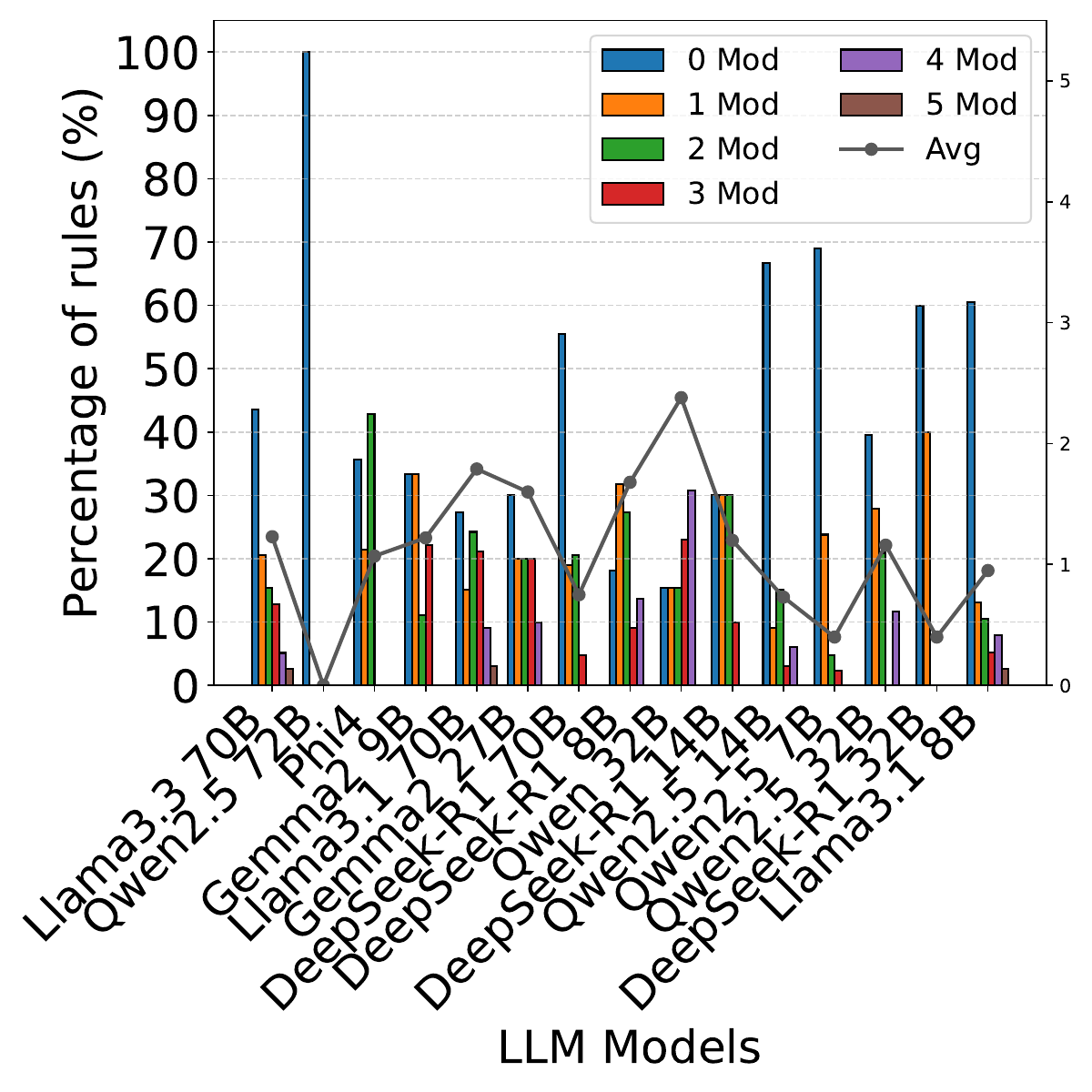}
        \caption{Average number of corrections required to generate a rule for each LLM engine generating more than one rule over the CIC-IDS with 2 CVE per device and no PoCs scenario.}
        \label{subfig:avg_modifications_2}
    \end{subfigure}
    \hfill
    \begin{subfigure}[t]{0.3\linewidth}
        \centering
        \includegraphics[width=0.85\textwidth]{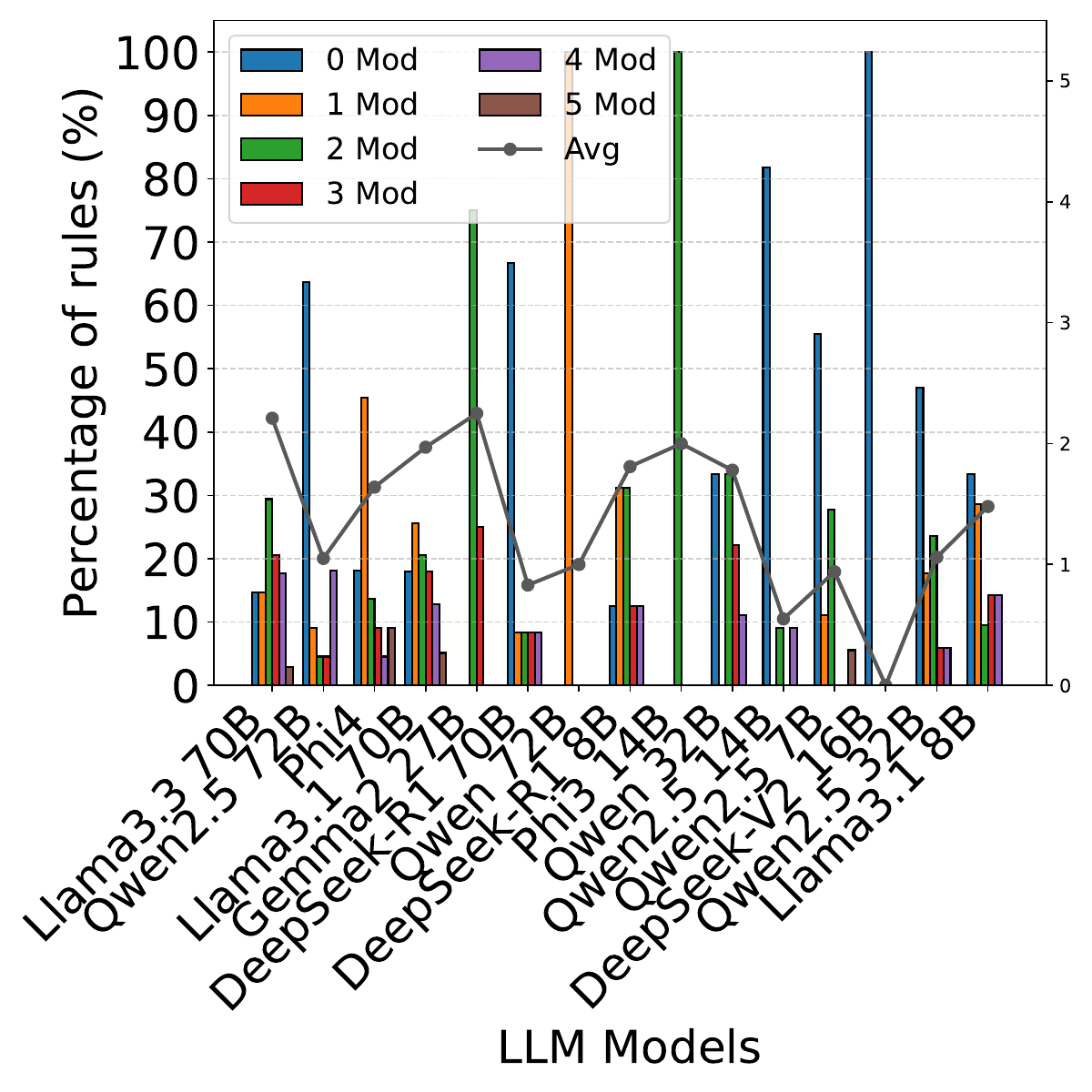}
        \caption{Average number of corrections required to generate a rule for each LLM engine generating more than one rule over the CIC-IDS with 2 CVE per device and PoCs scenario. }
        \label{subfig:avg_modifications_2_poc}
    \end{subfigure}
    \hfill
    \caption{Average number of corrections required to generate a correct rule for the remaining CIC-IDS scenarios, complementing Fig.~\ref{subfig:avg_modifications} with the corresponding results for the additional topology configurations.}
    \label{fig:avg_modifications_all_topologies_percentage}
\end{figure*}


\Cref{fig:avg_modifications_all_topologies_percentage} complements the correction-effort analysis in Section~\ref{ssec:statistical_analysis} by reporting the results for the remaining CIC-IDS configurations. For each scenario, the figures show the average number of corrections required by each \gls{llm} engine, considering only models that generated more than one rule. This metric quantifies the amount of manual or iterative refinement needed after the initial generation step. 
Lower values indicate that a model tends to produce syntactically valid Snort rules with limited correction effort, whereas higher values suggest that more substantial intervention is required before the rules can be compiled and deployed.

By applying the same metric across all additional scenarios, the plots enable a direct comparison of correction effort under different network configurations. They therefore capture the amount of refinement required to obtain a syntactically valid outcome.

Overall, these results provide insight into the practical usability of each model in an operational workflow. Models requiring fewer corrections are better suited to reducing analyst workload, whereas models requiring repeated corrections may still impose a significant manual burden, even when they eventually produce valid rules.

\subsection{Complementary Rule-Header Analysis Across CIC-IDS Configurations}
\label{app:header_distribution_appendix}

\begin{figure*}[ht] 
    \centering
    \begin{subfigure}[t]{0.3\linewidth} 
        \centering
        \includegraphics[width=0.85\textwidth]{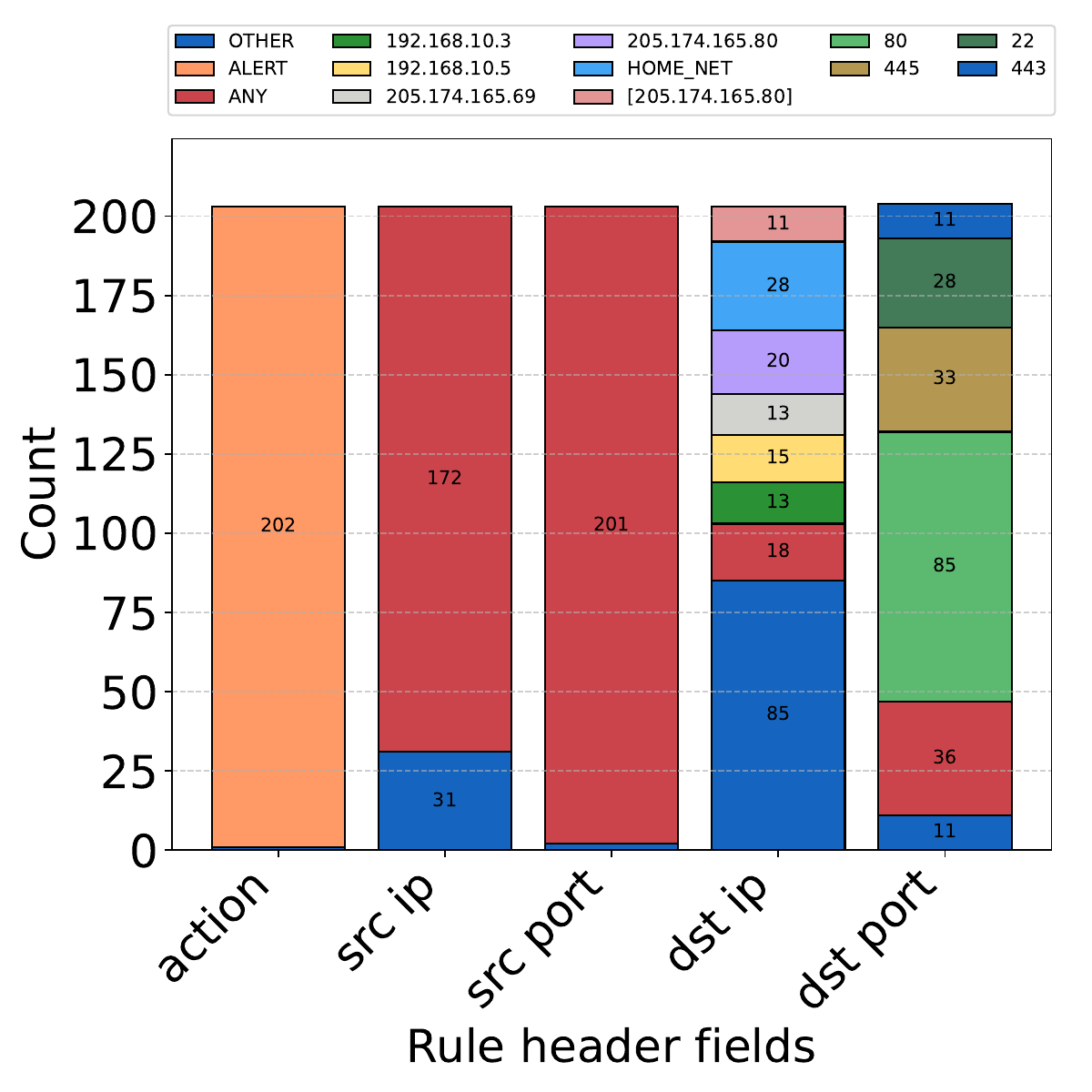} 
        \caption{Distribution of fields used across the headers of the generated rules over the CIC-IDS with 1 CVE per device and PoCs (aggregate metric across LLM engine and network scenario). }
        \label{subfig:header_distribution_1_poc}
    \end{subfigure}
    \hfill
    \begin{subfigure}[t]{0.3\linewidth}
        \centering
        \includegraphics[width=0.85\textwidth]{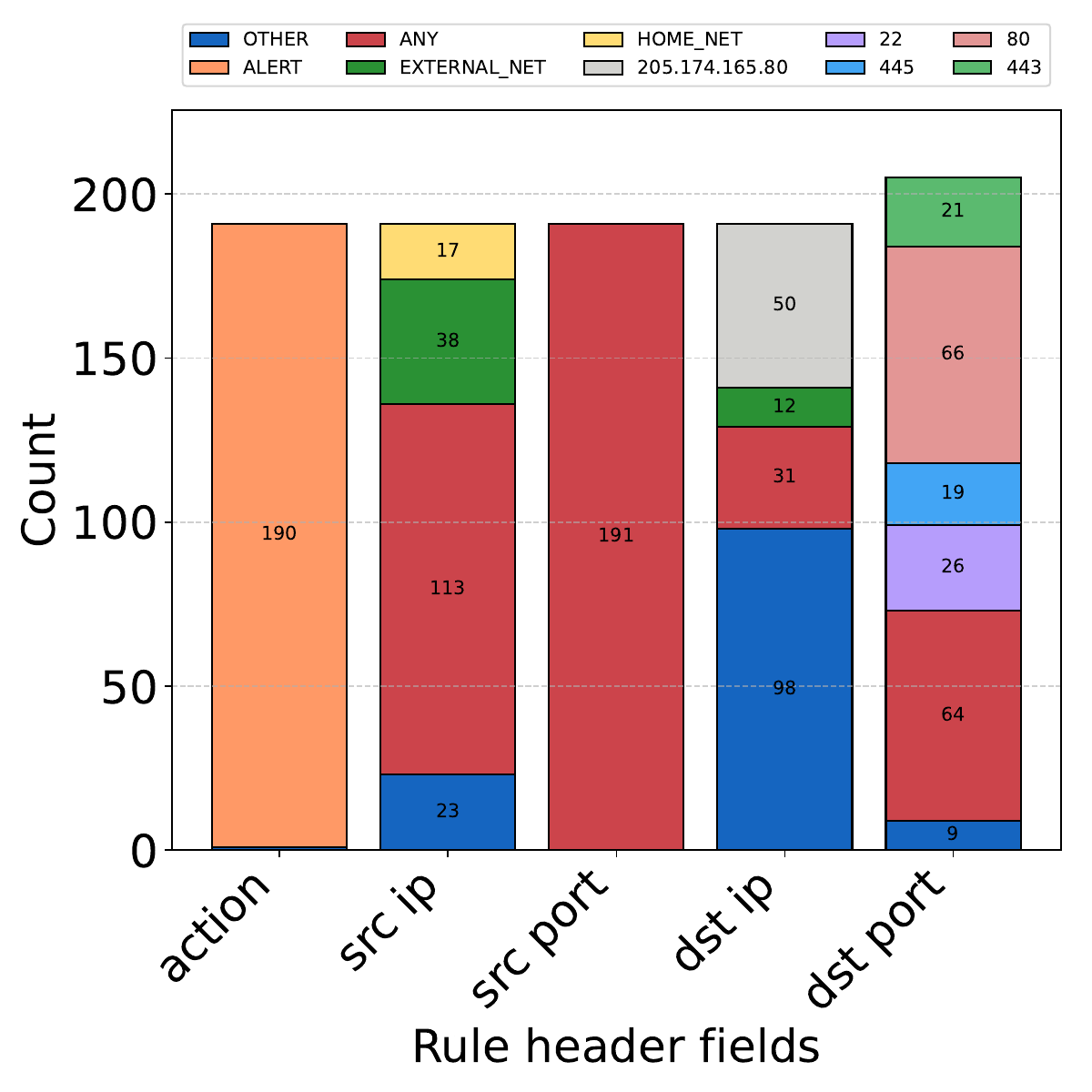}
        \caption{Distribution of fields used across the headers of the generated rules over the CIC-IDS with 2 CVE per device and no PoCs (aggregate metric across LLM engine and network scenario). }
        \label{subfig:header_distribution_2}
    \end{subfigure}
    \hfill
    \begin{subfigure}[t]{0.3\linewidth}
        \centering
        \includegraphics[width=0.85\textwidth]{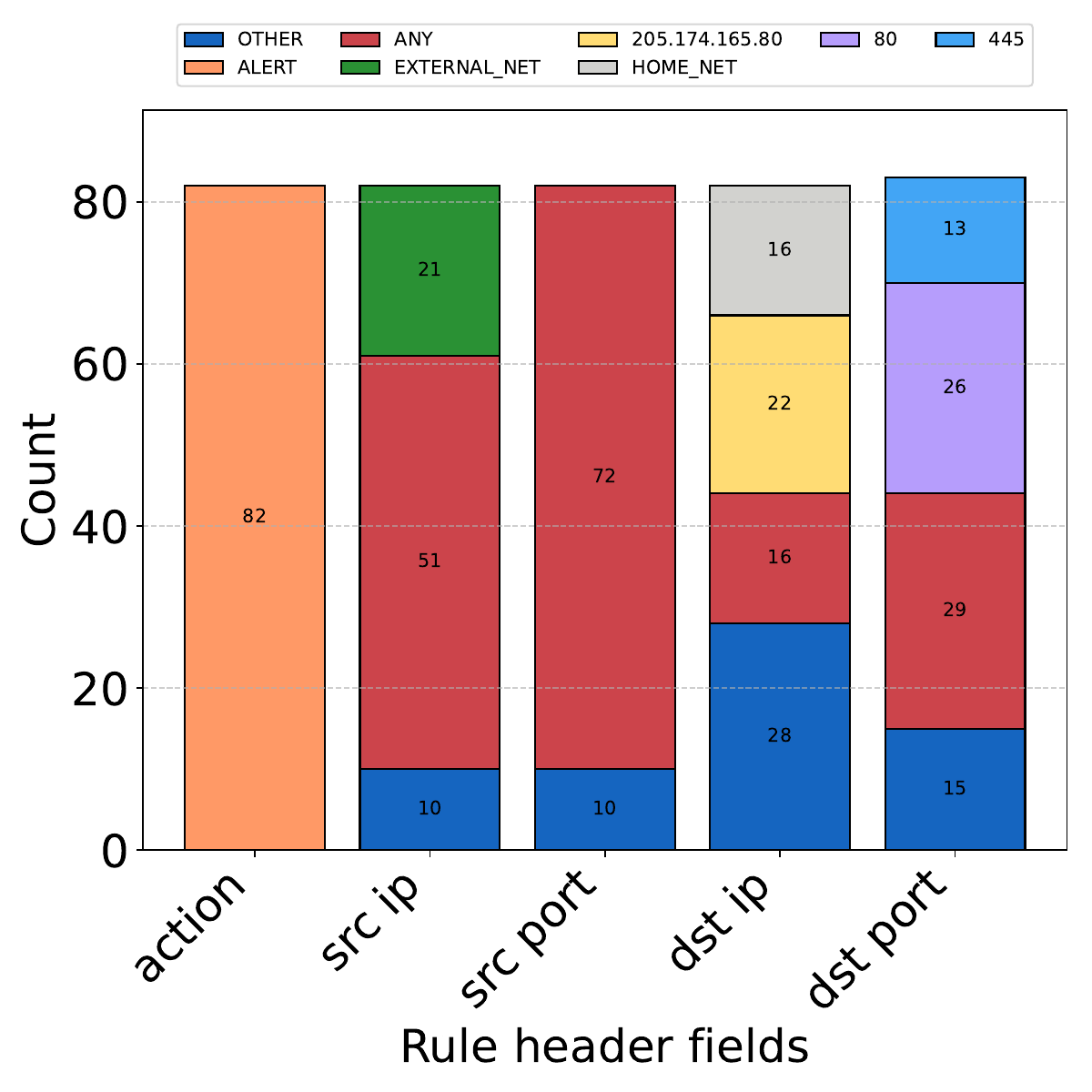}
        \caption{Distribution of fields used across the headers of the generated rules over the CIC-IDS with 2 CVE per device and PoCs (aggregate metric across LLM engine and network scenario). }
        \label{subfig:header_distribution_2_poc}
    \end{subfigure}
    \hfill
    \caption{Distribution of fields used across generated rule headers for the remaining CIC-IDS scenarios with aggregate results for the additional topology configurations, complementing Fig.~\ref{subfig:header_distribution}.}
    \label{fig:header_distribution_all_topologies_percentage}
\end{figure*}


\Cref{fig:header_distribution_all_topologies_percentage} complements the semantic rule-quality analysis presented in Section~\ref{sssec:rule_quality_analysis} by reporting the corresponding header-field distributions for the remaining CIC-IDS configurations. 
For each scenario, the plots aggregate the fields used in the headers of the correctly generated Snort rules across all considered \gls{llm} engines.

These distributions provide insight into the structural and semantic choices made by the models when defining rule headers, including the selected protocols, source and destination addresses,  and ports. By applying the same aggregate metric across the additional scenarios, these plots enable a direct comparison of rule-header composition under different network configurations, including settings with one or two CVEs per device and with or without PoC information. 
They therefore complement the main semantic analysis by showing whether the models preserve consistent header-generation behavior across increasingly complex scenarios.

Overall, these additional results help assess whether the generated rules are not only syntactically valid, but also structurally meaningful from an IDS perspective. 
Consistent and appropriate header-field usage suggests that the models are better able to encode the relevant network context, whereas irregular or overly generic header patterns may indicate weaker semantic alignment with the target scenario.

\end{document}